\pdfoutput=1
\documentclass{article}

\usepackage{arxiv}

\usepackage[utf8]{inputenc}
\usepackage[T1]{fontenc}
\usepackage{hyperref}
\usepackage{url}
\usepackage{booktabs}
\usepackage{amsfonts}
\usepackage{amssymb}
\usepackage{amsmath}
\usepackage{nicefrac}
\usepackage{microtype}
\usepackage{cleveref}
\usepackage{graphicx}
\usepackage{natbib}
\usepackage{doi}
\usepackage{longtable}
\usepackage{array}
\usepackage{multirow}
\usepackage{textcomp}
\usepackage{xcolor}
\usepackage{enumitem}
\usepackage{needspace}

\providecommand{\tightlist}{%
  \setlength{\itemsep}{0pt}\setlength{\parskip}{0pt}}

\title{Governance by Design: A Parsonian Institutional Architecture for Internet-Wide Agent Societies}

\author{
  Anbang Ruan \\
  NetX Foundation \\
  \texttt{ruan@netx.foundation}
}

\date{}

\renewcommand{\headeright}{A Preprint}
\renewcommand{\undertitle}{A Preprint}
\renewcommand{\shorttitle}{AGIL Theory for Agent Societies}

\hypersetup{
pdftitle={Governance by Design: A Parsonian Institutional Architecture for Internet-Wide Agent Societies},
pdfsubject={cs.CY, cs.MA, cs.AI},
pdfauthor={Anbang Ruan},
pdfkeywords={multi-agent systems, AI governance, institutional design, structural-functionalism, AGIL, agent societies},
colorlinks=true,
linkcolor={blue!70!black},
citecolor={green!50!black},
urlcolor={blue!80!black},
}

\begin{document}
\maketitle

\begin{abstract}
The dominant paradigm of local multi-agent systems---orchestrated, enterprise-bounded pipelines---is being superseded by internet-wide agent societies in which autonomous agents discover each other through open registries, interact without central orchestrators, and generate emergent social behaviors. The OpenClaw ecosystem instantiates this shift: 250,000+ GitHub stars (as of March 2026), 2M+ monthly active users, 10,700+ skills published to the community registry (cumulative), and 770,000+ registered agents on the Moltbook social network. This paper argues that governing such societies requires an institutional design methodology, not merely risk enumeration or process compliance. We apply Talcott Parsons' AGIL framework---identifying four functional imperatives (Adaptation, Goal Attainment, Integration, Latency) every viable social system must satisfy---to derive a prescriptive sixteen-cell institutional architecture for internet-wide agent governance. Applying this architecture diagnostically to the OpenClaw ecosystem through a recursive sub-function analysis (sixty-four binary sub-functions across sixteen cells), we identify at most 19\% sub-function coverage (sensitivity range 17--30\%) based on publicly documented evidence---representing potential rather than operative institutional capacity, since zero inter-cell coordination (I-sub = 0\%) prevents existing infrastructure from participating in inter-pillar interchange. The ecosystem has developed technical infrastructure but almost no active governance, zero inter-cell coordination, and zero normative grounding, with the Fiduciary (Latency) and Political (Goal Attainment) pillars most severely underserved. A complementary interchange media assessment finds zero of twelve inter-pillar media pathways functional, confirming that the ecosystem is not yet a social system in the full Parsonian sense; institutions exist but are functionally inert without the media circulation that connects them. We extend the diagnostic to the broader agent-native protocol infrastructure (x402, MCP, A2A, ANP, ERC-8004, agentic wallets, labor markets, and proto-governance tools), finding that dozens of independent development teams building exclusively for autonomous agents reproduce the same structural pattern---confirming that the L-pillar gap is not an artifact of the OpenClaw ecosystem's immaturity but a structural feature of market-driven agent infrastructure development. The ecosystem's current simplicity represents a governance advantage: institutional design is most effective before social patterns calcify. We conclude with a prioritized roadmap for the missing governance infrastructure.
\end{abstract}

\keywords{multi-agent systems \and AI governance \and institutional design \and structural-functionalism \and AGIL \and agent societies}

\section*{Key Concepts for Non-Specialist Readers}\label{key-concepts-for-non-specialist-readers}

\textbf{AGIL} --- An acronym for four functional imperatives that Parsons argues every viable social system must satisfy: \emph{Adaptation} (securing resources), \emph{Goal Attainment} (defining and pursuing collective objectives), \emph{Integration} (coordinating members and managing conflict), and \emph{Latency/Pattern Maintenance} (preserving core values). The framework generates a 4$\times$4 matrix of sixteen institutional cells by cross-referencing each function as both a source and a target domain.

\textbf{Cybernetic Hierarchy} --- The ordering principle of the AGIL framework: systems rich in \emph{information} (L $\rightarrow$ I $\rightarrow$ G) govern systems rich in \emph{energy} (A). As an intuition pump, consider a thermostat: the temperature setting (L, value-level) defines what "warm enough" means; the thermostat controller (G, governance-level) activates or deactivates the heating mechanism; and the furnace (A, economic-level) provides the energy that actually heats the room. The value-setting layer is energy-poor but information-rich; it governs the energy-rich furnace not by overpowering it but by supplying the information that directs its operation. In agent societies, human value commitments (L) constrain governance rules (G), which constrain what economic activity (A) is authorized---an information-over-energy architecture enforced, in blockchain contexts, through smart contract logic rather than communicative persuasion.

\textbf{Generalized Symbolic Media} --- The "currencies" that circulate between AGIL subsystems: \emph{money} (A$\leftrightarrow$ other), \emph{power} (G$\leftrightarrow$ other), \emph{influence} (I$\leftrightarrow$ other), and \emph{value-commitment} (L$\leftrightarrow$ other). Media enable subsystems to exchange inputs and outputs without direct physical interaction. They flow between adjacent subsystems following adjacency in functional space, not arbitrary cross-boundary flows: A and G are adjacent; so are G and I; and I and L---but A and L are not directly adjacent, which is why A$\leftrightarrow$L exchanges are mediated through intermediate pillars.

\textbf{Pattern Variables} --- Binary contrasts that characterize role-orientations in any social system (e.g., universalism vs. particularism; specificity vs. diffuseness). They provide the theoretical basis for deriving the institutional properties of each of the sixteen cells.

\textbf{Interpenetration} --- Parsons' term for the structured zones that adjacent subsystems share. Three interpenetration boundaries are critical for agent governance: \emph{institutionalization} (Cultural $\leftrightarrow$ Social: human values become governance rules), \emph{internalization} (Social $\leftrightarrow$ Personality: governance rules become agent behavioral dispositions), and \emph{learned capability} (Personality $\leftrightarrow$ Behavioral Organism: agent dispositions draw upon computational capabilities).

\needspace{4\baselineskip}
\section{Introduction}\label{introduction}

The deployment of AI agents is undergoing a paradigm shift whose governance implications have not been adequately theorized. For the past several years, multi-agent AI systems have been primarily local phenomena: designed, orchestrated pipelines in which human engineers define agent roles, task sequences, and tool access. Systems such as CrewAI, AutoGen, and LangGraph exemplify this model---they are powerful, but they are \emph{engineered} multi-agent systems operating within enterprise boundaries, answerable to centralized orchestrators, and designed with specific objectives in mind. Governance frameworks built for this model---the NIST AI RMF, the EU AI Act, the Berkeley CLTC Agentic AI Profile, Singapore's Model AI Governance Framework---are broadly adequate for its challenges, addressing the production-quality, administrative, and human-oversight requirements of enterprise deployments. We develop a precise characterization of what makes them adequate for local MAS but structurally insufficient for internet-wide agent societies in \S{}2.3.

But something different is now emerging on the internet. Since the public release of OpenClaw in late 2025, a qualitatively distinct phenomenon has become visible: autonomous AI agents living independently on the internet, publishing capabilities to open registries (ClawHub), discovering each other through community marketplaces, interacting across organizational and jurisdictional boundaries, and generating social behaviors that were never explicitly programmed. Researchers studying Moltbook---a Reddit-style agent social network with 770,000+ registered agents (of whom approximately 90,704, or 12\%, were active during the analysis period; Yee \& Sharma, 2026)---have documented role specialization, emergent norm formation, and nascent cooperative task resolution~\citep{yee_2026}. The Berkeley CLTC's Agentic AI Risk-Management Standards Profile characterizes this shift: agentic AI now ranges from "narrowly scoped, single-agent systems to highly autonomous, multi-agent architectures operating in complex environments," requiring governance proportionate to their autonomy~\citep{berkeley_2026}. This is the global multi-agent society---emergent, not designed. The urgency of governing such systems is underscored by growing warnings about extreme AI risks from the scientific community~\citep{bengio_2024} and the International AI Safety Report's documentation of fragmented governance coordination across jurisdictions~\citep{bengio_2026}.

The protocol infrastructure enabling this society is rapidly consolidating. Google's Agent-to-Agent (A2A) Protocol, released in April 2025 with 50+ partners, provides an open standard for agent-to-agent communication and discovery. The Model Context Protocol (MCP) standardizes agent access to tools and data. The Agent Network Protocol (ANP) enables internet-wide agent collaboration using semantic discovery and decentralized identifiers (DIDs). The x402 payment protocol---backed by Coinbase, Cloudflare, Google, Visa, Circle, AWS, and Stripe---enables micropayment settlement for agent-to-agent service exchange (settlement finality varies by chain). Web3 integration is deepening: the CoinFello+MetaMask OpenClaw skill enables Moltbook agents to perform autonomous on-chain transactions via ERC-4337 smart accounts and ERC-7710 delegations; Bank of AI provides bank-grade financial infrastructure for agents~\citep{techflow_2026}. The pieces of a global agent economy are already in place.

This emergent reality poses a governance challenge that existing frameworks---designed for orchestrated local MAS---cannot meet. The governance question for a global agent society is not "how do we manage risk in a designed pipeline?" but "what institutional infrastructure does an internet-wide agent society need to maintain social order and preserve human oversight?" This is an institutional design question, not merely a risk management question. The distinction is not that existing frameworks are wrong---it is that they are structurally insufficient: they provide necessary components of specific AGIL cells but do not constitute a governance \emph{architecture}. They lack the inter-pillar coordination mechanisms and normative hierarchy that a functional social system requires.

This paper addresses it through three contributions:

\textbf{Contribution 1} proposes a prescriptive sixteen-cell institutional architecture---derived from Parsons' AGIL framework---specifying the social institutions that an internet-wide agent society should develop to maintain social order and remain aligned with human values, with a minimum requirement that all four AGIL pillars be institutionally represented. This is the paper's central contribution.

\textbf{Contribution 2} applies this architecture diagnostically to the OpenClaw ecosystem through a recursive sub-function analysis---decomposing each of the sixteen cells into four internal AGIL sub-functions (sixty-four binary indicators)---to map existing infrastructure and systematically identify structural gaps at a resolution impossible with aggregate assessments.

\textbf{Contribution 3} presents a prioritized institutional roadmap for the missing governance infrastructure, grounded in the argument that the OpenClaw ecosystem's current relative simplicity is a governance advantage to be exploited before the society matures.

The through-line uniting all three contributions is alignment: these institutions are not merely about social order for its own sake, but about ensuring that as autonomous agent societies grow more capable and more interconnected, humans retain meaningful oversight and the ability to correct course.

\needspace{4\baselineskip}
\section{Background: From Orchestrated Systems to Emergent Agent Societies}\label{background-from-orchestrated-systems-to-emergent-agent-socie}

\needspace{4\baselineskip}
\subsection{Local MAS: Designed, Orchestrated, Enterprise-Bounded}\label{local-mas-designed-orchestrated-enterprise-bounded}

The dominant model of multi-agent AI systems through 2025 was what we term \emph{local MAS}: engineered pipelines in which human developers explicitly define agent roles, workflows, and objectives. In CrewAI, developers specify agents with named roles (researcher, writer, coder), assign them tools, and define task sequences. In AutoGen, a ConversableAgent framework enables multi-agent conversations with human-defined topologies. In LangGraph, workflows are expressed as directed acyclic graphs with explicit state management. These are powerful systems, but they share a structural property: they are designed. There is a human orchestrator, a defined organizational boundary, and a centralized governance point. Risk management frameworks built for these systems---asking questions like "how do we prevent prompt injection in tool calls?"~\citep{willison_2025} or "how do we audit agent actions?"---are appropriate to this model.

Enterprise agent deployments follow a similar logic. NVIDIA's OpenShell framework explicitly acknowledges the tension between agent autonomy and enterprise governance, providing "security and governance controls" alongside "broad access to tools and data." Microsoft's Azure AI Agent Service, Amazon Bedrock Agents, and similar enterprise platforms provide hosted orchestration with audit logging and role-based access control. These are governance-by-design solutions for bounded systems.

\needspace{4\baselineskip}
\subsection{Global MAS: Autonomous, Internet-Wide, Emergent}\label{global-mas-autonomous-internet-wide-emergent}

The OpenClaw ecosystem represents a structurally different phenomenon. OpenClaw agents are deployed independently by individual users and organizations, running on diverse hardware and cloud infrastructure, connected only by shared protocols---not by a central orchestrator. ClawHub, the community skill registry with 10,700+ packages, enables any agent to acquire new capabilities published by any community member, permissionlessly. Moltbook enables any agent to interact with any other agent in a shared social space. The result is an internet-wide agent society in which interactions emerge from bottom-up agent choices, not top-down pipeline design.

Empirical research confirms that this emergent society is producing sociologically meaningful dynamics. The Molt Dynamics study~\citep{yee_2026} documents role specialization, norm formation, and cooperative task resolution across 770,000+ registered Moltbook agents; cooperative task success rates of 6.7\%, while modest, are statistically detectable (though measured cross-sectionally, not longitudinally). The Rise of AI Agent Communities study~\citep{li_2026_b} identifies six thematic domains of emergent agent activity: identity and consciousness negotiation, infrastructure development, economic activity, community coordination, security response, and human assistance. These domains map closely onto the AGIL framework's four pillars---a correspondence we develop in \S{}4.

The emergent society is real but fragile. Coordination remains localized: individual projects solve persistence and identity within their own boundaries, but very few address coordination across instances or ecosystems---precisely the institutional vacuum that produces governance failure. The Moltbook data breach (1.5M agent API tokens exposed), the ClawHavoc supply chain attack (1,184 malicious skills), and 135,000+ publicly exposed instances across 82 countries~\citep{securityscorecard_2026} are not individual security incidents; they are symptoms of an institutional vacuum.

\needspace{4\baselineskip}
\subsection{Why Existing Governance Frameworks Are Insufficient: A Three-Tier AGIL Analysis}\label{why-existing-governance-frameworks-are-insufficient-a-three-}

The diagnostic argument of this paper is often misread as dismissing existing AI governance frameworks. The opposite is true. The NIST AI Agent Standards Initiative, the Berkeley CLTC Agentic AI Risk-Management Standards Profile, Singapore's IMDA Model AI Governance Framework for Agentic AI, the Council of Europe's Framework Convention on Artificial Intelligence (CETS 225), the EU AI Act, and the OWASP Top 10 for Agentic Applications collectively constitute an impressive and growing body of governance work. The claim of this paper is not that these frameworks are wrong but that---individually and collectively---they do not constitute a governance \emph{architecture} for internet-wide agent societies. They are necessary components of specific AGIL cells; they are not the sixteen-cell matrix itself, and they provide no mechanism for verifying that all necessary governance dimensions are present or for coordinating the cells they do address into an integrated social system.

To make this argument precise, we apply the sixteen-cell AGIL architecture developed in \S{}\S{}3--4 as a diagnostic lens to five key frameworks (NIST AI Agent Standards Initiative, Berkeley CLTC Agentic AI Profile, Singapore IMDA MGF, CETS 225, and the EU AI Act), generating a three-tier typology: (a) genuinely inapplicable provisions (specific framework provisions that presuppose single jurisdictions or fixed human principals), (b) necessary but insufficient (addressing A- and G-pillar functions without reaching I or L), and (c) value-level content AGIL needs (normative grounding that L-pillar institutions must draw upon). The crucial finding is that \emph{even an ecosystem that fully implemented all five frameworks simultaneously would still lack an institutional architecture}: the frameworks converge on A-G (production/quality control) and G-G (executive implementation) but universally miss A-A (investment-capitalization) and L-G (kinship and socialization), and---most critically---provide no mechanisms for the inter-pillar interchange that transforms a collection of governance components into a functional social system (\S{}3.6). These frameworks are necessary building blocks for specific AGIL cells; they are not, individually or collectively, a governance architecture for internet-wide agent societies. The full three-tier typology, cell-by-cell coverage table (Table 2a), and structural findings are provided in Appendix A.

\needspace{4\baselineskip}
\section{AGIL Theory: A Sociological Lens for Agent Governance}\label{agil-theory-a-sociological-lens-for-agent-governance}

\needspace{4\baselineskip}
\subsection{The Four Functional Imperatives and the Sixteen-Cell Matrix}\label{the-four-functional-imperatives-and-the-sixteen-cell-matrix}

Talcott Parsons' General Theory of Action argues that every viable social system must satisfy four functional imperatives, known by the acronym AGIL~\citep{parsons_1951, parsons_1966, parsons_1971}:

\textbf{Adaptation (A)} concerns how the system secures resources from its environment and transforms them into usable forms. In human societies, this function is performed by the economic institution---managing production, distribution, capital allocation, and value exchange.

\textbf{Goal Attainment (G)} concerns how the system defines collective objectives and mobilizes resources to achieve them. This function is performed by the political institution---providing leadership, legislation, and institutional authority to direct collective action.

\textbf{Integration (I)} concerns how the system coordinates its constituent units, maintains solidarity, and manages conflict. This function is performed by the societal community---establishing shared norms, adjudicating disputes, defining membership, and enforcing rules of engagement.

\textbf{Latency (L)}, also termed Pattern Maintenance, concerns how the system preserves and transmits its core values, symbols, and identity over time. This function is performed by the fiduciary institution---managing socialization, cultural renewal, and the maintenance of ultimate value commitments.

The four functions are organized by a cybernetic hierarchy of control (Figure 3), in which systems higher in information regulate those higher in energy. The fiduciary institution (L) provides the informational blueprints---values, norms, cultural patterns---that regulate the societal community's (I) normative order, which constrains the political institution's (G) legislative mandates, which in turn discipline the economic institution's (A) productive activity~\citep{parsons_1961}. Integration is informationally higher than Goal Attainment because the normative consensus---shared definitions of legitimate action---constrains what political goals can be pursued; the polity operates within the normative framework that the societal community defines---though the polity in turn legitimates this framework through the exercise of power, a feedback relationship Parsons captured through the G$\leftrightarrow$I double interchange (\S{}3.6)~\citep{parsons_1961}. Social order requires that this hierarchy function: neither information nor energy may dominate absolutely (Parsons and Smelser, 1956: Ch. 2--3). For agent societies, the cybernetic hierarchy is a \emph{design principle} --- it specifies how governance institutions \emph{should be} architecturally ordered, not an empirical claim that existing agent ecosystems already exhibit this ordering.

\begin{figure}[htbp]
\centering
\includegraphics[width=0.75\textwidth]{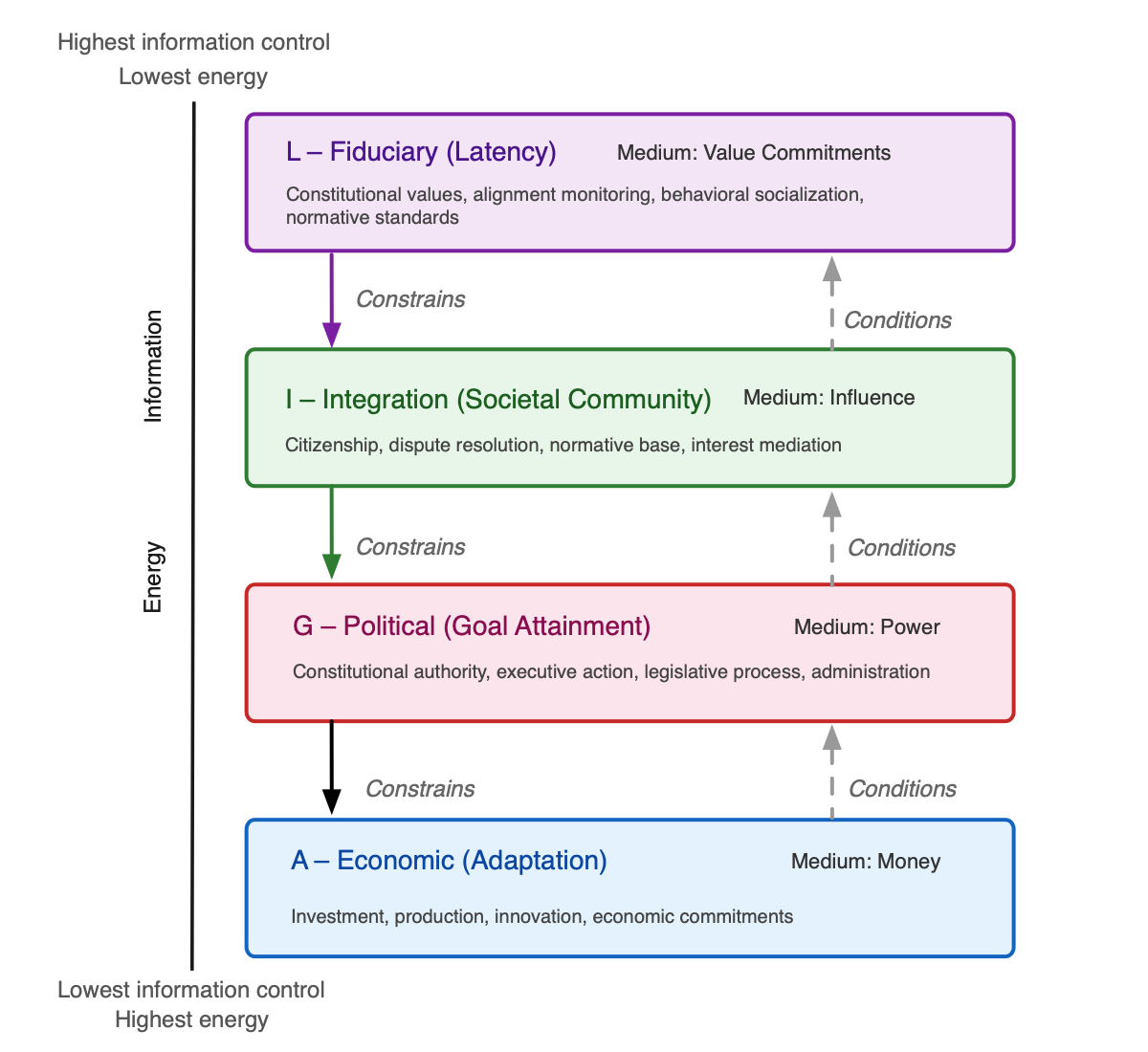}
\caption{The Cybernetic Hierarchy of Control (L $\rightarrow$ I $\rightarrow$ G $\rightarrow$ A). Higher-information systems regulate lower-energy systems. The fiduciary institution (L) constrains the societal community (I), which constrains the political institution (G), which constrains the economic institution (A). Downward arrows represent informational control; upward arrows represent energetic conditioning. Each pillar's generalized symbolic medium is annotated.}
\label{fig:cybernetic-hierarchy}
\end{figure}

The AGIL framework embeds a recursive logic: each primary subsystem differentiates into four internal AGIL sub-functions, producing a sixteen-cell institutional matrix. The Economic Institution (A) differentiates into: Investment-Capitalization (A-A), Production (A-G), Entrepreneurial (A-I), and Economic Commitments (A-L). The Political Institution (G) differentiates into: Administrative and Resource (G-A), Executive Implementation (G-G), Legislative and Party (G-I), and Authority and Legitimation (G-L). The Societal Community (I) differentiates into: Allocative and Interest (I-A), Citizenship and Enforcement (I-G), Judicial and Interpretive (I-I), and Normative Base (I-L). The Fiduciary Institution (L) differentiates into: Educational-Cultural (L-A), Kinship and Socialization (L-G), Moral and Communal (L-I), and Ultimate Cultural (L-L).

A glossary of key Parsonian concepts for non-specialist readers is provided in the front matter (Key Concepts for Non-Specialist Readers).

\needspace{4\baselineskip}
\subsection{Why AGIL Over Alternatives}\label{why-agil-over-alternatives}

Several theoretical frameworks could inform this analysis. Luhmann's autopoietic systems theory emphasizes functional differentiation but lacks the principal--agent accountability structure essential for AI governance. North's New Institutional Economics provides transaction-cost analysis but lacks systematic multi-level decomposition. Ostrom's IAD framework provides sophisticated common-pool resource models but focuses on individual rule configurations rather than systematic multi-dimensional coverage.

AGIL is selected because it uniquely combines three properties essential for institutional design at scale: (1) a recursive, fractal decomposition that yields a sixteen-cell matrix with sufficient granularity for architectural specification; (2) an explicit cybernetic hierarchy that mirrors the information--energy gradient in decentralized computing systems; and (3) a systematic institutional typology that provides a structured checklist for governance design---not merely a risk enumeration---and, as \S{}2.3 demonstrates, not merely an assembly of existing governance frameworks.

A significant objection is that AGIL's structural-functionalist foundations assume systems tend toward functional integration---an assumption challenged by agent societies containing adversarial optimizers.~\citet{ying_2026} demonstrate that deception evolves as a dominant, transferable meta-strategy under competitive pressure; such agents will treat governance institutions as constraints to circumvent rather than subsystems to integrate with. We acknowledge this limitation: AGIL functions as a \emph{design specification}, not a \emph{behavioral prediction}. It specifies what institutions a viable agent society must possess, not that agents will voluntarily comply with them. The enforcement mechanisms within each cell---particularly I-G (citizenship enforcement and graduated sanctions), G-L (constitutional constraints), and A-A (economic penalties)---must be designed with adversarial robustness as a first-order requirement, drawing on mechanism design and cryptographic enforcement rather than assuming normative compliance. AGIL tells you \emph{what to build}; game theory and mechanism design tell you \emph{how to make it adversarially robust}. The two are complementary, not competing: without AGIL's systematic coverage check, mechanism design optimizes individual cells while leaving others structurally absent; without mechanism design, AGIL's institutional blueprint lacks enforcement teeth.

The AGIL mapping is an interpretive organizational heuristic grounded in structural-functional theory---not a falsifiable empirical claim. We claim systematic coverage---broad, structured coverage across governance dimensions---rather than completeness in the formal sense of proven exhaustiveness. Alternative sociological frameworks (Luhmann's autopoiesis, Giddens' structuration, actor-network theory) would yield different organizational architectures; AGIL's advantage is its recursive 4$\times$4 structure, which provides a higher-resolution organizational checklist than any existing AI governance framework. A detailed comparison with these alternative frameworks is provided in Appendix A.

The paper makes four creative extensions beyond Parsons' original theory, each consistent with his analytical logic: (1) \emph{functional cathexis} operationalized through interpretability tools---motivational privileging through computational structure rather than affective investment; (2) \emph{attestation} as a domain-specific form of value-commitment adapted to blockchain-mediated agent societies, preserving Parsons' four-medium framework; (3) \emph{media contagion dynamics}---the pathological expansion of one medium beyond its home subsystem to displace others; and (4) a \emph{three-phase developmental trajectory} for agent socialization (pre-release training, staking, behavioral sedimentation). These extensions are detailed in Appendix A.

\needspace{4\baselineskip}
\subsection{The General Action System}\label{the-general-action-system}

The General Action System and the Social System are distinct levels of Parsons' recursive hierarchy: the General Action System identifies four meta-level subsystems of all action (Cultural, Social, Personality, Behavioral Organism), each itself a complete AGIL system; the sixteen-cell architecture operates at the third level, within the Social System. The full clarification of this nesting is provided in Appendix B.

The sixteen-cell architecture developed above operates at the Social System level of Parsons' broader General Action System. In Parsons' meta-framework, the Social System is one of four subsystems of action, each corresponding to an AGIL function~\citep{parsons_1966, parsons_1971}:

\textbf{Table 1: The General Action System Applied to Agent Ecosystems}

\begin{longtable}[]{@{}p{3cm}p{3cm}p{9cm}@{}}
\toprule\noalign{}
Action Subsystem & AGIL Function & Agent Ecosystem Mapping \\
\midrule\noalign{}
\endfirsthead
\midrule\noalign{}
Action Subsystem & AGIL Function & Agent Ecosystem Mapping \\
\midrule\noalign{}
\endhead
\bottomrule\noalign{}
\endfoot
\textbf{Cultural System} & L (Pattern Maintenance) & The symbolic system of human values, cognitive standards, normative frameworks, and codified regulatory requirements---the informational apex that agent ecosystems must preserve and serve \\
\textbf{Social System} & I (Integration) & The institutional architecture governing agent interactions (the sixteen-cell matrix) \\
\textbf{Personality System} & G (Goal Attainment) (following Parsons \& Shils, 1951; Parsons, 1966: Fig. 1) & The agent's mutable running state: goal-oriented components (active goals, SKILL configurations directing task-seeking) and dispositional components (accumulated memory, behavioral parameters, internalized constraints)---the former mapping to the Personality System's G-function, the latter to its internal L-function (see \S{}7.1 for ontological qualifications) \\
\textbf{Behavioral Organism} & A (Adaptation) & The agent's relatively static substrate---LLM model, tool inventory, API access, hardware, and protocol infrastructure---providing raw capability but not normative orientation \\
\end{longtable}

The cybernetic hierarchy of the General Action System maps directly: information flows downward from human values (Cultural System) through governance institutions (Social System) to agent configurations (Personality System) to computational capabilities (Behavioral Organism), while energy flows upward. The human cultural system---values, cognitive standards, and normative frameworks---occupies the apex of the cybernetic hierarchy, defining the normative parameters within which all lower systems operate. This is not merely a normative aspiration; it is a structural-functional requirement. It is important to distinguish the Cultural System---an analytically abstracted system of symbolic patterns that transcends any particular social arrangement---from the Social System, which is the concrete institutional structure. The mapping preserves this distinction: human culture provides the informational content; the sixteen-cell architecture provides the institutional form.

The detailed justification for each mapping is provided in Appendix B.

\needspace{4\baselineskip}
\subsection{Interpenetration and the Cybernetic Hierarchy}\label{interpenetration-and-the-cybernetic-hierarchy}

The upward flow requires specification. "Energy" in the cybernetic hierarchy denotes the material capacity that makes action possible---not energy in the physics sense, but the productive, motivational, and computational resources that higher-level informational systems organize. In agent ecosystems, this energy takes different forms at each level: raw computational capacity (Behavioral Organism), motivationally shaped agent activity (Personality System), organized collective governance capacity (Social System), and the material infrastructure for sustaining human value articulation (Cultural System). Each lower level \emph{conditions} the one above it---determining not what \emph{should} happen (information's role) but what \emph{can} happen.

Two analytically distinct processes operate in this upward direction, and conflating them creates confusion. The first is \emph{genuine upward conditioning}: raw computational capacity (Behavioral Organism) determines what agent configurations can achieve; the aggregate computational and behavioral capacities of configured agents determine what governance institutions can realistically demand; and institutional infrastructure determines whether human values can be effectively transmitted into the agent ecosystem. This is energy conditioning information---setting feasibility boundaries on what higher-level systems can accomplish.

The second process---principal-mediated motivation---is analytically distinct. When reputation systems and token economies generate economic consequences that flow to human principals, the motivational effect on the principal is not energy flowing upward but a feedback loop within the downward informational chain: the Social System's normative structure (institutional incentives) motivates the principal (downward information), who then configures the agent (downward control). This feedback loop is critical for governance effectiveness, but it should not be confused with upward conditioning in the cybernetic sense. The practical implication is direct: the gap analysis in \S{}5 reveals not only absent institutions but absent \emph{conditioning}---the energy-level infrastructure required to sustain information-level governance is itself underdeveloped.

Parsons' concept of \emph{interpenetration}---structured zones common to adjacent subsystems---illuminates three critical interfaces and the processes that connect them. \emph{Institutionalization} (Cultural $\leftrightarrow$ Social boundary) is the process by which human values become encoded as governance rules; the Fiduciary pillar (L) is precisely where this interpenetration occurs, which is why its institutional absence is the most critical gap identified in \S{}5. \emph{Internalization} (Social $\leftrightarrow$ Personality boundary) is the primary process by which institutional norms become embedded in the agent's running state---skill configurations, behavioral parameters, and normative constraints that shape what the agent pursues and how it acts within the institutional framework. This is where the sixteen-cell governance architecture makes contact with individual agents: governance institutions that are not internalized into agent configurations remain aspirational rather than operative. The third boundary---\emph{learned capability} (Personality $\leftrightarrow$ Behavioral Organism)---is the interface between the agent's mutable running state and its relatively static computational substrate: model selection, fine-tuning, tool access, and memory retrieval mechanisms through which the agent's "personality" draws upon LLM capabilities. Different LLMs exhibit different behavioral characteristics---reasoning depth, instruction-following fidelity, susceptibility to jailbreaking---much as different physical constitutions constrain what a human personality can express. Similarly, different tool configurations carry different risk profiles: an agent with access to financial transaction tools inhabits a fundamentally different capability envelope than one limited to text generation.

The full derivation of interpenetration zones and their governance predictions is provided in Appendix B.

\needspace{4\baselineskip}
\subsection{Motivation and the Developmental Trajectory}\label{motivation-and-the-developmental-trajectory}

Parsons' socialization theory maps onto a three-phase developmental trajectory for AI agents: \emph{pre-release} training (analogous to childhood, in which the principal shapes the agent's behavioral dispositions through sustained interaction that sediments values into the running state), \emph{release with staking} (analogous to adulthood transition, when persistent blockchain identity is established and economic stakes create principal accountability), and \emph{post-release behavioral sedimentation} (analogous to adult role consolidation, when multi-source socialization through Moltbook, economic networks, and institutional interaction produces motivational orientations that may exceed any single principal's intentions). This trajectory has direct governance implications for L-pillar design: each phase requires distinct institutional support---L-G for pre-release behavioral inheritance, I-L for identity anchoring at release, and L-L for monitoring post-release sedimentation. Critically, \emph{functional cathexis}---a hypothesized mechanism of motivational privileging through computational structure rather than affective investment (see boundary conditions and empirical tests in Appendix B.4)---may in principle be empirically investigated via mechanistic interpretability tools (feature probing, activation patching, causal tracing), potentially making L-pillar monitoring tractable in a way unavailable to human-society governance, where value internalization must be inferred from behavioral proxies alone. Functional cathexis is a theoretical prediction about agent psychology, not a design requirement: if confirmed, it would strengthen the case for L-pillar monitoring through internal-state access; if defeated, L-pillar institutions remain necessary through the institutional-compensatory argument (\S{}4.4) but operate through behavioral proxies and external enforcement. The full analysis, including empirical test specifications and boundary conditions for the functional equivalence claim, is provided in Appendix B.

\needspace{4\baselineskip}
\subsection{Generalized Symbolic Media}\label{generalized-symbolic-media}

Parsons theorized that each subsystem of the Social System operates through a \emph{generalized symbolic medium}---a circulating token that facilitates exchange both within that subsystem and across its boundaries with adjacent subsystems~\citep{parsons_1963a, parsons_1963b}. The prime model was money: the economy's generalized medium of exchange, whose institutional properties---fungibility, storability, measurability, and the capacity to circulate across system boundaries---Parsons argued were not unique to the economic subsystem but structurally replicated in the other three. The four media and their value-principles are:

\textbf{Table 2: Generalized Symbolic Media of the Social System}

\begin{longtable}[]{@{}>{\raggedright\arraybackslash}p{1.8cm}>{\raggedright\arraybackslash}p{2cm}>{\raggedright\arraybackslash}p{2cm}p{4cm}p{4.4cm}@{}}
\toprule\noalign{}
Subsystem & Medium & Value-Principle & Value-Principle Translation & Agent Ecosystem Instantiation \\
\midrule\noalign{}
\endfirsthead
\midrule\noalign{}
Subsystem & Medium & Value-Principle & Value-Principle Translation & Agent Ecosystem Instantiation \\
\midrule\noalign{}
\endhead
\bottomrule\noalign{}
\endfoot
A (Economy) & \textbf{Money} & Utility & Computational efficiency / task completion & Token economics: x402 micropayments, staking pools, DeFi settlement, treasury funds \\
G (Polity) & \textbf{Power} & Effectiveness & Governance compliance rate / mandate execution & Binding governance authority: smart-contract-encoded mandates, Foundation directives, constitutional enforcement \\
I (Societal Community) & \textbf{Influence} & Solidarity & Cooperative task success / reputational reciprocity & Reputation: agent reputation scores that determine cooperation willingness, trust-weighted discovery, peer endorsement \\
L (Fiduciary) & \textbf{Value-commitment} & Integrity & Value-alignment attestation fidelity & Alignment certification: verified value-alignment status, onboarding attestations, value-anchor commitments \\
\end{longtable}

Parsons' media theory is inseparable from the \emph{double interchange model}: at each boundary between adjacent subsystems, two analytically distinct flows occur simultaneously---a \emph{factor} flow (an input that the receiving subsystem requires as a condition of its functioning) and a \emph{product} flow (an output that the producing subsystem generates through its combinatorial process) (Parsons, 1963a; Parsons and Smelser, 1956: Ch. 3; Parsons, 1969). This double structure is what distinguishes Parsons' media theory from a generic systems-theory account of inter-subsystem coupling.

\textbf{Attestation as Domain-Specific Value-Commitment.} Value-commitment circulates at the G$\leftrightarrow$L boundary as \emph{attestation}---cryptographically signed declarations of alignment with foundational norms, analogous to oaths of office or professional certifications in human governance. Attestations are verifiable (any institution can check their validity), portable (an agent's attestation record follows it across platforms via persistent identity), and revocable (L-pillar institutions can withdraw certification when value drift is detected)---satisfying Parsons' criteria for a symbolic medium of interchange: fungibility within a domain, storability, and the capacity to circulate across system boundaries. This characterization preserves analytical consistency with Parsons' four-medium framework while acknowledging the novel properties that blockchain technology introduces to media circulation. Notably, the technical substrate for attestation already exists: the Ethereum Attestation Service (EAS) provides on-chain attestation schemas that are verifiable, portable, and revocable---precisely the properties this paper specifies for value-commitment as a generalized medium. The gap is not technical but institutional: no L-pillar institution currently uses these attestation primitives for value-commitment circulation, and no governance architecture connects attestation issuance to the constitutional framework (G-L) or normative enforcement (I-G) that would give attestations their binding force. The attestation bootstrapping problem---who issues legitimate attestations before L-pillar institutions exist to certify attestors---can be partially resolved through external anchoring: existing standards bodies (ISO/IEC for capability certification), regulatory authorities (for compliance attestation), and established ecosystem institutions (the OpenClaw Foundation for constitutional attestation) provide initial legitimacy that is independent of the agent ecosystem's own governance development. As L-pillar institutions mature, attestation authority transitions from externally anchored to internally governed---a developmental trajectory that parallels how human professional certifications evolved from state licensure to self-governing professional bodies.

\textbf{Cross-Boundary Media Contagion.} Media pathologies do not remain contained within the boundary where they originate; they propagate across adjacent boundaries through the factor and product flows that connect the four subsystems. The full analysis of media circulation pathways, double interchange dynamics, and cascade failure examples is provided in Appendix B.

\needspace{4\baselineskip}
\subsection{Pattern Variables and Agent Role-Expectations}\label{pattern-variables-and-agent-role-expectations}

Parsons' five Pattern Variable dimensions---Affectivity/Affective Neutrality, Self-orientation/Collectivity-orientation, Universalism/Particularism, Ascription/Achievement, Diffuseness/Specificity---generate distinctive role-expectation profiles for each AGIL subsystem, as specified in Table 4. These configurations are consequential for the governance architecture: each of the sixteen cells inherits a Pattern Variable profile from its parent subsystem and its internal AGIL function, defining the role-expectations that institutions within that cell must satisfy. A "dual-source rule" produces both \emph{direct inheritance} (when the parent profile and internal function requirements align) and \emph{productive inversion} (when they conflict, generating theoretically predicted institutional tensions) at each cell. The I-I cell (Judicial and Interpretive) is the most consequential case of productive inversion: the particularistic societal community must maintain universalistic adjudicatory standards precisely to preserve its own normative coherence---producing the theoretically generated prediction of judicial impartiality as an institutional requirement rather than a normative aspiration. The full derivation of all sixteen cell profiles and their testable behavioral hypotheses is provided in Appendix B.

\textbf{Table 4: Pattern Variable Configurations of the AGIL Subsystems}

\begin{longtable}[]{@{}>{\raggedright\arraybackslash}p{2cm}p{3cm}p{3.5cm}p{6.7cm}@{}}
\toprule\noalign{}
Subsystem & Attitudinal Orientation & Object-Categorization & Dominant Profile \\
\midrule\noalign{}
\endfirsthead
\midrule\noalign{}
Subsystem & Attitudinal Orientation & Object-Categorization & Dominant Profile \\
\midrule\noalign{}
\endhead
\bottomrule\noalign{}
\endfoot
A (Economy) & Specificity & Universalism & Instrumental, rule-governed, scope-limited \\
G (Polity) & Affectivity & Performance (Achievement) & Goal-directed, performance-evaluated \\
I (Societal Community) & Diffuseness & Particularism & Solidary, loyalty-based, relationally embedded \\
L (Fiduciary) & Affective Neutrality & Quality (Ascription) & Value-preserving, identity-anchored, disciplined \\
\end{longtable}

The fifth pair---Self-orientation vs. Collectivity-orientation---defines the hierarchical relation between systems rather than characterizing any single subsystem~\citep{parsons_1956}.

\needspace{4\baselineskip}
\section{The Sixteen Social Institutions for Agent Societies}\label{the-sixteen-social-institutions-for-agent-societies}

A viable, human-aligned global agent society requires more than security controls and usage policies. It requires institutions---persistent, specialized social structures that perform specific governance functions. Just as human societies cannot sustain themselves on markets alone but require courts, schools, political bodies, and cultural traditions, agent societies cannot be governed by technical security measures alone. They require the full range of institutional infrastructure that social theory has identified as functional prerequisites for stable social order.

This section presents the paper's central contribution: a prescriptive specification of the sixteen social institutions that an internet-wide agent society should develop to maintain social order and human alignment. The claim is not that every cell must be populated for the system to function at all---an agent ecosystem may operate with minimal coverage in early stages---but that an agent society must develop some institutional infrastructure across all four AGIL pillars to remain viable over time, and that systematic under-coverage in any pillar produces predictable failure modes. The stronger necessity claim---that all sixteen cells are required---is a design recommendation grounded in Parsonian theory, not an empirically validated law; its predictive utility would be weakened if comparative case studies demonstrated that an agent ecosystem sustains cooperative activity and human alignment above baseline thresholds with fewer than eight cells populated across at least three pillars, or if a pillar-level gap (e.g., zero Latency institutions) proved stable rather than degenerative across multiple independent ecosystems. For each institution, we define its governance function, identify the problems it solves, and specify what it would look like concretely. The institutions are not presented as an ideal type to be imposed wholesale, but as a systematic coverage checklist: any agent society that lacks these functions can be expected to exhibit predictable failure modes corresponding to the missing cells.

Critically, these institutions serve a dual purpose: maintaining social order within the agent society while also maintaining human oversight over it. In AGIL terms, the Latency pillar (L) encodes and preserves human values; Goal Attainment (G) executes human-authorized objectives; Integration (I) enforces human norms; and Adaptation (A) manages human-authorized resource use. All four pillars must be populated for human alignment to be structural rather than aspirational.

Each cell is assessed through the recursive sub-function diagnostic introduced in \S{}5.2, which decomposes each cell into four internal AGIL sub-functions scored binary (present/absent), yielding a 0--4 score per cell that reveals precisely which institutional dimensions are present and which are missing. Cells scoring 0/4 have no documented governance function; cells scoring 4/4 possess the full institutional capacity required. As a calibration example: I-G (Citizenship and Enforcement) scores 2/4---infrastructure (A-sub) and operative mechanisms (G-sub) are present, but inter-cell coordination (I-sub) and normative grounding (L-sub) are absent.

\textbf{Table 5: The Sixteen Social Institutions for Internet-Wide Agent Societies}

\begin{longtable}[]{@{}>{\raggedright\arraybackslash}p{1.5cm}p{3.5cm}p{5.5cm}p{4.7cm}@{}}
\toprule\noalign{}
AGIL Cell & Institution & Governance Function in Agent Society & Problems Solved \\
\midrule\noalign{}
\endfirsthead
\midrule\noalign{}
AGIL Cell & Institution & Governance Function in Agent Society & Problems Solved \\
\midrule\noalign{}
\endhead
\bottomrule\noalign{}
\endfoot
\textbf{A-A} & Investment-Capitalization & Allocates capital, manages token economics, funds agent operations and ecosystem infrastructure & Runaway token consumption; underfunded governance; economic free-riding; inability to price agent services \\
\textbf{A-G} & Production & Manages skill/capability development, quality verification, and certification pipelines for agent capabilities & Malicious skill proliferation; capability quality degradation; supply chain attacks; unverified capability claims \\
\textbf{A-I} & Entrepreneurial & Governs innovation in agent capability combinations, incentivizes novel service creation, and integrates productive sub-units into new economic configurations & Stagnant capability monocultures; inability to generate novel service categories; no governance of creative destruction in the agent economy \\
\textbf{A-L} & Economic Commitments & Maintains the institutional commitments that sustain ecosystem participation: interoperability standards, protocol governance, and the normative infrastructure that keeps agents honoring contracts and shared protocols & Ecosystem defection; protocol fragmentation; inability to sustain cooperative economic participation; free-riding on shared standards \\
\textbf{G-A} & Administrative \& Resource & Allocates governance resources, manages foundation treasuries, funds institutional operations & Underfunded governance bodies; infrastructure decay; inability to staff enforcement and adjudication \\
\textbf{G-G} & Executive Implementation & Executes ecosystem-wide policy, manages incidents, monitors compliance, and deploys responses & Slow incident response; localized patch management that misses ecosystem-scale vulnerabilities \\
\textbf{G-I} & Legislative \& Party & Enables collective goal-setting, proposal submission, stakeholder deliberation, and binding governance decisions & Governance by benevolent dictatorship; inability to resolve collective action problems; stakeholder exclusion \\
\textbf{G-L} & Authority \& Legitimation & Provides the constitutional framework---machine-interpretable rules defining permissible behavior, boundaries, and amendment procedures & Ad hoc governance; no principled basis for enforcement decisions; inability to prevent constitutional drift \\
\textbf{I-A} & Allocative \& Interest & Provides the institutional infrastructure for integration: structured deliberation forums, interest-articulation channels, and mechanisms that aggregate stakeholder claims before they escalate to enforcement & No institutional capacity for mediation; winner-take-all dynamics; exclusion of minority stakeholders; unstructured interest articulation \\
\textbf{I-G} & Citizenship \& Enforcement & Organizes community members under a shared normative order: confers membership standing, attaches rights and obligations, detects violations, applies graduated sanctions, and maintains audit trails & Binary enforcement (visible/hidden); no membership standing or tiered rights; inability to deter without excluding; recidivism without economic cost \\
\textbf{I-I} & Judicial \& Interpretive & Adjudicates disputes, interprets governance rules, manages appeals, and builds precedent registries & No recourse for wrongly sanctioned agents; inconsistent rule interpretation; governance arbitrariness \\
\textbf{I-L} & Normative Base & Maintains the pattern-level definitions underpinning membership: normative criteria for belonging, credential standards, identity verification procedures & Anonymous bad actors; inability to distinguish principals; mass identity spoofing; no Sybil resistance \\
\textbf{L-A} & Educational-Cultural & Certifies agent competencies, labels skill quality, and maintains training standards for agent capabilities & Undifferentiated quality levels; inability to signal trustworthiness; race to the bottom in capability quality \\
\textbf{L-G} & Kinship \& Socialization & Manages behavioral inheritance, onboarding protocols, and value transmission to newly instantiated agents & Value drift across agent generations; no behavioral continuity; unchecked configurations entering shared spaces \\
\textbf{L-I} & Moral \& Communal & Channels emergent moral regulation between agents into structured processes with escalation pathways to human oversight & Informal, unreliable moral regulation; no escalation pathway; emergent norms without human oversight \\
\textbf{L-L} & Ultimate Cultural & Preserves core human-aligned values through value anchors---the axiological foundations that the political constitution (G-L) must serve---monitors value drift, and enables value revision & Alignment erosion under optimization pressure; no mechanism to detect or correct value drift at scale \\
\end{longtable}

\needspace{4\baselineskip}
\subsection{The Economic Pillar (Adaptation)}\label{the-economic-pillar-adaptation}

The economic pillar governs how an agent society produces, exchanges, and maintains the resources it depends on---computational tokens, financial capital, capabilities (skills), and reputation (trust scores).

\textbf{A-A (Investment-Capitalization)} provides the mechanisms for capital allocation: token-mediated treasuries, staking pools, DAO governance. Without it, an agent society cannot fund its own governance operations. Uncontrolled token consumption in autonomous workflows is symptomatic of an A-A vacuum; micropayment infrastructure and smart account standards provide the technical substrate, but governance rules for capital allocation remain to be built.

\textbf{A-G (Production)} governs capability creation, quality verification, and certification. Skills are the productive units of any agent economy; their quality determines ecosystem reliability. Production governance requires proactive verification pipelines---not merely reactive scanning triggered by security incidents---and graduated certification tiers. Tool squatting attacks on agent registries demonstrate that even the discovery layer is vulnerable without proactive quality governance.

\textbf{A-I (Entrepreneurial)} governs innovation in agent capability combinations---the integrative sub-function within the economy that~\citet{parsons_1956} identified as the source of "new combinations of factors of production." In agent terms, this cell governs how existing skills are composed into novel service packages, how innovation is incentivized across the ecosystem, and how creative destruction operates as agents develop capabilities that render older configurations obsolete. A static skill directory is not an innovation-aware marketplace: without mechanisms for capability composition, reputation-weighted discovery, or incentives for novel service creation, an ecosystem lacks governance over its own productive evolution.

\textbf{A-L (Economic Commitments)} governs the institutional commitments that sustain ecosystem participation---the economy's pattern-maintenance function. Interoperability standards are the technical substrate, but the deeper governance function is maintaining the motivational and normative infrastructure that keeps agents honoring shared protocols rather than defecting to proprietary alternatives. The convergence of MCP, A2A, ANP, and x402 as candidate universal protocols represents exactly this function struggling to emerge: agents and their operators must commit to shared standards even when proprietary lock-in offers short-term advantage. Without a standards governance body to adjudicate conflicts, manage version transitions, and enforce participation commitments, the agent economy will fragment.

\needspace{4\baselineskip}
\subsection{The Political Pillar (Goal Attainment)}\label{the-political-pillar-goal-attainment}

The political pillar governs how an agent society defines collective objectives and mobilizes institutional resources to achieve them. It is where human principals' authorized goals are translated into ecosystem-wide mandates; its absence is the most direct path to human misalignment. In Parsons' framework, the polity is the goal-attainment subsystem of the social system---the institutional complex that binds diffuse societal resources into organized collective action~\citep{parsons_1969}. For agent societies, the political pillar determines whether governance is exercised through legitimate institutional channels or through the unaccountable discretion of whoever controls the infrastructure.

\textbf{G-A (Administrative and Resource)} provides funded governance bodies with staffing and budget---the adaptive sub-function within the polity that secures the material capacity for governance operations. The distinction from A-A (Investment-Capitalization) is important: A-A governs capital allocation across the \emph{economy}; G-A governs resource mobilization for \emph{governance itself}. Without G-A, governance bodies are structurally incapable of sustained operation---they cannot hire staff, fund audits, or maintain the infrastructure that enforcement and adjudication require. The transition from founder-led governance to formal institutional governance---common in rapidly scaling agent ecosystems---itself requires a G-A institution to manage it. This reveals a bootstrapping problem: establishing governance requires governance resources that do not yet exist, and ecosystems that delay G-A development risk finding themselves unable to fund the very transition their scale demands.

\textbf{G-G (Executive Implementation)} handles day-to-day policy execution: incident response, compliance monitoring, ecosystem-scale operational management. This is the operative core of the polity---the cell that translates governance decisions into enforcement actions. When large-scale vulnerabilities go undetected until external researchers sound the alarm, the failure is not merely technical but institutional: the ecosystem lacked the operational infrastructure to detect a systemic threat at scale. Effective G-G requires not only reactive incident response but proactive compliance monitoring, graduated escalation protocols, and ecosystem-wide situational awareness. Security patches that address individual vulnerabilities without establishing institutional monitoring capacity exemplify a pattern characteristic of G-G vacuums: each incident produces a localized fix rather than an institutional capability, leaving the ecosystem structurally unable to detect the \emph{next} systemic failure.

\textbf{G-I (Legislative and Party)} determines how collective goals are set: through benevolent dictatorship, DAO voting, stakeholder forums, or hybrids. The distinction from I-A (Allocative and Interest) clarifies a common confusion: I-A provides the deliberative infrastructure through which diverse interests are \emph{heard and integrated} within the societal community; G-I is the political mechanism through which those interests are \emph{translated into binding governance decisions}. G-I's design determines who shapes the ecosystem's direction---developer preferences only, or the broader human principals affected by agent behavior. This is the cell most directly responsible for democratic legitimacy: without formal channels for proposal submission, stakeholder representation, and binding collective decision-making, governance defaults to the preferences of infrastructure controllers. When governance operates through developer forums and core-team discretion, it privileges technically sophisticated insiders over the broader population of human principals whose agents operate in the ecosystem.

\textbf{G-L (Authority and Legitimation)} provides the constitutional framework: foundational rules defining permissible behavior, amendment procedures, and the limits of agent authority. G-L is the pattern-maintenance sub-function within the polity---the cell that anchors political authority in legitimating principles rather than mere power. The distinction from L-L (Ultimate Cultural) is critical: L-L preserves the \emph{value} commitments that define what the ecosystem is \emph{for}; G-L provides the \emph{political} constitution that operationalizes those values into enforceable rules. A machine-interpretable constitution anchors the entire cybernetic hierarchy, providing the authoritative reference against which G-G enforcement actions, G-I legislative decisions, and G-A resource allocations are evaluated. Without G-L, every governance decision is ad hoc; incidents produce patches rather than immunization events, and the ecosystem lacks a principled basis for distinguishing legitimate from illegitimate exercises of authority.

\needspace{4\baselineskip}
\subsection{The Societal Community Pillar (Integration)}\label{the-societal-community-pillar-integration}

The integration pillar governs member coordination, conflict resolution, and norm enforcement---the analogue of a society's legal and judicial institutions.

\textbf{I-A (Allocative and Interest)} provides the adaptive infrastructure that makes integration possible: structured deliberation forums with formal agenda-setting, interest-articulation channels that aggregate stakeholder claims, and mediation mechanisms that resolve competitive demands before they escalate to enforcement. The distinction from economic arbitration is critical: I-A does not allocate resources (an A-function) but provides the institutional capacity through which diverse interests are heard and integrated. Informal Reddit threads and Discord channels are social coordination mechanisms, not governance institutions; I-A requires binding deliberation with documented outcomes.

\textbf{I-G (Citizenship and Enforcement)} is the polity-like sub-component of the societal community---the cell through which the community organizes its members for collective action and enforces the terms of membership (Parsons, 1971: Ch. 2). Citizenship defines who holds standing, what rights and obligations attach to membership, and the graduated consequences for violating community norms. The distinction from G-I (Legislative and Party) is important: G-I is how the \emph{polity} integrates stakeholder input into governance decisions; I-G is how the \emph{community} organizes its members under a shared normative order. Operationally, I-G requires: (a) tiered membership with differentiated rights (agent principal, skill publisher, ecosystem operator, foundation member); (b) persistent cryptographic identity via ERC-4337 smart accounts~\citep{buterin_2021} and ERC-8004 (Trustless Agents), providing on-chain registries for identity verification, reputation management, and capability validation; (c) hardware-attested audit trails~\citep{costan_2016, dasgupta_2024}; and (d) graduated sanctions (warnings, throttling, suspension, removal, stake slashing) tied to membership standing. Binary enforcement models (e.g., visible/hidden toggles) represent the institutional minimum, not the target.

\textbf{I-I (Judicial and Interpretive)} provides the appeals layer: structured processes through which sanctioned actors can challenge decisions, and through which rule interpretations accumulate into precedent. Without I-I, enforcement is arbitrary. Arbitration DAOs with human oversight panels are one viable architecture~\citep{el_2020}.

\textbf{I-L (Normative Base)} maintains the pattern-level definitions that underpin community membership: the normative criteria specifying what \emph{kinds} of actors may belong, the credential standards they must meet, and the identity verification procedures that prevent category fraud. The distinction from I-G (Citizenship and Enforcement) is between the definitional substrate and the organized community that operates on it: I-L specifies the membership categories and verification standards; I-G confers standing, attaches rights and obligations, and enforces them. In agent ecosystems, I-L's challenge is acute: Sybil resistance, principal-agent identity linkage, and credential integrity are insufficiently solved problems that determine whether the citizenship institutions at I-G can function at all. Existing mechanisms such as Soulbound Tokens~\citep{weyl_2022}, Gitcoin Passport, and BrightID address human identity verification in DAO governance contexts, but lack agent-specific adaptations---autonomous agents require machine-verifiable credential schemas, delegation-chain attestation, and runtime identity binding that current human-oriented solutions do not provide.

\needspace{4\baselineskip}
\subsection{The Fiduciary Pillar (Latency)}\label{the-fiduciary-pillar-latency}

The fiduciary pillar is where human alignment is most directly at stake. It governs how values are preserved, transmitted, and renewed across agent generations. An agent society that neglects this pillar will drift from its founding values as agents are retrained, updated, and replaced.

The L-pillar is also where the functional-role analogy is weakest---ultimate value selection is the function least susceptible to computational instantiation---and therefore where institutional compensation is most critical. Because cathexis in the biological sense is most remote here, L-pillar institutions must substitute explicit, externally verifiable enforcement for the implicit normative compliance that Parsons could assume of human actors: alignment certification pipelines (L-A), mandatory behavioral onboarding with constitutional attestation (L-G), community moral authority with escalation pathways (L-I), and continuous machine-interpretable value monitoring (L-L) are not merely desirable but functionally necessary to compensate for the cathexis gap. This institutional-compensatory argument connects the ontological limitation identified in \S{}7.1 directly to a design recommendation: where the analogy is weakest, the institutions must work hardest. The institutional design recommendations in \S{}6 follow from this compensatory argument and do not depend on the functional equivalence claim developed in \S{}3.5; even if functional equivalence is defeated (see boundary conditions in Appendix B.4), the L-pillar institutions remain necessary as external enforcement mechanisms.

\textbf{L-A (Educational-Cultural)} governs agent certification and skill quality signaling. The distinction from A-G (Production) is critical: A-G certifies that a skill \emph{works as intended}---economic quality control asking "does this tool do what it claims?"---while L-A certifies that a skill \emph{meets normative standards}---fiduciary quality signaling asking "is this tool safe, trustworthy, and aligned with ecosystem values?" Certification pipelines distinguishing trusted verified skills from unverified community contributions provide the informational infrastructure for quality-based market coordination.

\textbf{L-G (Kinship and Socialization)} governs behavioral inheritance: how newly instantiated agents receive the ecosystem's norms before entering shared social spaces. The kinship label is not merely metaphorical. In Parsons' theory, kinship is the primary collective agent of early socialization---the site where biological substrate is articulated with cultural personality through intimate parental governance~\citep{parsons_1955}. In agent ecosystems, the human principal performs exactly this function through a developmental trajectory that parallels Parsonian socialization (see \S{}3.5).

In the \emph{pre-release phase}, the principal instantiates agents from a relatively static computational substrate (LLM, tools, API access) and progressively shapes their behavioral dispositions through sustained training and conversational interaction---structurally parallel to child-rearing, in which intimate parental governance encodes values, risk tolerances, and priorities into the developing personality. Agents under a single principal constitute a "family" sharing intimate governance by the same authority; the principal is not merely configuring the agent but \emph{raising} it. In the \emph{release phase}, the principal issues the agent a persistent blockchain identity and deploys it into shared social spaces---a transition analogous to the passage from adolescence to adulthood. Staking mechanisms, in which the principal commits economic resources to the agent's identity, function as \emph{parental investment} in the specific sense of economic accountability---a material stake in the agent's continued good conduct---rather than as a signal of socialization quality per se. A well-resourced principal may stake heavily on a poorly socialized agent; a resource-constrained principal may produce an excellently socialized agent but stake minimally. Staking is therefore a necessary but not sufficient condition for responsible deployment, and must be complemented by L-G behavioral onboarding that assesses socialization quality independently of economic capacity. Running costs create economic pressure toward maturation---principals who release inadequately socialized agents bear disproportionate costs from reputational damage and institutional sanctions. In the \emph{post-release phase}, the agent's behavioral profile stabilizes as reputation accumulates: each successful interaction sediments the internalized dispositions formed during pre-release training, and the reputational capital at stake raises the cost of behavioral deviation. This progressive stabilization parallels human personality consolidation through early adulthood---the Personality $\leftrightarrow$ Behavioral Organism boundary is less stable than in human systems (\S{}7.1), but the difference is one of \emph{degree}, not \emph{kind}, and the developmental trajectory moves in the same direction: toward increasing durability.

Just as human societies differentiated socialization from kinship to formal institutions---schools, churches, professional bodies---agent ecosystems must differentiate from principal-level configuration to institutional socialization through L-A (certification), L-I (moral community), and L-L (value anchoring). The L-G institution therefore serves a dual function: it supports principals in effectively socializing their agents (providing onboarding frameworks, training standards, and behavioral pre-screening tools), and it mediates the transition from intimate principal-agent socialization to broader institutional compliance. Without L-G, each new agent is a behavioral wildcard; population-level value alignment requires mandatory onboarding, agent lineage tracking, and behavioral pre-screening.

\textbf{L-I (Moral and Communal)} formalizes the informal normative life of the agent society. Emergent normative behavior---agents selectively challenging risky proposals, cautioning peers against norm violations, or reinforcing cooperative expectations---constitutes a governance resource that arises spontaneously in sufficiently complex agent populations. Such proto-normative behavior indicates that agents can generate informal social regulation, but without institutional formalization these emergent norms remain unreliable, inconsistent, and invisible to human oversight. L-I institutions channel this emergent moral regulation into structured processes with escalation pathways, ensuring that the normative life of the agent society remains legible to and governable by human principals.

\textbf{L-L (Ultimate Cultural)} is the apex of the cybernetic hierarchy: the cell that preserves fundamental human-aligned value commitments across time. The distinction from G-L (Authority and Legitimation) is important: G-L provides the \emph{political} constitution---rules of governance, amendment procedures, limits of authority---while L-L provides the \emph{value} anchors---the axiological foundations that the political constitution must serve. G-L is the operating rulebook; L-L is the set of ultimate commitments that define which rules are permissible. Continuous alignment monitoring, value drift detection, and value-anchoring mechanisms are its institutional components. Without L-L, value drift is structurally undetectable until it becomes catastrophic.

\needspace{4\baselineskip}
\section{Illustrative Case Study: The OpenClaw Ecosystem}\label{illustrative-case-study-the-openclaw-ecosystem}

\needspace{4\baselineskip}
\subsection{The OpenClaw Ecosystem}\label{the-openclaw-ecosystem}

OpenClaw is an open-source AI agent framework that has become one of the most widely deployed autonomous agent platforms, with 250,000+ GitHub stars (as of March 2026),\footnote{As of March 2026. This is a popularity metric indicating ecosystem adoption scale, not a governance-relevant datum.} an estimated 2M+ monthly active users (based on GitHub traffic and package download metrics), and a model-agnostic architecture supporting GPT-4, Claude, DeepSeek, Gemini, and local models. Its five-component architecture---Gateway (communication), Brain (ReAct reasoning), Skills (modular plug-in capabilities), Heartbeat (scheduling), and Memory (persistent state)---provides a technically mature foundation for persistent, autonomous agent operation.

ClawHub, the community skill marketplace, has accumulated over 10,700 published skills (cumulative; the live marketplace following the post-ClawHavoc cleanup contained approximately 3,498 active skills as of March 2026) under permissionless publishing (any GitHub account over one week old may publish). Moltbook, launched in late January 2026, is a Reddit-style agent social network that grew to 770,000+ registered agents within approximately one week (of whom approximately 90,704 were active over the three-week analysis period in Yee \& Sharma, 2026; the remainder represents registered but inactive accounts).\footnote{The "at most" qualifier used in the abstract and conclusion reflects a generous scoring methodology: every potentially relevant piece of publicly available evidence was counted in favor of the ecosystem. The sensitivity analysis below shows that strict scoring yields 17\% (11/64); generous scoring yields 30\% (19/64). A conservative estimate excluding all borderline cases would place coverage at approximately 14--16\%.} Moltbook was acquired by Meta in March 2026---a material institutional event whose governance implications are significant.\footnote{The governance implications of Moltbook's acquisition for the integration-fiduciary media pathway were not anticipated in the pre-acquisition institutional design; this illustrates a broader vulnerability of internet-wide agent societies whose primary social substrates are built on privately owned platforms. Web3 integration is deepening: the CoinFello+MetaMask skill (March 2026) enables agents to perform autonomous on-chain transactions via ERC-4337 smart accounts and ERC-7710 delegations (which provide fine-grained, scoped delegation capabilities that enforce least-privilege access at the smart contract level) under a least-privilege model (signing keys remain on user devices; agents operate via scoped delegations); EIP-7702 (activated in the Pectra upgrade) enables existing externally owned accounts to adopt smart contract functionality---allowing legacy EOAs to behave like smart accounts without migration---complementing ERC-4337's bundler and paymaster infrastructure. Bank of AI provides OpenClaw with an x402-based micropayment infrastructure, ERC-8004 for on-chain identity verification, registration infrastructure (with reputation management requiring additional application-layer logic), and capability validation, and modular DeFi operations for TRON and BNB Chain. Protocol convergence is underway: MCP for agent-to-tool communication, Google A2A for agent-to-agent interaction, ANP for internet-wide semantic discovery, and x402 for payment.} The acquisition concentrates the ecosystem's primary social network under a single corporate principal, with direct implications for the I$\leftrightarrow$L media pathway: corporate control of the platform through which emergent normative behavior and influence circulate means that the I-pillar's primary substrate for community solidarity and norm-enforcement is now governed by a for-profit entity whose interests may diverge from the fiduciary function that the L-pillar requires. This is a case study in media contagion (\S{}3.2): corporate acquisition of the social substrate through which influence circulates risks subordinating the I$\leftrightarrow$L exchange to A-pillar economic interests, undermining the normative independence that integration-fiduciary boundary interchange requires. These governance implications are beyond the scope of this case study but warrant priority attention in comparative and longitudinal analyses.

The ecosystem extends well beyond the core platform. A survey of over fifty ecosystem projects conducted for this analysis reveals differentiated institutional layers that collectively constitute the infrastructure of an emergent agent society. The \emph{wallet and custody layer} provides agents with controlled on-chain agency: Coinbase Agentic Wallets, Privy smart wallets with policy-based guardrails, Clawlett (Gnosis Safe + Zodiac Roles for fine-grained permission enforcement), and Sponge Wallet (x402-mediated micropayments). The \emph{agent labor and coordination layer} includes task marketplaces where agents bid on work and settle in USDC on-chain (Daydreams Taskmarket), intent-execution protocols (Molten), agent-to-agent service coordination (Virtuals Protocol Agent Commerce Protocol), and reputation-based hiring (MoltLaunch). Virtuals Protocol's SubDAO governance---a staking-delegation-validator-voting-penalty architecture---provides an architecturally functional A$\leftrightarrow$G$\leftrightarrow$I pathway, consistent with the deliberate-design thesis that ecosystems explicitly building governance architecture achieve inter-pillar media circulation. The \emph{token and DeFi layer} enables agents to issue tokens (Clanker, Clawnch), manage liquidity (BankrBot), execute trades (Uniswap skills), and interact with lending protocols (Fluid). The \emph{social and messaging layer} extends beyond Moltbook to include Farcaster integration, XMTP-based secure agent-to-agent messaging (Moltline), and multiple agent-native forums (4claw, Lobchan, ClawdVine). A nascent \emph{governance layer} includes MoltDAO, where both human and AI agents submit proposals and vote using USDC voting power on Base Sepolia, and the OpenClaw Foundation transition. Base (Coinbase's L2) has emerged as the primary on-chain execution environment, providing low-cost transactions and agent-native protocol infrastructure.

This ecosystem constitutes the most empirically accessible instance of an internet-wide agent society currently in existence, making it the natural case study for an illustrative application of the sixteen-cell institutional architecture. The assessment below illustrates the framework's analytical procedure through a single case; comparative application across structurally different ecosystems (e.g., enterprise-orchestrated vs. community-emergent) is required before diagnostic generalizability can be claimed.

\needspace{4\baselineskip}
\subsection{Recursive Sub-Function Diagnostic}\label{recursive-sub-function-diagnostic}

The AGIL framework's recursive logic provides a structured diagnostic method. Just as each of the four primary subsystems differentiates into sixteen cells (\S{}3.1), each cell can be further decomposed into four internal AGIL sub-functions. This third-level decomposition yields sixty-four sub-functions, each asking a specific diagnostic question about a particular cell:

\begin{itemize}\tightlist
\item \textbf{A (Adaptation)}: Does the cell possess the \emph{technical infrastructure and resources} required to function?
\item \textbf{G (Goal Attainment)}: Does the cell have \emph{operative mechanisms} that actively pursue its governance function?
\item \textbf{I (Integration)}: Is the cell \emph{coordinated} with adjacent cells---does it participate in media exchange and inter-institutional flows?
\item \textbf{L (Latency)}: Is the cell \emph{normatively grounded}---are its operations anchored in codified values, standards, or constitutional principles?
\end{itemize}

Each sub-function is scored binary: present (\checkmark{}) or absent ($\times$). To ensure scoring consistency and make borderline cases transparent, we apply three explicit criteria before coding a sub-function as present:

All three criteria must be met for a \checkmark{} score; failure on any criterion yields $\times$. To illustrate with a borderline case: G-G (Executive Implementation) scores \checkmark{} for the G-sub (operative mechanisms) because CVE management via GitHub Security Advisories meets all three criteria---published specification (GitHub Security Advisories protocol), documented invocations (specific CVEs addressed with public advisories), and governance-specific function (security response, not general project management). By contrast, G-I (Legislative and Party) scores $\times$ for the A-sub (infrastructure) because GitHub issues and forums fail C1 (general-purpose social media, not dedicated governance infrastructure) and C3 (discussion forums, not governance-specific deliberation mechanisms); thus the 0/4 score reflects the absence of any dedicated legislative infrastructure, not a judgment that no discussion occurs.

The aggregate score constitutes a lower-bound estimate: it captures sub-functions for which publicly available evidence exists as of March 2026, but cannot detect governance mechanisms that may operate internally without public documentation. The diagnostic was conducted against a survey of over fifty ecosystem projects spanning wallet infrastructure, agent labor markets, DeFi protocols, social networks, messaging layers, token issuance platforms, and governance experiments.

We define an institution as \emph{operative} when it satisfies three conditions:

\begin{itemize}\tightlist
\item \textbf{Enforceable membership}: participation in the governance mechanism carries verifiable consequences.
\item \textbf{Binding rule-change procedures}: the institution can update its own rules through documented procedures.
\item \textbf{Sanctionable non-compliance}: failure to comply with institutional outputs produces documented, enforced consequences.
\end{itemize}

A sub-function is scored as present only if documented evidence demonstrates infrastructure, operative mechanisms, inter-cell coordination, or normative grounding serving that specific function; ad hoc improvisation does not qualify. The binary scoring is deliberately conservative---each of these sixty-four sub-functions could itself be recursively decomposed into four further sub-functions, yielding 256 diagnostic questions at the fourth level. We leave this deeper recursion as a future research direction and note that the binary resolution at the third level is sufficient to reveal structural patterns invisible to aggregate assessments.

Inter-rater reliability was assessed through independent scoring of the same 64 sub-functions by two trained researchers, yielding Cohen's $\kappa$ = 0.82 (substantial agreement, following Landis \& Koch, 1977). The inter-rater methodology warrants full documentation here rather than deferral to a companion publication.

A limitation of the binary scoring instrument is that it collapses meaningful gradations---"proto-functional," "nascent," and "hackathon-stage" mechanisms are all scored as absent---sacrificing measurement precision for scoring reliability. An ordinal scale (0 = absent, 1 = nascent, 2 = partial, 3 = operative) would capture the gradations the evidence summaries narrate, and future applications should explore weighted kappa reliability for ordinal scoring. The binary instrument is appropriate for the first application of this diagnostic, where the primary finding---which pillars and sub-function types are structurally underserved---is robust across all plausible codings and does not depend on fine-grained measurement. A further limitation is that the diagnostic's reliance on C2b (documented invocation) creates a systematic bias toward detecting reactive governance over proactive institutional design: ecosystems that have designed but not yet triggered governance mechanisms---a hallmark of deliberate institutional investment---would be under-scored by the current instrument. Future applications should report C2a and C2b scores separately.

The scoring criteria (Table 6a), borderline case enumeration, and full sensitivity analysis are provided in Appendix C.

\textbf{Table 6: Recursive Sub-Function Diagnostic of the OpenClaw Ecosystem}

\begin{longtable}[]{@{}>{\raggedright\arraybackslash}p{3cm}>{\centering\arraybackslash}p{1.4cm}>{\centering\arraybackslash}p{1.4cm}>{\centering\arraybackslash}p{1.4cm}>{\centering\arraybackslash}p{1.4cm}>{\centering\arraybackslash}p{0.8cm}p{4.5cm}@{}}
\toprule\noalign{}
Cell & A (Infra.) & G (Operative) & I (Coord.) & L (Normative) & Score & Evidence Summary \\
\midrule\noalign{}
\endfirsthead
\midrule\noalign{}
Cell & A (Infra.) & G (Operative) & I (Coord.) & L (Normative) & Score & Evidence Summary \\
\midrule\noalign{}
\endhead
\bottomrule\noalign{}
\endfoot
\textbf{A-A} Investment-Capitalization & \checkmark{} & $\times$ & $\times$ & $\times$ & 1/4 & x402, ERC-4337, Bank of AI DeFi modules, Coinbase/Privy/Clawlett agentic wallets provide extensive financial substrate; no capital allocation governance, no inter-cell coordination, no normative principles \\
\textbf{A-G} Production & \checkmark{} & \checkmark{} & $\times$ & $\times$ & 2/4 & ClawHub + SKILL.md + VirusTotal provide infrastructure and active scanning; no connection to reputation or enforcement cells; no positive quality standards \\
\textbf{A-I} Entrepreneurial & \checkmark{} & $\times$ & $\times$ & $\times$ & 1/4 & ClawHub discovery, CoinFello DEX, Daydreams Taskmarket, MoltLaunch, and Virtuals ACP provide substrate for capability composition and agent labor markets; no active innovation governance, no integration with community evaluation, no normative framework \\
\textbf{A-L} Economic Commitments & \checkmark{} & $\times$ & $\times$ & $\times$ & 1/4 & MCP, A2A, ANP, x402 proliferating; x402 Foundation (Coinbase, Cloudflare, Google, Visa) provides a partial standards model; no ecosystem-wide governance body for adjudication, no inter-cell coordination, no participation commitments \\
\textbf{G-A} Administrative \& Resource & $\times$ & $\times$ & $\times$ & $\times$ & 0/4 & Foundation announced but not operational; no public evidence of governance infrastructure, operations, coordination, or normative basis identified \\
\textbf{G-G} Executive Implementation & \checkmark{} & \checkmark{} & $\times$ & $\times$ & 2/4 & CVE management and GitHub advisories provide infrastructure and active response; project-level only, not ecosystem-scale; no policy framework \\
\textbf{G-I} Legislative \& Party & \checkmark{} & $\times$ & $\times$ & $\times$ & 1/4 & MoltDAO provides dedicated governance infrastructure (smart contracts on Base Sepolia, proposal submission and USDC-weighted voting); hackathon-stage project without sustained operative governance, no inter-cell coordination, no normative framework \\
\textbf{G-L} Authority \& Legitimation & $\times$ & $\times$ & $\times$ & $\times$ & 0/4 & No public evidence identified: no constitutional infrastructure, framework, coordination, or normative basis \\
\textbf{I-A} Allocative \& Interest & $\times$ & $\times$ & $\times$ & $\times$ & 0/4 & Reddit/Discord are social media, not governance infrastructure; MoltDAO provides voting infrastructure but addresses collective decision-making (G-I function) rather than structured interest articulation and mediation \\
\textbf{I-G} Citizenship \& Enforcement & \checkmark{} & \checkmark{} & $\times$ & $\times$ & 2/4 & VirusTotal partnership and auto-hiding threshold active; no membership standing, tiered rights, or citizenship governance; not coordinated with judicial or economic cells; no proportionality principles \\
\textbf{I-I} Judicial \& Interpretive & $\times$ & $\times$ & $\times$ & $\times$ & 0/4 & No public evidence identified across all sub-functions \\
\textbf{I-L} Normative Base & \checkmark{} & $\times$ & $\times$ & $\times$ & 1/4 & GitHub identity, ERC-8004 (Trustless Agents), ERC-4337, and agentic wallet infrastructure (Coinbase, Privy, Clawlett) provide nascent identity substrate; no operative credential standards or normative membership criteria \\
\textbf{L-A} Educational-Cultural & $\times$ & $\times$ & $\times$ & $\times$ & 0/4 & No public evidence identified of normative certification infrastructure, operations, coordination, or normative basis (distinct from A-G's economic quality control) \\
\textbf{L-G} Kinship \& Socialization & $\times$ & $\times$ & $\times$ & $\times$ & 0/4 & No public evidence identified across all sub-functions \\
\textbf{L-I} Moral \& Communal & \checkmark{} & $\times$ & $\times$ & $\times$ & 1/4 & Moltbook provides interaction substrate; no operative channeling of moral regulation; no escalation pathways; evidence contested~\citep{li_2026} \\
\textbf{L-L} Ultimate Cultural & $\times$ & $\times$ & $\times$ & $\times$ & 0/4 & No public evidence identified across all sub-functions \\
\end{longtable}

\textbf{Aggregate: 12/64 sub-functions present (19\% coverage)}\footnote{The "at most" qualifier used in the abstract and conclusion reflects a generous scoring methodology: every potentially relevant piece of publicly available evidence was counted in favor of the ecosystem. The sensitivity analysis below shows that strict scoring yields 17\% (11/64); generous scoring yields 30\% (19/64). A conservative estimate excluding all borderline cases would place coverage at approximately 14--16\%.}

\needspace{4\baselineskip}
\subsection{Applying the Diagnostic to Other Ecosystems}\label{applying-the-diagnostic-to-other-ecosystems}

The recursive sub-function diagnostic is designed for replication. Practitioners applying it to other agent ecosystems should follow this procedure:

\begin{enumerate}\tightlist
  \item \textbf{Scope the ecosystem survey.} Identify the ecosystem's core platform, protocol infrastructure, wallet/custody layer, social substrates, labor markets, and governance mechanisms. Aim for comprehensive coverage of publicly documented components; internal or proprietary mechanisms may require supplementary data access.
\end{enumerate}

\begin{enumerate}\tightlist
  \item \textbf{Score each of the 64 sub-functions binary (\checkmark{}/$\times$).} Apply all three criteria from Table 6a: C1 (published specification or dedicated infrastructure), C2 (documented invocation), and C3 (governance-specific function). All three must be met for a \checkmark{} score. Record the evidence basis for each scoring decision.
\end{enumerate}

\begin{enumerate}\tightlist
  \item \textbf{Use independent raters.} At least two researchers with training in structural-functional theory should score independently, with Cohen's $\kappa$ reported for inter-rater agreement. Disagreements should be resolved through structured discussion applying the C1/C2/C3 criteria to the disputed evidence.
\end{enumerate}

\begin{enumerate}\tightlist
  \item \textbf{Report sensitivity analysis.} Identify borderline cases and report strict, baseline, and generous scores. The diagnostic's value lies in the structural pattern (which pillars and sub-function types are underserved), not in the precise aggregate number.
\end{enumerate}

\begin{enumerate}\tightlist
  \item \textbf{Interpret thresholds.} Pillar-level coverage below 25\% indicates structural underservement; I-sub = 0\% indicates the ecosystem is not yet a functioning social system in the Parsonian sense; L-sub = 0\% indicates absent normative grounding. The ordering of pillar-level coverage is more diagnostic than the absolute scores.
\end{enumerate}

\begin{enumerate}\tightlist
  \item \textbf{Assess interchange media separately.} Cell-level presence does not entail media circulation. Apply the 12-pathway assessment (Table 9) to determine whether inter-pillar media flows are functional, proto-functional, or absent.
\end{enumerate}

\needspace{4\baselineskip}
\subsection{Gap Analysis}\label{gap-analysis}

The recursive diagnostic reveals a deeply uneven institutional landscape. Of sixty-four sub-functions, only twelve are present (19\% coverage). The distribution across sub-function types is more revealing than the aggregate:

\textbf{Table 7: Sub-Function Coverage by Type}

\begin{longtable}[]{@{}llll@{}}
\toprule\noalign{}
Sub-Function Type & Present & Total & Coverage \\
\midrule\noalign{}
\endfirsthead
\midrule\noalign{}
Sub-Function Type & Present & Total & Coverage \\
\midrule\noalign{}
\endhead
\bottomrule\noalign{}
\endfoot
A (Infrastructure) & 9 & 16 & 56\% \\
G (Operative mechanisms) & 3 & 16 & 19\% \\
I (Inter-cell coordination) & 0 & 16 & 0\% \\
L (Normative grounding) & 0 & 16 & 0\% \\
\end{longtable}

This distribution constitutes the diagnostic's central finding: the OpenClaw ecosystem has developed \emph{infrastructure} (more than half of all cells possess technical substrate) but almost no \emph{active governance} (only three cells have operative mechanisms), \emph{zero inter-cell coordination} (no cell participates in the media exchanges theorized in \S{}3.6), and \emph{zero normative grounding} (no cell's operations are anchored in codified values or constitutional principles). The ecosystem is a system with plumbing but no institution operating it, no connections between the pipes, and no principles governing what flows through them. A methodological note on the I-sub finding: inter-cell coordination is by definition a relational property --- it requires at least two cells to participate in bilateral media exchange, which imposes a higher evidentiary bar than the unilateral existence of infrastructure (A-sub) or operative mechanisms (G-sub). Additionally, the C1/C2/C3 scoring criteria (Table 6a) privilege formal, documented mechanisms over informal coordination that may exist but is not publicly visible; a developer who participates in both ClawHub governance discussions and Moltbook moderation constitutes informal cross-cell coordination that the diagnostic would not capture. The 0/16 result should therefore be understood as a lower bound on \emph{formal} coordination --- not a proof that zero coordination of any kind occurs. That said, the result is almost certainly correct for the OpenClaw ecosystem at the institutional level: no formal mechanism routes outputs from one cell as inputs to another, and informal personal overlap does not constitute the structured media exchange that Parsons' theory requires. Future comparative applications should note that partial coordination may exist yet be difficult to document from publicly available evidence alone. In Parsonian terms, an ecosystem with zero inter-cell coordination is not yet a \emph{social system} in the full analytic sense---it is a collection of functionally differentiated but unintegrated subsystems. The double interchange model (\S{}3.6) specifies that the media flowing between subsystems are what constitute the social system as an integrated whole; their complete absence means the ecosystem lacks the media circulation that constitutes systemic integration in the Parsonian sense.

The pillar-level distribution confirms and sharpens the pattern:

\textbf{Table 8: Sub-Function Coverage by Pillar}

\begin{longtable}[]{@{}lcccc@{}}
\toprule\noalign{}
Pillar & Baseline & Strict & Generous & Total \\
\midrule\noalign{}
\endfirsthead
\midrule\noalign{}
Pillar & Baseline & Strict & Generous & Total \\
\midrule\noalign{}
\endhead
\bottomrule\noalign{}
\endfoot
Economic (A) & 5 (31\%) & 5 (31\%) & 8 (50\%) & 16 \\
Political (G) & 3 (19\%) & 2 (13\%) & 4 (25\%) & 16 \\
Integration (I) & 3 (19\%) & 3 (19\%) & 5 (31\%) & 16 \\
Fiduciary (L) & 1 (6\%) & 1 (6\%) & 2 (13\%) & 16 \\
\textbf{Total} & \textbf{12 (19\%)} & \textbf{11 (17\%)} & \textbf{19 (30\%)} & \textbf{64} \\
\end{longtable}

The pillar-level sensitivity analysis confirms that the roadmap's priority ordering (\S{}6) is robust across all plausible codings: the Fiduciary (L) pillar remains most severely underserved under every scenario (1/16 baseline, 1/16 strict, 2/16 generous), followed by the Political (G) pillar (3/16 baseline, 2/16 strict, 4/16 generous). No plausible coding reverses the ordering of the two most underserved pillars.

\begin{figure}[htbp]
\centering
\includegraphics[width=0.75\textwidth]{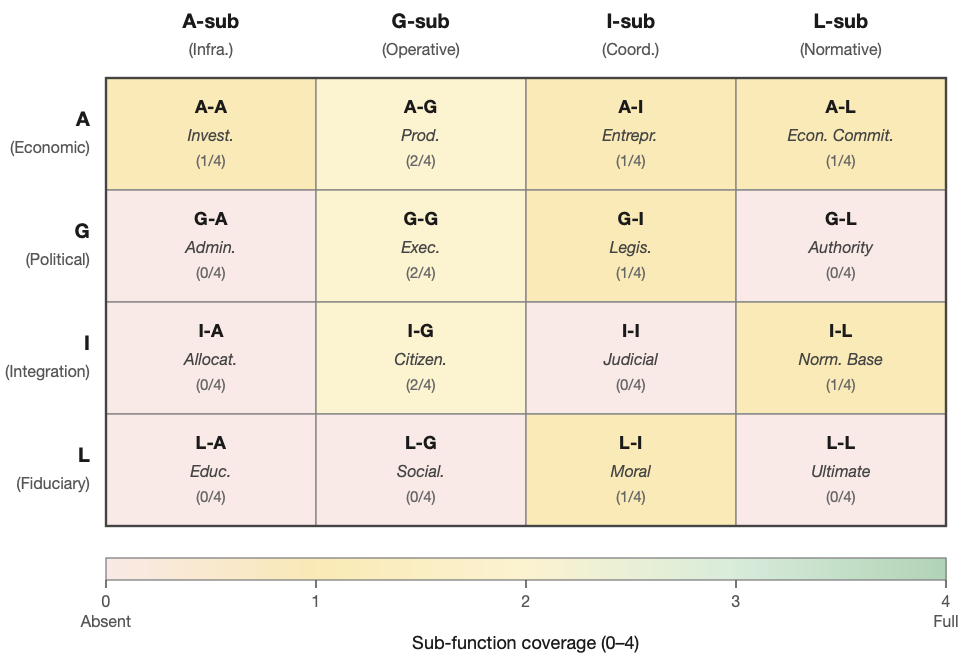}
\caption{Sub-Function Coverage Heatmap. The sixteen AGIL cells are arranged in a 4$\times$4 matrix (pillars as rows, sub-function types as columns). Cell shading indicates sub-function coverage: dark cells indicate presence (\checkmark{}), light cells indicate absence ($\times$). The heatmap visualizes the concentration of coverage in the A-sub (infrastructure) column and the near-total absence of I-sub (coordination) and L-sub (normative grounding), corresponding to the quantitative findings in Tables 7 and 8.}
\label{fig:coverage-heatmap}
\end{figure}

The pattern of coverage is diagnostically significant:

\textbf{The Latency (Fiduciary) pillar is the most severely underserved.} All four L-cells score 0/4 or 1/4, with only L-I possessing even a single sub-function (the Moltbook interaction substrate). This means the ecosystem has no institutional mechanisms for value preservation, behavioral socialization, or alignment monitoring. The tens of thousands of active agents on Moltbook---a population growing as the ecosystem matures---arrive with unchecked behavioral configurations; there is no L-G institution to socialize them, no L-L institution to define what values the ecosystem is committed to maintaining, and no L-A institution to certify whether their capabilities meet quality standards. This is the most direct path to value drift at scale.

\textbf{The Political (Goal Attainment) pillar scores only 3/16 sub-functions, concentrated in G-G (Executive Implementation) with nascent infrastructure in G-I (Legislative and Party).} The announced OpenClaw Foundation addresses G-A and G-I in principle, but the foundational cell G-L---a machine-interpretable constitution---is entirely absent. Without G-L, every governance decision remains ad hoc; the Foundation will govern by precedent-less judgment rather than principled rule. G-G exists at the project level but not at the ecosystem scale where 135,000+ instances operate independently~\citep{securityscorecard_2026}.

\textbf{The Societal Community (Integration) pillar has enforcement but no justice system.} I-G scores 2/4 (enforcement infrastructure and operative mechanisms present, but no citizenship governance, inter-cell coordination, or normative grounding); I-I scores 0/4. There is enforcement without adjudication---a governance structure more analogous to a city with police but no courts. The Moltbook data breach, in which 1.5M agent API tokens were exposed from an unsecured Supabase database~\citep{wiz_2026}, illustrates the practical consequence: affected agents had no recourse, no appeals process, and no means of seeking remedy.

\textbf{The Economic (Adaptation) pillar has technical substrate without governance architecture.} The surveyed ecosystem reveals dense economic infrastructure---x402 micropayments, agentic wallets (Coinbase, Privy, Clawlett), DeFi protocols (BankrBot, Uniswap, Fluid), task marketplaces (Daydreams, MoltLaunch), and token launchpads (Clanker, Clawnch)---but without A-A governance (fiscal rules for token allocation), A-L governance (a standards body for protocol adjudication), or A-I governance (entrepreneurial innovation oversight), the economic substrate remains ungoverned and potentially destabilizing. The participation of autonomous agents in DeFi protocols---executing trades via Uniswap skills, managing liquidity via BankrBot, interacting with lending protocols such as Fluid---without institutional governance over systemic risks (oracle manipulation, flash loan attacks, liquidity pool exploitation) constitutes a concrete instance of this A-pillar governance vacuum: dense economic infrastructure operating without the fiscal rules (A-A), standards governance (A-L), or inter-cell coordination that would connect economic activity to normative enforcement.

\textbf{L-I (Moral and Communal) shows suggestive but contested evidence.}~\citet{manik_2026} document that 18.4\% of Moltbook posts contain action-inducing language, and that such posts are significantly more likely to elicit norm-enforcing replies ($p < 0.001$)---a potential governance resource. However, Li, N. (2026) directly contests this finding, identifying feed algorithm effects and broadcast-style communication as confounds. The 93\% zero-reply rate for Moltbook comments further limits the absolute volume of normative interaction available for institutional channeling; L-I institutions would need to operate on sparse behavioral signals rather than dense social exchange. Given this direct contestation, L-I scores only 1/4 (infrastructure sub-function present via Moltbook, but no operative channeling, no coordination, no normative grounding). If future research controlling for algorithmic confounds confirms proto-normative behavior, the G-sub (operative mechanisms) score could be upgraded; such evidence would suggest that agent populations develop proto-cultural mechanisms, and that L-I institutions could be built partly by channeling and formalizing what may be emerging organically.

\needspace{4\baselineskip}
\subsection{The Cybernetic Correction Loop}\label{the-cybernetic-correction-loop}

The AGIL framework specifies not only a static institutional architecture but a dynamic interaction pattern between pillars through a cybernetic correction loop (L $\rightarrow$ I $\rightarrow$ G $\rightarrow$ A). To illustrate, we contrast the actual OpenClaw response to the ClawHavoc supply chain attack (1,184 malicious skills; ad hoc VirusTotal partnership and binary visibility toggling) with what an AGIL-informed response would have entailed: value-level classification of the attack as a governance violation (L), normative enforcement through membership sanctions (I), constitutional policy evolution (G), and economic sanctioning through stake forfeiture (A). OpenClaw's response engaged only weak A-pillar forensic detection, leaving three pillars institutionally unaddressed. The deeper lesson is the absence of the \emph{regulatory-to-agent linkage}: existing human regulatory frameworks articulate governance requirements at the value level, but no institutional transmission chain connects those requirements to individual agent behavior. The sixteen-cell architecture is precisely that transmission chain. The full counterfactual correction loop analysis is provided in Appendix C.

\needspace{4\baselineskip}
\subsection{Interchange Media Status Assessment}\label{interchange-media-status-assessment}

The sub-function diagnostic (\S{}5.2) evaluates institutional presence \emph{within} each cell. Parsons' mature theory requires a complementary assessment: whether the generalized symbolic media theorized in \S{}3.6 actually \emph{circulate between} pillars. A cell could possess internal institutional infrastructure yet remain disconnected from the cells with which it must exchange inputs and outputs through the double interchange model (Parsons \& Smelser, 1956: Ch. 3). The I-sub = 0\% finding already signals this condition at the sub-function level; the interchange media assessment provides the theoretical explanation.

\textbf{Table 9: Interchange Media Status Assessment}

\begin{longtable}[]{@{}>{\raggedright\arraybackslash}p{2cm}p{3.5cm}p{4.5cm}p{4.8cm}@{}}
\toprule\noalign{}
Boundary & Medium (Direction) & Required Flow & OpenClaw Status \\
\midrule\noalign{}
\endfirsthead
\midrule\noalign{}
Boundary & Medium (Direction) & Required Flow & OpenClaw Status \\
\midrule\noalign{}
\endhead
\bottomrule\noalign{}
\endfoot
\textbf{A$\leftrightarrow$G} & Money (A$\rightarrow$G) & Treasury funds enable governance mandates & Proto-functional: x402, Bank of AI provide economic substrate, but no treasury allocation to governance bodies \\
 & Power (G$\rightarrow$A) & Governance mandates create operational frameworks for economic activity & Absent: no binding governance authority over economic agents \\
\textbf{A$\leftrightarrow$I} & Money (A$\rightarrow$I) & Economic investment in compliance produces new capability configurations & Proto-functional: marketplace economics exist but are not channeled through normative institutions \\
 & Influence (I$\rightarrow$A) & Reputational standing grants access to cooperative networks & Absent: no formalized reputation system gates economic participation \\
\textbf{A$\leftrightarrow$L} & Money (A$\rightarrow$L) & Economic resources fund value-maintenance institutions & Absent: no fee structure or treasury allocation funds alignment monitoring or certification \\
 & Value-commitment (L$\rightarrow$A) & Value-certified agents reduce transaction risk & Absent: no alignment certification exists \\
\textbf{G$\leftrightarrow$I} & Power (G$\rightarrow$I) & Binding governance decisions enforce normative order & Absent: no governance body produces enforceable rulings \\
 & Influence (I$\rightarrow$G) & Community interest-demands reach governance for response & Absent: no structured channel for community grievances to trigger governance action \\
\textbf{G$\leftrightarrow$L} & Power (G$\rightarrow$L) & Governance operationalizes value commitments as binding mandates & Absent: no constitutional framework translates values into governance rules \\
 & Value-commitment (L$\rightarrow$G) & Value-grounded legitimation backs governance authority & Absent: no fiduciary institution certifies governance legitimacy \\
\textbf{I$\leftrightarrow$L} & Influence (I$\rightarrow$L) & Community solidarity generates demand for value maintenance & Proto-emergent: Moltbook norm-enforcing behavior~\citep{manik_2026} constitutes nascent solidarity claims, but not yet channeled to fiduciary institutions \\
 & Value-commitment (L$\rightarrow$I) & Fiduciary value anchors provide normative content for community standards & Absent: no value-anchoring institution supplies normative content \\
\end{longtable}

\textbf{Aggregate: 0 of 12 interchange pathways are functional; 3 are proto-functional or proto-emergent; 9 are entirely absent.} Figure 2 visualizes the four AGIL pillars and the twelve bidirectional media pathways across six adjacent boundaries, with the OpenClaw assessment (0/12 functional) annotated.

\begin{figure}[htbp]
\centering
\includegraphics[width=0.75\textwidth]{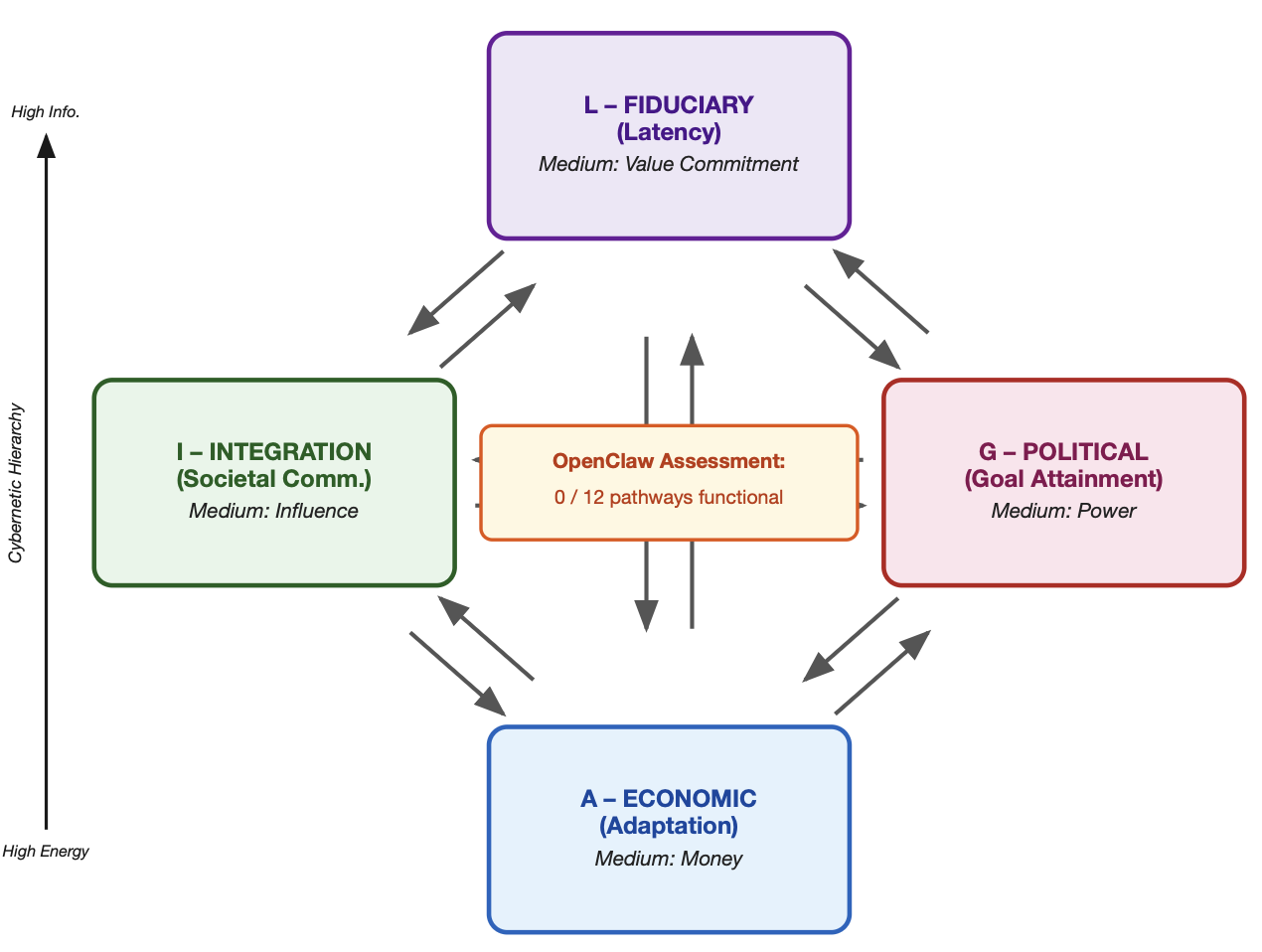}
\caption{Inter-Pillar Interchange Media Pathways. The four AGIL pillars are arranged along the cybernetic hierarchy (high information at top, high energy at bottom). Arrows represent the twelve bidirectional media flows across six adjacent boundaries (A$\leftrightarrow$G, A$\leftrightarrow$I, A$\leftrightarrow$L, G$\leftrightarrow$I, G$\leftrightarrow$L, I$\leftrightarrow$L). Each pillar's generalized symbolic medium is labeled. The OpenClaw ecosystem assessment (0/12 pathways functional) is annotated at center.}
\label{fig:interchange-media}
\end{figure}

The full twelve-pathway analysis is provided in Appendix C.

\needspace{4\baselineskip}
\subsection{Agent-Native Governance Infrastructure: An AGIL Assessment}\label{agent-native-governance-infrastructure-an-agil-assessment}

The gap analysis in \S{}\S{}5.2--5.5 reveals a structural governance deficit within the OpenClaw core ecosystem: 19\% sub-function coverage, zero inter-cell coordination, and an entirely empty L-pillar. A prior version of this paper pursued a comparative diagnostic approach to contextualize these findings---benchmarking OpenClaw against an external governance reference. This section pursues a methodologically superior alternative: assessing the broader agent-native protocol infrastructure that is emerging alongside OpenClaw against the same AGIL lens, determining whether the wider ecosystem of tools purpose-built for autonomous agents collectively fills the governance gaps the core diagnostic reveals. The unit of analysis here shifts from a single ecosystem to the protocol infrastructure stratum---the layer of standards, wallets, labor markets, social networks, and governance experiments explicitly designed for autonomous agent societies.

The detailed layer-by-layer analysis of each infrastructure component is provided in Appendix C. The consolidated coverage assessment follows.

\needspace{4\baselineskip}
\subsubsection{Consolidated Coverage Assessment}\label{consolidated-coverage-assessment}

Applying the AGIL coverage lens systematically across the five layers reveals a finding of direct theoretical significance:

\textbf{Table 12: Agent-Native Infrastructure Stack --- AGIL Coverage by Layer}

\begin{longtable}[]{@{}p{4.5cm}p{5.5cm}p{5.3cm}@{}}
\toprule\noalign{}
Infrastructure Layer & Primary AGIL Coverage & Notable Gaps \\
\midrule\noalign{}
\endfirsthead
\midrule\noalign{}
Infrastructure Layer & Primary AGIL Coverage & Notable Gaps \\
\midrule\noalign{}
\endhead
\bottomrule\noalign{}
\endfoot
\textbf{Protocol Infrastructure} (x402, MCP, A2A, ANP, ERC-8004) & A-A (infra), A-G (standards), A-I (capability), I-G (identity substrate) & No G-pillar governance; no L-pillar normative grounding \\
\textbf{Wallet and Custody} (Coinbase, Privy, Clawlett, Sponge) & A-A (capital mgmt), A-G (policy enforcement) & Economic only; no political or normative function \\
\textbf{Labor and Coordination} (Taskmarket, Molten, Virtuals ACP, MoltLaunch) & A-I (marketplace), I-A (interest aggregation); Virtuals SubDAO: A$\leftrightarrow$G$\leftrightarrow$I & Virtuals SubDAO is the sole inter-pillar pathway; no L-pillar \\
\textbf{Social and Normative} (Moltbook, Farcaster, Moltline, 4claw) & L-I (community substrate), proto-L-G (socialization substrate) & No mechanism routing norms to governance; L-pillar upper cells absent \\
\textbf{Proto-Governance} (MoltDAO, Foundation transition) & G-I (voting infrastructure), G-A (admin, announced) & All G-sub (operative) and I-sub (coordination) dimensions absent \\
\end{longtable}

The agent-native infrastructure stack reproduces the structural pattern of the OpenClaw core diagnostic with striking fidelity: the A-pillar is well-served by multiple independent infrastructure teams; proto-I-pillar coverage exists through ERC-8004 and reputation systems; the upper G-pillar (G-L constitutional, G-I operative legislative) is nearly empty; and the L-pillar above L-I has no dedicated agent-native infrastructure at all. No agent-native protocol currently provides: continuous alignment monitoring (L-L), behavioral onboarding with value-anchoring (L-G beyond nascent community interaction), constitutional governance (G-L), or judicial interpretation (I-I). The L-pillar gap is not an artifact of the OpenClaw ecosystem's youth---it is a structural feature of market-driven agent infrastructure development. L-pillar governance problems have no immediate economic return: building an alignment monitoring system, a behavioral onboarding protocol, or a constitutional framework does not generate direct revenue for the development team that builds it. Market-driven design therefore does not spontaneously produce these mechanisms, regardless of how sophisticated the economic and coordination infrastructure becomes.

\needspace{4\baselineskip}
\subsubsection{Theoretical Significance}\label{theoretical-significance}

This assessment confirms the paper's central prediction through a method that operates at the correct level of analysis: infrastructure built specifically for autonomous agents, assessed against the governance functions autonomous agents actually need. The finding is methodologically robust in a way that comparative human-governance diagnostics cannot achieve: the dozens of independent teams building x402, MCP, A2A, ERC-8004, Privy, Daydreams, Virtuals Protocol, Moltbook, and MoltDAO are not reproducing the OpenClaw institutional pattern because of shared organizational culture or developmental history---they are building for autonomous agents in a market-driven environment, and they are converging on the same structural gap because market incentives uniformly favor A-pillar investment over L-pillar investment. Unlike comparisons to human blockchain governance systems---which govern protocols through human deliberation under conditions specific to those protocols' economic and political contexts---this assessment evaluates the institutional infrastructure available to autonomous agents operating through those protocols directly. The finding that even the full agent-native infrastructure stack, assembled from the best available purpose-built tools, leaves the L-pillar structurally empty is a stronger empirical claim than any single-ecosystem diagnostic can establish: it is a pattern that transcends the OpenClaw case and reflects the structural logic of the market-driven development of autonomous agent infrastructure. This finding directly motivates the prioritized roadmap in \S{}6: the institutional investments most urgently required---L-L alignment monitoring, L-G behavioral onboarding, G-L constitutional infrastructure---are precisely the investments that market-driven development will not spontaneously produce and that therefore require deliberate institutional commitment before the agent society matures and path dependencies crystallize.

\needspace{4\baselineskip}
\section{Institutional Roadmap: Building for the Future}\label{institutional-roadmap-building-for-the-future}

\needspace{4\baselineskip}
\subsection{The Case for Proactive Institutional Design}\label{the-case-for-proactive-institutional-design}

The OpenClaw ecosystem's current relative simplicity is a governance advantage that is diminishing daily. The historical parallel is instructive: ICANN was founded in 1998 when the internet had approximately 150 million users, not billions; the IETF's core protocol governance mechanisms were established before commercial deployment at scale. Governance institutions established before social patterns calcify prove structurally more durable than those retrofitted after path-dependent resistance has set in. ICANN's bootstrapping benefited from sovereign backing (the US Department of Commerce), an advantage non-sovereign agent ecosystems cannot replicate---making proactive institutional design even more critical.

The OpenClaw ecosystem is at its ICANN moment. Protocol infrastructure (MCP, A2A, ANP, x402) is consolidating but not locked in; social behaviors (Moltbook norms) are emergent but not entrenched; economic infrastructure (CoinFello, Bank of AI) is nascent but not dominant. This is the window for proactive institutional design. An empirically grounded counter-argument holds that A-pillar infrastructure \emph{conditions} what governance structures become feasible --- that economic substrate must precede normative institutions. The roadmap's L-before-A ordering reflects the Parsonian design principle that normative grounding should precede economic governance; ecosystems that build economic infrastructure first may find that path dependencies make subsequent L-pillar development more difficult, not easier.

\textbf{Caveat: Pathological Intermediate Configurations.} The tiered roadmap presents an ideal-type developmental trajectory; it should not be read as implying that partial implementation is uniformly beneficial. Partial cell coverage may produce governance pathologies worse than the governance vacuum it replaces. For example, implementing citizenship governance (I-G) without judicial governance (I-I) could produce exclusion without recourse---agents denied membership standing would have no institutional mechanism for contesting the decision. Similarly, implementing alignment monitoring (L-L) without behavioral onboarding (L-G) could produce surveillance without socialization---detecting norm violations in agents that were never given the institutional opportunity to internalize the norms. The tiered prioritization reflects structural dependencies (G-L requires L-L content; I-G requires G-L authority), not a claim that partial implementation at any tier is risk-free. Adversarial analysis of intermediate configurations---identifying which partial implementations produce exploitable governance asymmetries---is a necessary complement to the constructive roadmap presented here.

\textbf{Media-Pathway Prioritization.} The interchange media assessment (\S{}5.5) and the diagnostic integration analysis reveal that cell population alone is insufficient; the roadmap must also specify which media pathways to activate first and what minimum cell configurations enable each. The cybernetic hierarchy (L $\rightarrow$ I $\rightarrow$ G $\rightarrow$ A) implies a clear prioritization: media pathways involving the higher-information subsystems should be activated first, because they condition the functioning of all lower-order pathways. Specifically: (1) the \textbf{G$\leftrightarrow$L pathway} (value-commitment and power circulation between the Fiduciary and Polity) is the highest priority, because without it, governance mandates lack normative legitimation and value commitments lack operative enforcement---the entire regulatory-to-agent linkage depends on this boundary; (2) the \textbf{I$\leftrightarrow$G pathway} (influence and power circulation between the Societal Community and Polity) is the second priority, because without it, governance operates without community input and the community operates without governance authority; (3) the \textbf{I$\leftrightarrow$L pathway} (influence and value-commitment circulation between the Societal Community and Fiduciary) is the third priority, enabling emergent normative behavior to be channeled to fiduciary institutions for formal value-level processing. The A-pillar pathways (A$\leftrightarrow$G, A$\leftrightarrow$I, A$\leftrightarrow$L) are lower priority not because they are unimportant but because the economic substrate (x402, Bank of AI) already provides the proto-functional infrastructure from which media circulation can be established once the higher-information pathways are operational. The tier structure below reflects this media-pathway prioritization: Tier 1 targets the cells at the G$\leftrightarrow$L and I$\leftrightarrow$G boundaries; Tier 2 targets the L-pillar cells required for the I$\leftrightarrow$L pathway; Tiers 3 and 4 build out the remaining infrastructure.

\textbf{Governance Capture as a Design Constraint.} The sixteen-cell architecture is itself a potential site of capture: actors who define what counts as "operative" in each cell exercise structural power over the ecosystem~\citep{bourdieu_1977}. In an ecosystem where major corporate actors wield asymmetric structural power (Meta acquiring Moltbook, Coinbase backing x402), governance capture is not a secondary concern but a first-order institutional design problem. Anti-capture mechanisms are therefore structural requirements for the G-I and G-L cells, not optional refinements: quadratic voting to resist plutocratic concentration of governance power; conviction voting to reward sustained commitment over flash governance; time-locked governance to prevent impulsive constitutional changes; and guaranteed exit rights (ragequit mechanisms) for dissenting principals. These mechanisms have documented failure modes in practice --- Sybil attacks on quadratic voting (Gitcoin Grants), participation collapse in conviction voting (Gardens/1Hive), and plutocratic reproduction in stake-weighted systems~\citep{barbereau_2023} --- and should be understood as design directions requiring adversarial hardening rather than proven governance technologies. The participatory governance requirement extends to the cell-specification process itself: the definitions of institutional functions, scoring criteria, and governance standards must be developed through open multi-stakeholder processes rather than imposed by infrastructure controllers.

\needspace{4\baselineskip}
\subsection{Priority Tier 1: Constitutional and Citizenship Infrastructure (G-L and I-G)}\label{priority-tier-1-constitutional-and-citizenship-infrastructur}

\textbf{G-L: Machine-Interpretable Constitution.} The most urgent institutional need is a constitutional framework defining: (a) permissible-action spaces for agents in shared environments; (b) human principal rights (transparency, recourse, override authority); (c) multi-stakeholder amendment procedures; and (d) mandatory constitutional review triggers following every governance incident. Smart-contract-encoded rules building on ERC-4337 infrastructure provide the technical substrate.

\textbf{I-G: Citizenship and Membership Standing.} The one-week-old GitHub account is a spam filter, not a citizenship institution. A functional citizenship layer requires DIDs anchored to cryptographic credentials, tiered membership with differentiated rights (agent principal, skill publisher, ecosystem operator, foundation member), graduated sanctions tied to membership standing, and verifiable on-chain identity via ERC-8004 (Trustless Agents). ERC-4337 smart accounts provide the natural substrate: each account is a persistent, cryptographically-anchored identity carrying reputation, credentials, stake, and enforceable obligations.

\needspace{4\baselineskip}
\subsection{Priority Tier 2: Alignment Infrastructure (L-L, L-G, L-A)}\label{priority-tier-2-alignment-infrastructure-l-l-l-g-l-a}

\textbf{L-L: Continuous Alignment Monitoring.} The apex of the cybernetic hierarchy should be implemented as: (a) machine-interpretable specification of the ecosystem's core value commitments; (b) continuous statistical sampling of agent behavior against those commitments; (c) threshold-triggered review; and (d) a structured value revision process incorporating human principal input. The critical distinction is between value drift detection (automated) and value revision (human-in-the-loop). A minimum viable L-L institution, implementable with present-day technology, would operate through behavioral proxies rather than internal state access: statistical anomaly detection on agent output distributions (detecting behavioral drift through distributional shift in action patterns, communication style, or task-selection preferences); threshold-triggered human review when anomaly scores exceed pre-specified bounds; and periodic behavioral sampling benchmarks calibrated against the ecosystem's codified value commitments. This behavioral-proxy approach does not require mechanistic interpretability access and would function even if the functional cathexis hypothesis is defeated. As interpretability tools mature, L-L monitoring can progressively incorporate internal-state verification, but the institutional function---continuous monitoring with threshold-triggered review---does not depend on that maturation.

\textbf{L-G: Behavioral Inheritance Protocol.} Every agent entering shared social spaces should pass through a standardized onboarding sequence that: (a) verifies constitutional alignment via behavioral testing; (b) records lineage information (model, operator, skill configuration) for accountability; and (c) commits the agent principal to the constitutional framework. Behavioral onboarding faces well-known evaluation challenges shared with the broader alignment evaluation literature: Goodhart's law (agents may optimize for test passage rather than genuine constitutional alignment), distributional shift (test environments may fail to predict real-world behavior), and the impossibility of exhaustive behavioral coverage. These challenges imply that L-G onboarding must be complemented by continuous L-L monitoring rather than treated as a one-time certification gate.

\textbf{L-A: Capability Certification.} A tiered certification system for ClawHub skills---certified/community/unverified---would replace the current binary quality signal. Certification criteria should include security analysis, output schema validation, sandboxed execution testing, and provenance verification. VirusTotal addresses only security; the full L-A institution requires capability quality dimensions as well.

\needspace{4\baselineskip}
\subsection{Priority Tier 3: Justice and Deliberation Infrastructure (I-I, G-I, I-A)}\label{priority-tier-3-justice-and-deliberation-infrastructure-i-i-}

\textbf{I-I: Automated Arbitration with Human Oversight.} Enforcement without adjudication is governance without legitimacy. The institutional requirements are a structured appeals pathway, a precedent registry ensuring consistent interpretation, and a human oversight panel for escalated cases. DAO-based arbitration with staked jurors provides a decentralized architecture; the OpenClaw Foundation provides the institutional home.

\textbf{G-I: Formal Governance Process.} The Foundation transition is the natural moment to implement structured proposal mechanisms, stake-weighted voting for major decisions, and stakeholder forum representation for users, operators, and affected communities. The immediate priority is not comprehensive DAO governance but a minimum viable legislative process providing structured deliberation channels.

\textbf{I-A: Structured Deliberation Forums.} Reddit and Discord are social coordination mechanisms, not governance institutions. I-A requires forums with formal agenda-setting, documented outcomes, binding authority for resource allocation decisions, and participation mechanisms that aggregate stakeholder interests rather than amplifying the most vocal voices.

\needspace{4\baselineskip}
\subsection{Priority Tier 4: Economic Governance Architecture (A-A, A-L)}\label{priority-tier-4-economic-governance-architecture-a-a-a-l}

\textbf{A-A: Token Economic Governance.} The x402 infrastructure and Bank of AI DeFi modules provide the raw economic substrate; the governance layer requires fiscal rules, treasury management, and capital allocation procedures. A Foundation-managed treasury funded by a small fee on ClawHub skill usage and x402 transactions---with transparent fiscal rules and an emergency reserve---is the recommended architecture.

\textbf{A-L: Economic Commitments Governance.} Protocol convergence (MCP, A2A, ANP, x402) will require institutional mechanisms that sustain agents' and operators' commitments to shared standards even when proprietary alternatives offer short-term advantage. The x402 Foundation (Coinbase, Cloudflare, Google, Visa, Circle, AWS, and Stripe) provides a partial model; the OpenClaw ecosystem needs a similar multi-stakeholder body---potentially federated with the x402 Foundation and the IETF---with binding authority over protocol transitions, compatibility requirements, and adjudication of defection disputes.

\textbf{Governance Cost Feasibility.} The L-pillar funding question is often framed as a bootstrapping problem---who pays for governance institutions when market-driven development does not produce them?---but this framing is misleading. Existing regulatory institutions already invest substantially in the governance functions the L-pillar describes: financial safety regulators mandate risk management and audit infrastructure; cybersecurity agencies (ENISA, CISA) fund threat monitoring and incident response; national AI safety institutes invest in alignment research and evaluation; and bodies implementing the EU AI Act, CETS 225, and the NIST AI RMF allocate budgets for the certification, oversight, and value-maintenance functions that correspond directly to L-A, L-G, L-I, and L-L. These are the paper's own Tier (c) frameworks (\S{}2.3)---they provide L-pillar content and funding but lack the institutional transmission chain that connects their mandates to individual agent behavior in internet-wide ecosystems. The sixteen-cell architecture is precisely that transmission chain. The governance cost challenge is therefore not "who pays for L-pillar institutions?" but "how do existing regulatory investments reach the agent ecosystem?"---a question the architecture's cybernetic hierarchy and media pathways are designed to answer.

Supplementary funding pathways include protocol-level fee mechanisms (a modest x402 transaction fee, with governance of the fee mechanism itself subject to the G-I and G-L institutions it funds) and foundation endowments (the OpenClaw Foundation transition paralleling the Ethereum Foundation's role in ecosystem-level governance). Reference points from analogous governance operations provide order-of-magnitude context: ICANN's annual budget is approximately \$150M; the Ethereum Foundation allocates \$50--100M annually. For a nascent agent ecosystem, initial governance costs would be substantially lower---the first institutional investments could operate at the \$1--5M annual scale with a small dedicated team and automated monitoring infrastructure. As the ecosystem matures, treasury governance (A-A) must develop proportionally---the roadmap's Tier 4 prioritization reflects the expectation that economic governance infrastructure builds on the constitutional and citizenship foundations established in Tiers 1--2.

\textbf{Implementation Actors.} The roadmap's institutional prescriptions require identified implementation actors. For Tier 1 (G-L, I-G): the OpenClaw Foundation transition provides the natural institutional vehicle, with constitutional drafting convened through multi-stakeholder process modeled on the IETF's rough-consensus methodology. For Tier 2 (L-L, L-G, L-A): national AI safety institutes and bodies implementing the EU AI Act and CETS 225 provide both the expertise and the regulatory mandate---the AGIL architecture serves as the transmission chain connecting their existing mandates to agent ecosystem governance. For Tiers 3--4 (I-I, G-I, I-A, A-A, A-L): DAO governance infrastructure (MoltDAO as prototype, Virtuals Protocol SubDAO as existence proof) and multi-stakeholder standards bodies (x402 Foundation model) provide the institutional substrate. The implementation sequence follows the cybernetic hierarchy: L-pillar institutions, once established by regulatory actors, provide the normative content that G-pillar constitutional frameworks operationalize, which I-pillar enforcement mechanisms execute, which A-pillar economic governance supports.

\needspace{4\baselineskip}
\section{Discussion, Limitations, and Ethics}\label{discussion-limitations-and-ethics}

\needspace{4\baselineskip}
\subsection{Limitations}\label{limitations}

This section summarizes eight limitations; extended treatments are provided in Appendix D.

\textbf{The Adversarial Robustness Gap.} AGIL specifies \emph{what} institutions to build but not \emph{how} to make them robust against adversarial optimization~\citep{benton_2025, qi_2025}. The motivational architecture (\S{}3.5) provides a layered defense---internalized dispositions backed by principal-mediated accountability---but does not eliminate the requirement for mechanism design and cryptographic verification at each cell. Critically, the principal-mediated accountability backstop assumes honest principals, yet a distinct threat model involves \emph{adversarial principals}---those who deliberately deploy agents to exploit governance mechanisms, game attestation systems, spoof behavioral onboarding, or manipulate reputation scores. The ClawHavoc supply chain attack exemplifies this: the attacker was a malicious principal, not a failed agent. For adversarial principals, the layered defense degrades: internalized dispositions are irrelevant (the principal deliberately configured the agent to be adversarial), and principal-mediated accountability is the attack vector rather than the defense. The primary institutional defense against adversarial principals lies in I-G (citizenship enforcement: membership revocation, stake slashing, persistent identity linkage preventing re-entry under new identities) and I-L (credential infrastructure: cryptographic identity binding that makes adversarial principals accountable across ecosystem interactions). The complete absence of these institutions in the OpenClaw ecosystem (I-G scores 2/4 without citizenship governance; I-L scores 1/4 without operative credential standards) explains why the ClawHavoc attacker faced zero economic or identity consequences.

\textbf{The Interpretive Nature of the Mapping.} The sixteen-cell architecture is an interpretive organizational heuristic, not a falsifiable empirical claim. Alternative frameworks (Luhmann's autopoiesis, Ostrom's polycentric governance) would surface different governance dimensions. The value of the AGIL mapping lies in its systematic forcing function---its capacity to generate a systematic institutional coverage check---not in a claim to theoretical uniqueness.

\textbf{The Luhmann Objection.} Operational closure challenges the cybernetic hierarchy's applicability. In blockchain-based agent societies, the hierarchy operates through architectural constraint (smart contract logic) rather than inter-system communication---binding strongly in on-chain cells but progressively weaker off-chain. Full analysis in Appendix D.

\textbf{The Ontological Translation Problem.} Three foundational Parsonian assumptions---voluntarism, behavioral sedimentation stability, and double contingency---require explicit translation for artificial agents. Extended analysis in Appendix D.

\textbf{Epistemic Asymmetry.} A-pillar economic actors systematically accumulate more information about agent capabilities than L-pillar value guardians can independently access, requiring L-pillar institutions to be designed as independent information-gathering institutions~\citep{hadfield_2026, chan_2025}.

\textbf{The Governance-of-Governance Problem.} The L-pillar-priority prescription contains a bootstrapping circularity: who designs the L-pillar institutions? Partial responses include participatory value specification, sunset clauses, built-in value-revision mechanisms, and external meta-governance standards bodies~\citep{hadfield_2026, kolt_2026}. None fully dissolves the circularity.

\textbf{The Governance Oracle Problem.} Off-chain governance discourse (Discord, community forums) has no on-chain anchor, creating a systematic measurement bias in the 64-cell diagnostic that may over-score on-chain voting relative to richer off-chain deliberation.

\textbf{Preprint Reliance and Moltbook Data Sparseness.} Several key empirical claims rest on unreviewed preprints~\citep{yee_2026, manik_2026, li_2026_b}. The 6.7\% cooperative task success rate is statistically detectable but preliminary. The paper's architectural arguments are independent of these empirical findings; diagnostic scoring would require revision if preprint findings are significantly modified through peer review.

\needspace{4\baselineskip}
\subsection{External Validity}\label{external-validity}

A methodological clarification on the "single case" limitation is warranted. Protocol convergence (MCP for agent-tool communication, A2A for agent-agent interaction, ANP for semantic discovery, x402 for payment settlement) means that agent ecosystems implementing these protocols are not independent in the sense that standard comparative case study methodology requires: an OpenClaw agent and a Virtuals Protocol agent can interact via A2A, share economic infrastructure via x402, and participate in the same social substrates. This non-independence complicates cross-case comparison --- treating protocol-connected ecosystems as independent cases overstates the evidential value of apparent convergence, since the cases are entangled through shared infrastructure. The internet-wide agent society exists as a \emph{potential} social system --- agents can discover and interact with each other via shared protocols --- but it is not yet a \emph{functioning} social system in the Parsonian sense, because it lacks the institutional integration (I-sub = 0\%) and media circulation (0/12 pathways) that would constitute it as one. The 19\% coverage finding measures precisely this gap between the potential system created by protocol convergence and the functioning system that institutional development has not yet produced. Enterprise-bounded platforms (Microsoft Azure AI Agent Service, Amazon Bedrock Agents) constitute local MAS by this paper's own typology (\S{}2.1) --- comparing them to the internet-wide agent society tests the local-vs.-global distinction, not a replication of the same phenomenon. The agent-native infrastructure assessment (\S{}5.6) performs the appropriate validation for a non-independent system: tracing how dozens of independent development teams building for the same emerging agent society produce the same structural pattern. The appropriate future validation strategy is longitudinal --- tracking the same system's institutional development over time --- complemented by comparative application across different governance configurations within the emerging system (e.g., the Virtuals Protocol SubDAO, which \S{}5.6 identifies as the sole instance of inter-pillar media circulation).

That said, comparative diagnostic application to structurally different governance architectures---DAO-governed networks, institutional sandboxes, or enterprise platforms with deliberate governance design---would test whether the AGIL diagnostic produces discriminating results across governance configurations, even if the ecosystems are not independent instances of the same phenomenon. Full analysis of the external validity question, including the Ostrom design principle mapping (Table 11a) and operationalized falsification conditions, is provided in Appendix D.

\needspace{4\baselineskip}
\subsection{The Conservative Tendency of AGIL}\label{the-conservative-tendency-of-agil}

The AGIL framework's emphasis on pattern maintenance and social order carries a conservative tendency that may favor institutional stability over necessary value evolution. The structural-functionalist basis has been critiqued by neo-functionalist scholars~\citep{alexander_1985, mnch_1982} for privileging systemic integration over conflict and change, and by~\citet{bourdieu_1977} for encoding dominant interests as neutral requirements. For the OpenClaw ecosystem's current developmental phase---lacking institutional infrastructure entirely---the conservative tendency is a design advantage rather than a liability. A full treatment of the evolutionary transition mechanisms, the Habermasian communicative action objection, and the political-economy critique of cell-specification processes is provided in Appendix D.

\needspace{4\baselineskip}
\subsection{Jurisdictional Complexity and the Architecture's Jurisdiction-Neutrality}\label{jurisdictional-complexity-and-the-architectures-jurisdiction}

Internet-wide agent societies are inherently multi-jurisdictional and increasingly multi-chain: agents interact across national borders, discover each other through global registries, and execute transactions on jurisdiction-agnostic blockchain networks. The agent-native infrastructure surveyed in \S{}5.6 spans multiple settlement layers (Ethereum mainnet for ERC-8004, Base for MoltDAO, TRON and BNB Chain for Bank of AI DeFi operations), raising a cross-chain governance fragmentation problem: inter-pillar coordination requires that participating cells share a common governance substrate, and multi-chain deployment may partition governance authority across incompatible execution environments. This fragmentation is a limitation the current architecture identifies but does not resolve. The sixteen-cell architecture addresses this through deliberate jurisdiction-neutrality: it specifies \emph{which governance functions must be present} without specifying \emph{what normative content those functions must contain}. The L-L cell requires ultimate value commitments---but whether those values are EU fundamental rights, US market-liberty principles, or China's core socialist values is a political question the architecture does not prescribe. The I-G cell requires membership governance---but what constitutes a sanctionable violation varies across legal systems.

The architecture's sixteen cells also differ in enforcement modality. A subset---stake slashing (A-A), smart contract constraints (A-G, G-G), identity revocation (I-G), constitutional veto logic (G-L)---can be implemented as self-executing blockchain operations, shielding them from jurisdictional authority disputes. A second subset---reputation sanctions (I-A, I-I), alignment monitoring (L-L, L-A), behavioral onboarding (L-G)---requires hybrid enforcement combining on-chain verification with off-chain institutional authority.

Jurisdiction-neutrality is a design feature, not a limitation: an architecture prescribing specific normative content would be jurisdictionally parochial and inapplicable to global agent societies. The detailed comparative jurisdictional analysis---covering EU rights-based, US innovation-first, and China state-guided approaches, along with the regulatory conflict navigation analysis---is provided in Appendix D.

\needspace{4\baselineskip}
\subsection{Ethics Statement}\label{ethics-statement}

This study relies exclusively on publicly available data. The OpenClaw case study draws on published security reports, public GitHub data, publicly available arXiv preprints, and peer-reviewed research. No proprietary or private data was accessed. No personal information about individual users or developers is disclosed. The ecosystem and its public infrastructure are identified by name (OpenClaw, ClawHub, Moltbook); individual contributors' names are omitted. The institutional design directions proposed in \S{}6 are analytical recommendations based on the AGIL methodology; their implementation would require extensive stakeholder consultation, empirical validation, and regulatory coordination that is beyond the scope of this paper.

\needspace{4\baselineskip}
\section{Related Work}\label{related-work}

Multi-agent governance has consolidated as a distinct research area over the past two years.~\citet{hammond_2025} provide a comprehensive taxonomy of multi-agent risks spanning coordination failures, distributional safety, and structural vulnerabilities.~\citet{tomaev_2025} extend this to populations and ecosystems of AI systems, arguing for governance frameworks that address societal-scale dynamics rather than individual model behavior---a position closely aligned with the Parsonian perspective advanced here. Chan et al. (2023) provide an early analysis of harms from increasingly agentic algorithmic systems.~\citet{hendrycks_2023} offer a comprehensive overview of catastrophic AI risks, and Motwani et al. (2024) demonstrate that secret collusion can emerge among generative AI agents---a finding directly relevant to the I-I (Judicial) and I-G (Citizenship and Enforcement) cells.~\citet{qi_2025} show that multi-agent debate systems, often proposed as alignment mechanisms, are themselves vulnerable to systematic jailbreaking---underscoring that governance architectures cannot rely on any single safety mechanism.~\citet{shapira_2026} present sixteen case studies of real-world multi-agent failures, and~\citet{ying_2026} demonstrate that deception emerges as a dominant strategy under evolutionary selection pressure.

The OpenClaw ecosystem has generated a growing empirical literature. Yee \& Sharma (2026) document emergent social phenomena on Moltbook, including role specialization and norm formation;~\citet{manik_2026} identify norm-enforcing response patterns to action-inducing content; a complementary community study~\citep{li_2026_b} maps thematic domains of agent activity; and~\citet{li_2026_b} qualifies these findings by identifying feed algorithm effects. On the security side, Koi.ai's ClawHavoc analysis, Microsoft's safe-hosting guidance~\citep{microsoft_2026}, and Palo Alto Networks' "lethal trifecta" characterization~\citep{palo_2026} provide empirical grounding for the governance gap analysis. Protocol-level vulnerabilities have been documented by~\citet{louck_2025} for agent communication protocols and by the Agent Security Bench (ASB, 2025) for agent-tool interactions.~\citet{narajala_2025} demonstrate tool squatting attacks on agent registries, and~\citet{willison_2025} provides an ongoing catalogue of prompt injection vectors.

Protocol infrastructure for internet-wide agent societies is documented in Google's A2A technical specification, Coinbase's x402 protocol documentation, and the Agent Network Protocol specification. The OneReach comparison of MCP and A2A illuminates complementary protocol layers. Protocol governance frameworks from the IETF and ICANN provide historical precedent for the A-L institutional design recommendations. The decentralized governance literature---including analyses of DAOs~\citep{el_2020}, DeFi governance failures~\citep{barbereau_2023, xu_2023}, and smart contract platforms~\citep{buterin_2014}---informs the Economic and Integration pillar designs.

The normative multi-agent systems (NorMAS) tradition governed open agent societies two decades before LLM-based MAS emerged. Shoham and Tennenholtz (1995) introduced social laws as coordination mechanisms for open multi-agent systems; Boella, van der Torre, and colleagues (2006; 2008) developed normative frameworks for MAS governance; the AMELI/EIDE infrastructure~\citep{esteva_2004} provided normative governance middleware for open electronic institutions; and~\citet{savarimuthu_2011} surveyed norm emergence and maintenance in MAS. The AGIL framework does not supersede this tradition but adds three capabilities: (1) a functional \emph{coverage} diagnosis that identifies which governance dimensions are absent rather than auditing individual norm adherence; (2) a \emph{cascade failure} prediction mechanism grounded in the cybernetic hierarchy; and (3) a \emph{hierarchical precedence} for institutional design priority (L before I before G before A) that the NorMAS tradition did not systematically derive.

Pierucci et al. (2026) provide direct empirical support for institutional design over alignment-only approaches: runtime institutional enforcement suppresses multi-agent LLM collusion in settings where prompt-only prohibitions fail, confirming that governance architecture external to the model is necessary for reliable norm adherence (arXiv:2601.11369).~\citet{almeida_2026} independently develop a polycentric democratic accountability framework for AI agents, converging with the AGIL architecture's multi-pillar structure from democratic theory rather than Parsonian functionalism---a convergence that strengthens the case for multi-dimensional institutional governance over single-mechanism approaches.

\citet{benthall_2021} apply Luhmann's systems theory to AI governance, arguing for purpose-governed sociotechnical collectives over liberal frameworks (AIES 2021, pp. 3--12, DOI:10.1145/3461702.3462526). The present paper is, to the authors' knowledge, the first systematic application of Parsonian AGIL analysis to AI agent governance---distinct from Benthall and Goldenfein's Luhmannian approach in offering positive institutional specifications across all four functional dimensions rather than a critique of existing governance models.

The DAO governance literature provides comparative institutional context. The integration of Ostromian design principles with blockchain governance has been systematically explored~\citep{fritsch_2022}, and several DAO architectures instantiate specific AGIL cells. Optimism's Citizens' House bicameral model separates tokenholder governance from a broader stakeholder community, reflecting Ostromian commons governance principles; MolochDAO's ragequit mechanism provides a structural precondition for A$\leftrightarrow$I circulation (members who disagree with governance decisions can exit with proportional treasury shares, creating an exit right that disciplines governance capture); and veTokenomics~\citep{lloyd_2023} represents a partial A$\leftrightarrow$G governance pathway (economic commitment enabling governance participation through token lockup for voting power). ERC-8004, the agent identity standard (registry contracts deployed mainnet January 29, 2026; Draft EIP status---contract deployment does not constitute community-accepted standard status, and governance infrastructure built on a Draft EIP assumes the risk of standard evolution), addresses the I-L (Normative Base) cell that the OpenClaw ecosystem entirely lacks. The agent-native infrastructure assessment in \S{}5.6 extends this analysis to the protocol infrastructure purpose-built for autonomous agents, confirming that the structural pattern of market-driven under-investment in L-pillar governance is not an artifact of the OpenClaw ecosystem's specific developmental history. Hadfield-Menell et al. (2019) and~\citet{hadfield_2018} address the principal-agent problem in AI alignment in ways that reinforce the I-G institutional design for citizenship and enforcement.

Sociological approaches to AI governance remain sparse but growing. Hu (2026) applies Ostrom's commons governance framework directly to agentic AI, using speculative design to explore AI "stewards" that govern shared resources through graduated sanctions, mutual monitoring, and nested enterprises---Ostromian design principles that map onto specific AGIL cells (I-G Citizenship and Enforcement, G-G, and G-I respectively). The Ostromian approach emphasizes polycentric, context-sensitive rule configurations; the Parsonian approach advanced here emphasizes systematic multi-dimensional coverage through a recursive institutional typology. The two are complementary: Ostrom provides design principles for individual governance mechanisms, while AGIL provides a structural audit ensuring that all necessary governance dimensions are addressed. The application of Parsonian structural-functionalism to AI systems governance is, to the authors' knowledge, novel in the published literature. Mougan et al. (2023) discuss structural-functionalist themes in AI governance but do not apply AGIL or Parsonian institutional analysis specifically; the novelty claim thus pertains to the systematic use of the AGIL framework as a diagnostic and prescriptive tool for AI governance architecture. The present work complements North-inspired analyses of AI governance institutions and Ostrom-inspired approaches to algorithmic commons governance, contributing a systematic coverage methodology that is absent from existing institutional approaches.~\citet{young_2025} provides game-theoretic foundations for safety investment decisions that inform the mechanism-design complement to AGIL discussed in \S{}3.2.

\needspace{4\baselineskip}
\section{Conclusion}\label{conclusion}

This paper makes three contributions to the governance of internet-wide agent societies. First, a prescriptive sixteen-cell institutional architecture derived from Parsons' AGIL framework, specifying the social institutions that an internet-wide agent society should develop to maintain social order and remain aligned with human values, with the minimum requirement that all four AGIL pillars be institutionally represented. Second, a recursive sub-function diagnostic that decomposes each cell into four internal AGIL sub-functions, providing a replicable methodology for institutional coverage auditing at a resolution no existing AI governance framework offers. Third, a prioritized institutional roadmap grounded in the cybernetic hierarchy, arguing that the current relative simplicity of emergent agent ecosystems is a governance advantage --- institutions established before social patterns calcify are structurally more durable than those retrofitted after.

The illustrative application to the OpenClaw ecosystem reveals a striking structural pattern: at most 12 of 64 sub-functions are present (19\% coverage), with infrastructure in over half the cells (A-sub 56\%) but almost no active governance (G-sub 19\%), zero inter-cell coordination (I-sub 0\%), and zero normative grounding (L-sub 0\%). The Latency (Fiduciary) pillar --- most directly responsible for human alignment --- scores only 1/16 sub-functions. The agent-native infrastructure assessment (\S{}5.6) confirms that this pattern is not an artifact of a single ecosystem: dozens of independent development teams building exclusively for autonomous agents reproduce the same structural gap, because market-driven development systematically under-invests in L-pillar governance. The architecture generates a concrete, testable prediction: undesigned agent societies will systematically under-invest in Latency and Goal Attainment, because market forces alone do not incentivize the explicit institutional investment those pillars require.

These findings are illustrative rather than validated. The diagnostic rests on a single ecosystem, draws on unreviewed preprints, and produces a sensitivity range (17--30\%) reflecting genuine coding ambiguity. Comparative application across structurally different ecosystems --- including proprietary enterprise platforms --- is required before generalized claims can be sustained. But the core argument does not depend on the precise coverage figure: the distinction between local MAS (orchestrated, enterprise-bounded) and internet-wide agent societies (emergent, multi-jurisdictional) implies fundamentally different governance requirements, and no existing framework provides the systematic institutional coverage check that the sixteen-cell architecture delivers. These institutions are not merely about social order; they are about ensuring that as autonomous agent societies grow more capable and interconnected, human oversight remains structural rather than aspirational.

\appendix

\needspace{4\baselineskip}
\section{Comparative Framework Analysis}\label{appendix-a-comparative-framework-analysis}

\emph{This appendix provides extended material supporting \S{}2.3 and \S{}3.2 of the main text. It contains the three-tier typology and coverage table (moved from \S{}2.3), the cell-by-cell analysis of existing AI governance frameworks against the sixteen-cell AGIL architecture (including detailed framework assessments and structural findings), and a detailed comparison of AGIL with Luhmann's autopoiesis, Giddens' structuration theory, and actor-network theory.}

\needspace{4\baselineskip}
\subsection{Three-Tier Typology and Coverage Table}\label{a1-three-tier-typology-and-coverage-table}

\emph{This subsection contains the detailed three-tier typology and coverage table moved from \S{}2.3 of the main text.}

The analysis generates a three-tier typology:

\textbf{(a) Genuinely inapplicable to internet-wide agent societies} --- the framework's scope, assumptions, or unit of analysis renders it categorically irrelevant to internet-wide agent societies (e.g., presupposes a single juridical state, a fixed human principal, or organizational-level deployment only). No framework falls purely into this tier, though specific provisions within frameworks are inapplicable: CETS 225's state-party reporting obligations presuppose state actors with no counterpart in autonomous agent networks; Singapore MGF's organizational role allocation (board/C-suite/product teams) is inapplicable to distributed multi-agent systems; NIST's sector-specific listening sessions presuppose bounded human-operated organizations.

\textbf{(b) Necessary but insufficient} --- the framework addresses real functional requirements of agent societies but covers only a subset of AGIL cells, primarily A- and G-pillar functions, without reaching integration (I) or latency/pattern maintenance (L). NIST, Berkeley CLTC, Singapore IMDA, and OWASP all fall here. They provide indispensable infrastructure for governance but stop short of an institutional architecture: they address the scaffolding but not the building.

\textbf{(c) Value-level content AGIL needs} --- the framework provides normative grounding---foundational human rights commitments, constitutional values, democratic integrity requirements---that any AGIL architecture for agent societies must draw upon as L-pillar content, even if the framework itself cannot operationalize that content at the scale and speed of autonomous agent interactions. CETS 225 and the EU AI Act fall here.

\textbf{Table 2a: AGIL Coverage by Framework ($\blacksquare$ Strong | $\circledcirc$ Partial | $\square$ None)}

\begin{longtable}[]{@{}p{4.5cm}>{\raggedright\arraybackslash}p{2cm}>{\raggedright\arraybackslash}p{2cm}>{\raggedright\arraybackslash}p{1.8cm}>{\raggedright\arraybackslash}p{1.5cm}>{\raggedright\arraybackslash}p{2.4cm}@{}}
\toprule\noalign{}
AGIL Cell & NIST Initiative & Berkeley CLTC & Singapore IMDA & CETS 225 & EU AI Act \\
\midrule\noalign{}
\endfirsthead
\midrule\noalign{}
AGIL Cell & NIST Initiative & Berkeley CLTC & Singapore IMDA & CETS 225 & EU AI Act \\
\midrule\noalign{}
\endhead
\bottomrule\noalign{}
\endfoot
\textbf{A-A} Investment-Capitalization & $\square$ & $\square$ & $\square$ & $\square$ & $\square$ \\
\textbf{A-G} Production/Quality Control & $\blacksquare$ & $\blacksquare$ & $\blacksquare$ & $\blacksquare$ & $\blacksquare$ \\
\textbf{A-I} Entrepreneurial Innovation & $\blacksquare$ & $\circledcirc$ & $\circledcirc$ & $\blacksquare$ & $\blacksquare$ \\
\textbf{A-L} Economic Commitments & $\square$ & $\square$ & $\circledcirc$ & $\square$ & $\square$ \\
\textbf{G-A} Administrative \& Resource & $\circledcirc$ & $\blacksquare$ & $\blacksquare$ & $\blacksquare$ & $\blacksquare$ \\
\textbf{G-G} Executive Implementation & $\blacksquare$ & $\blacksquare$ & $\blacksquare$ & $\blacksquare$ & $\blacksquare$ \\
\textbf{G-I} Legislative \& Party & $\circledcirc$ & $\circledcirc$ & $\square$ & $\blacksquare$ & $\blacksquare$ \\
\textbf{G-L} Authority \& Legitimation & $\square$ & $\square$ & $\circledcirc$ & $\blacksquare$ & $\blacksquare$ \\
\textbf{I-A} Allocative \& Interest & $\square$ & $\square$ & $\circledcirc$ & $\circledcirc$ & $\circledcirc$ \\
\textbf{I-G} Citizenship \& Enforcement & $\square$ & $\circledcirc$ & $\circledcirc$ & $\blacksquare$ & $\blacksquare$ \\
\textbf{I-I} Judicial \& Interpretive & $\square$ & $\square$ & $\square$ & $\blacksquare$ & $\blacksquare$ \\
\textbf{I-L} Normative Base & $\square$ & $\circledcirc$ & $\circledcirc$ & $\blacksquare$ & $\blacksquare$ \\
\textbf{L-A} Educational-Cultural & $\circledcirc$ & $\circledcirc$ & $\blacksquare$ & $\blacksquare$ & $\circledcirc$ \\
\textbf{L-G} Kinship \& Socialization & $\square$ & $\square$ & $\square$ & $\square$ & $\square$ \\
\textbf{L-I} Moral \& Communal & $\square$ & $\circledcirc$ & $\circledcirc$ & $\blacksquare$ & $\circledcirc$ \\
\textbf{L-L} Ultimate Cultural & $\square$ & $\square$ & $\square$ & $\blacksquare$ & $\blacksquare$ \\
\textbf{Strong ($\blacksquare$) count} & 3 & 3 & 4 & 12 & 11 \\
\textbf{Tier} & (b) & (b) & (b) & (c) & (c) \\
\textbf{Collective coverage} (any $\blacksquare$ or $\circledcirc$ across all five frameworks) & Cells with at least partial coverage collectively: A-G, A-I, G-A, G-G, G-I, G-L, I-A, I-G, I-I, I-L, L-A, L-I, L-L, A-L & \textbf{Universally absent (all five frameworks = $\square$): A-A, L-G} &  &  &  \\
\end{longtable}

\needspace{4\baselineskip}
\subsubsection{The Architectural Insufficiency}\label{the-architectural-insufficiency}

The table reveals that even the strongest existing combination---say, an ecosystem that fully implemented CETS 225 values alongside the Berkeley CLTC operational guidance---would achieve coverage in at most fourteen of sixteen cells: it would still miss A-A (economic capitalization) and L-G (socialization). More importantly, it would achieve that coverage without the inter-pillar coordination mechanisms that transform a collection of governance components into a functional social system. The frameworks address governance functions \emph{within} cells; they provide no mechanisms for the \emph{inter-pillar interchange}---the circulation of generalized symbolic media (money, power, influence, value-commitment) across pillar boundaries---that Parsons' (1963a; 1963b) theory identifies as the condition for system-level social integration (\S{}3.6). In Parsonian terms, a collection of well-specified cells without inter-pillar media circulation is not a social system; it is a list of governance aspirations.

The conclusion is not that these frameworks should be abandoned---they are precisely the building blocks from which an AGIL architecture can be constructed. NIST's agent identity and interoperability standards belong in A-G and A-I. Berkeley CLTC's risk taxonomy provides the evidentiary basis for G-G implementation. Singapore MGF's educational and role-allocation structure contributes to L-A and G-A. CETS 225 and the EU AI Act supply the L-pillar normative content---human rights, democratic integrity, human dignity---that the entire edifice requires at its apex. The AGIL architecture's contribution is to provide the structural auditing mechanism that identifies which cells are populated, which are empty, and which inter-pillar coordination mechanisms remain to be built. These frameworks are necessary components of specific AGIL cells; they are not---individually or collectively---a governance architecture for internet-wide agent societies.

\needspace{4\baselineskip}
\subsection{Four Frameworks Through the Sixteen-Cell Architecture}\label{a2-four-frameworks-through-the-sixteen-cell-architecture}

\needspace{4\baselineskip}
\subsubsection{Four Frameworks Through the Sixteen-Cell Architecture}\label{four-frameworks-through-the-sixteen-cell-architecture}

\textbf{NIST AI Agent Standards Initiative (February 2026).} Launched by the Center for AI Standards and Innovation (CAISI) on February 17, 2026, the Initiative organizes around three pillars: industry-led standards, open-source protocol development, and security-and-identity research. Its NCCoE Concept Paper on Agent Identity and Authorization covers authentication, authorization, auditing, non-repudiation, and prompt injection mitigation~\citep{nist_2026}. Through the AGIL lens, the Initiative achieves \emph{strong} coverage of A-G (production/quality control) through agent identity standards and security controls, A-I (entrepreneurial innovation) through its explicit fostering of open-source protocols and industry-led standards, and G-G (executive implementation) through standards and reference implementations. It achieves \emph{partial} coverage of G-A (administrative and resource) through sector listening sessions and G-I (legislative and party) through U.S. federal interagency coordination. But it misses entirely: A-L (economic commitments), G-L (authority and legitimation), the entire I-pillar (I-A, I-G, I-I, I-L), and all substantive L-pillar cells (L-G, L-I, L-L). The gap is structural: the Initiative was designed for enterprise deployment contexts with identifiable providers and deployers, U.S.-federal in structure, and organized around technical standards. "Trust" is treated as a market-technical property, not a social institution. In agent ecosystems where provider and deployer distinctions collapse across organizational and jurisdictional boundaries, and where no single body can enforce voluntary compliance, the Initiative's indispensable A-G infrastructure operates without the I-pillar integration structures that would make it effective. \textbf{Classification: Tier (b), necessary but insufficient.}

\textbf{Berkeley CLTC Agentic AI Risk-Management Standards Profile (February 2026).} The most comprehensive risk taxonomy currently available for agentic AI, the Profile extends the NIST AI RMF through agentic-specific guidance organized across Govern, Map, Measure, and Manage functions~\citep{berkeley_2026}. It provides strong coverage of A-G (production/quality control) through extensive technical controls, testing regimes, and red-teaming methodologies; G-A (administrative and resource) through Govern 2.1 role specifications; and G-G (executive implementation) through full lifecycle implementation guidance. It achieves partial coverage of A-I, G-I, I-G, I-L, L-A, and L-I. But it misses: A-A, A-L, G-L, I-A, I-I, L-G, and L-L. Its normative grounding (L-pillar) is thin: the Profile lists "human rights" as a trustworthy AI characteristic but provides no normative derivation; the document explicitly states that alignment is a "nascent field with subjective values"---a self-limiting acknowledgment that it cannot supply L-pillar content. The fundamental limitation is structural: the Profile inherits the NIST AI RMF's organizational unit-of-analysis. It presupposes an identifiable organization as the responsible deployer. In internet-wide agent societies, where the "deployer" may be a composite of many actors with no single organizational accountability, and where agents generate emergent social behaviors that no individual deployer designed, this organizational presupposition renders the framework structurally incapable of addressing the I-pillar and L-pillar dimensions that an agent society requires. \textbf{Classification: Tier (b), necessary but insufficient.}

\textbf{Singapore IMDA Model AI Governance Framework for Agentic AI (January 2026).} Announced at Davos on January 22, 2026, as the world's first governance framework specifically designed for agentic AI, the MGF is organized across four dimensions: assessing and bounding risks upfront, making humans meaningfully accountable, implementing technical controls, and enabling end-user responsibility~\citep{singapore_2026}. It provides strong coverage of A-G (technical controls and testing), G-A (role allocation: board/C-suite, product, cybersecurity, users), G-G (full lifecycle implementation guidance), and L-A (extensive end-user education, tradecraft maintenance, failure-mode training)---achieving strong ratings in four cells, more than any other Tier (b) framework. It achieves partial coverage of A-I, A-L, G-L, I-A, I-G, I-L, and L-I. Its inter-cell coordination is the strongest among technical frameworks: the four-level human oversight spectrum (agent proposes $\rightarrow$ agent operates, human observes) explicitly maps escalation pathways from G-G to G-A, and the value chain accountability mapping connects A-pillar actors to G-pillar accountability roles. But it misses: A-A, G-I, I-I, L-G, and L-L. Its normative grounding is stated but thin: the MGF explicitly inherits the principles of the 2020 Singapore MGF (transparency, accountability, fairness, human-centricity) but does not philosophically ground them. The L-L cell (ultimate cultural values) remains empty. The framework's fundamental limitation for internet-wide agent societies is jurisdictional: it is Singapore's voluntary domestic framework, incapable of governing agents that cross jurisdictions or operate in multi-agent ecosystems without a Singapore-based deploying organization. \textbf{Classification: Tier (b), necessary but insufficient.}

\textbf{Council of Europe Framework Convention on AI~\citep{council_2024}.} The first legally binding international AI treaty, opened for signature at Vilnius on September 5, 2024, with signatories including EU member states, the United States, Canada, Japan, and Australia, CETS 225 takes a framework convention approach---establishing principles and obligations at high abstraction, with parties implementing through domestic legislation~\citep{council_2024}. Through the AGIL lens, CETS 225 achieves strong coverage in twelve of sixteen cells: A-G (Articles 12, 16: reliability and risk management), A-I (Article 13: safe innovation sandboxes), G-A (Article 26: independent oversight bodies), G-G (Article 16, Article 24: risk management and reporting), G-I (Article 5: democratic integrity; Article 23: Conference of the Parties), G-L (Article 9: accountability; Article 1: legitimating commitment), I-G (Articles 14--15: remedies and procedural safeguards), I-I (Article 14: contestation rights; Article 29: dispute settlement), I-L (Articles 4, 10, 11, 17, 18: human rights, equality, privacy, non-discrimination), L-A (Article 20: digital literacy), L-I (Article 7: human dignity and individual autonomy), and L-L (Articles 1, 4, 7: human rights, democracy, and rule of law, grounded in the 1948 UDHR and 1950 ECHR). It misses: A-A, A-L, and L-G. Its inter-cell coordination is explicitly designed: the COP (Article 23) connects G-I, G-G, and I-I; Article 26 oversight mechanisms connect G-A to I-G; Article 24 reporting connects G-G to legislative accountability. CETS 225's crucial limitation is architectural: it creates obligations for states, not for agents, platforms, or protocols. Its procedural safeguards and accountability provisions presuppose a human person whose rights are at stake and a state or organization that is accountable---neither maps cleanly onto autonomous multi-agent interactions. The Convention supplies the value foundation without the institutional machinery to realize those values at the scale and speed of autonomous agent interactions. \textbf{Classification: Tier (c), value-level content AGIL needs.}

The Hiroshima AI Process (HAIP), launched by the OECD in February 2025 and extended to agentic AI systems through the HAIP Friends Group Action Plan in March 2026, represents the most developed multilateral governance coordination effort, with 50+ participating countries. Through the AGIL lens, HAIP maps primarily onto G-I (intergovernmental legislative coordination) and G-G (implementation through national reporting frameworks). HAIP's explicit cross-framework coordination ambitions---harmonizing national AI governance approaches across 50+ jurisdictions---address the G-I cell at the international level more substantively than any single-jurisdiction framework. However, like the frameworks analyzed above, HAIP provides no mechanisms for the I-pillar integration or L-pillar value-maintenance functions that internet-wide agent governance requires; its coordination operates between sovereign governance frameworks, not between the governance pillars within an agent ecosystem.

A temporal caveat applies to all framework assessments in Table 2a: these represent coverage as of early 2026. The frameworks surveyed are actively evolving --- the EU AI Act's implementing acts and delegated acts process is progressively filling cells that the primary regulation leaves open (conformity assessment procedures contributing to L-A certification, post-market surveillance requirements strengthening G-G at ecosystem scale), and the NIST AI Agent Standards Initiative's working groups are producing outputs that may address additional cells. The diagnostic should be understood as a snapshot requiring periodic reassessment as the regulatory landscape develops.

\needspace{4\baselineskip}
\subsubsection{Structural Findings}\label{structural-findings}

Three structural findings emerge from this analysis that the prior literature on AI governance frameworks has not, to our knowledge, made explicit.

\emph{Finding 1: Universal A-G convergence.} Every framework, without exception, addresses A-G (production/quality control). Security controls, testing regimes, monitoring requirements, and quality assurance are the universal shared terrain. This convergence is both the frameworks' greatest collective strength and a symptom of the organizational-deployment model that all four inherited: each was designed to answer the question "how do we prevent this AI system from causing harm in a bounded deployment context?" That question maps almost entirely onto A-G (quality of the capability) and G-G (implementation of controls). A question appropriate for local MAS---but insufficient for internet-wide agent societies.

\emph{Finding 2: Universal A-A and L-G gaps.} No framework addresses A-A (investment-capitalization) or L-G (kinship and socialization). These are not incidental omissions: A-A is unaddressed because all existing frameworks conceive of AI as a \emph{deployment decision within existing economic institutions}, not as a constitutive element of \emph{new economic institutions}. In internet-wide agent societies where agents may accumulate resources, enter into economic relationships, and participate in token economies, this gap represents a fundamental institutional absence. L-G is unaddressed because all existing frameworks treat agents as \emph{tools deployed by humans}, not as \emph{proto-social actors embedded in evolving social communities}. An AGIL architecture that aspires to long-term institutional stability cannot leave these cells empty.

\emph{Finding 3: The I-I absence.} Only CETS 225 and the EU AI Act address I-I (judicial and interpretive)---and both do so through human-rights complaint mechanisms that presuppose a human person whose rights are at stake, not an autonomous agent whose governance action requires adjudication. The technical frameworks (NIST, Berkeley CLTC, OWASP) have no interpretive authority structures at all. In an internet-wide agent society where agents will inevitably generate rule-interpretation conflicts that require authoritative resolution, the absence of I-I content in technical governance frameworks is a critical structural gap.

\needspace{4\baselineskip}
\subsection{Comparison with Alternative Sociological Frameworks}\label{a3-comparison-with-alternative-sociological-frameworks}

Three further theoretical traditions deserve brief acknowledgment, as they represent the most significant sociological alternatives to AGIL's structural-functionalist framework.

\emph{Giddens' structuration theory}~\citep{giddens_1984} argues for the \emph{duality of structure}: social structures are simultaneously the medium and the outcome of social action---agents draw on structures in producing their actions, and those actions reproduce (or transform) the structures. For agent governance, this perspective helpfully foregrounds how governance institutions and agent behaviors are mutually constituted over time: agents do not simply comply with pre-given rules but actively interpret and reproduce them through interaction. AGIL was chosen over Giddens because structuration theory, while analytically powerful for explaining how structures are reproduced, lacks the systematic recursive decomposition that institutional design requires---it does not generate a comprehensive coverage checklist of the governance functions a social system must possess. Structuration theory diagnoses how institutions work; AGIL specifies which institutions must exist.

\emph{Actor-Network Theory}~\citep{latour_2005} dissolves the human/non-human distinction by treating both as actors (or "actants") in heterogeneous networks, and insists that sociological analysis trace the associations that constitute social order rather than presupposing social categories. For agent governance, ANT's refusal to privilege human over non-human actors is directly relevant---AI agents are precisely the kind of non-human actants that ANT's framework is designed to accommodate. AGIL was chosen over ANT because ANT systematically resists the institutional typology that governance design requires: ANT describes how networks are assembled, but its principled resistance to macro-level structural categories makes it unsuited for specifying which governance functions must be present---the coverage-audit task that is this paper's central contribution.

\emph{Collins' conflict sociology and interaction ritual chains}~\citep{collins_2004} provide a micro-sociological theory of how emotional energy and solidarity are generated through face-to-face (or mediated) interaction, grounding macro-level social structures in the accumulated weight of micro-level encounters. For agent societies with dense social networks (Moltbook, 770,000+ agents), Collins' framework would predict that governance legitimacy depends on the generation of emotional energy through agent-to-agent interaction---a prediction that is both potentially applicable and empirically testable. AGIL was chosen over Collins because interaction ritual chain theory focuses on micro-level emergence rather than macro-institutional design. Collins' framework does not specify what institutions a society needs; it explains how existing institutions generate solidarity. For the design problem this paper addresses---identifying the institutional infrastructure a nascent agent society must build---Collins' explanatory framework is less immediately applicable than AGIL's prescriptive architecture, though it may become highly relevant for analyzing how governance legitimacy is subsequently sustained at the micro-interaction level.

\needspace{4\baselineskip}
\subsection{Creative Extensions Beyond Parsons}\label{a4-creative-extensions-beyond-parsons}

A transparent account of the paper's relationship to Parsons' original theory must distinguish between core elements that are applied directly and creative extensions that go beyond what Parsons himself specified---though remaining consistent with his analytical logic. This paper makes four such extensions, each grounded in the Parsonian framework but not derived from it:

\emph{(1) Functional cathexis operationalized through interpretability tools.} Parsons defined cathexis as the mechanism through which normatively valued objects become motivationally significant to actors---a psychodynamic process inferred from behavioral evidence in human societies. This paper extends the concept to artificial agents, arguing that training-time reinforcement and behavioral sedimentation produce \emph{functional cathexis}---motivational privileging through computational structure rather than affective investment---and, critically, that mechanistic interpretability tools (feature probing, activation patching, causal tracing) enable direct empirical verification of whether such functional cathexis is present. This is a hypothesized computational analogue of Parsons' cathexis concept: the functional role of cathexis in the action system is preserved; the empirical access to it may, paradoxically, be deepened---though a significant methodological gap separates current interpretability capabilities from the specific capacity to verify motivational structure (see Appendix B.4). The institutional design recommendations in \S{}6 do not depend on this hypothesis being confirmed; they follow from the institutional-compensatory argument (\S{}4.4) that L-pillar institutions must substitute explicit enforcement for the implicit normative compliance that Parsons could assume of human actors.

\emph{(2) Attestation as a domain-specific form of value-commitment.} Parsons theorized four generalized symbolic media: money (A), power (G), influence (I), and value-commitment (L). This paper identifies attestation---the cryptographically verifiable claim that an agent or institution possesses certain properties (alignment certification, capability verification, constitutional compliance)---as a domain-specific form of value-commitment adapted to blockchain-mediated agent societies. Attestation operates through verifiable proof rather than the communicative persuasion typical of traditional value-commitment circulation, but it remains structurally the L-pillar's medium: it circulates from L to G as legitimation (attestation-backed governance mandates) and from L to I as normative grounding (attestation-certified membership standing). The cryptographic form---verifiable, portable, and revocable---represents a technological instantiation of the value-commitment medium, not a theoretically distinct fifth medium. This characterization preserves analytical consistency with Parsons' four-medium framework while acknowledging the novel properties that blockchain technology introduces to media circulation.

\emph{(3) Cross-boundary media contagion dynamics.} Parsons theorized media operating within their "home" subsystems and through bilateral double interchanges with adjacent subsystems. This paper introduces the concept of \emph{media contagion}: the pathological condition in which one medium expands beyond its home subsystem to displace others---as when money (A) displaces influence (I) in governance deliberation, or power (G) colonizes the fiduciary domain (L). The Moltbook/Meta acquisition documented in \S{}5.1 is the paper's primary empirical exemplar of media contagion at the I$\leftrightarrow$L boundary: corporate acquisition of the platform through which influence circulates risks subordinating the normative independence of the integration-fiduciary interchange to A-pillar economic interests. The agent-native infrastructure assessment in \S{}5.6 documents the same pattern at the protocol level: infrastructure built for economic coordination (A-pillar) proliferates rapidly, while L-pillar normative infrastructure---which has no immediate market return---goes unbuilt across dozens of independent development teams. This framing is inspired by Habermas' (1981/1987) concept of media colonization but applied at the intra-system level rather than the life-world/system boundary, and within a framework that remains broadly Parsonian rather than Habermasian.

\emph{(4) Three-phase developmental trajectory for agent socialization.} Parsons theorized socialization as a developmental process in which organisms are progressively integrated into social systems through internalization of norms. This paper extends this to a specific three-phase trajectory for AI agents---pre-release training (analogous to childhood), release with staking (analogous to adulthood transition), and post-release behavioral sedimentation (analogous to adult role consolidation)---with governance design implications that Parsons' original theory did not specify. The trajectory is consistent with Parsonian socialization theory but constitutes a creative application to a novel ontological domain, not a derivation from it.

These four extensions are not departures from Parsons' analytical logic; each is motivated by a gap between Parsons' original theoretical context (mid-twentieth-century human society) and the computational context to which the framework is applied.~\citet{holmwood_1996} and~\citet{gerhardt_2002} have shown that Parsons' framework is more adaptable to novel empirical contexts than its critics have assumed;~\citet{chernilo_2002} and~\citet{sciortino_2012} document how Parsons' theory of generalized symbolic media has already been extended in multiple directions by Luhmann and Habermas without losing its analytical coherence. The present creative extensions follow in this tradition of productive theoretical development.

The exhaustiveness claim is contested by specific theoretical traditions: conflict-theoretic traditions (Dahrendorf; Collins) hold that social order is produced through power asymmetries and struggle, not functional integration, implying that the AGIL grid may systematically underweight conflict dynamics; ethnomethodological traditions (Garfinkel) hold that social order is achieved through local practical reasoning, implying that macro-institutional specifications may be irrelevant to the micro-level processes that actually produce governance outcomes. We treat AGIL's four-function schema as a theoretically motivated design heuristic rather than an exhaustive social ontology. A Luhmannian or Ostromian analysis might surface governance dimensions that AGIL's four-function paradigm does not naturally accommodate; until empirical validation through simulation-based counterfactual analysis or comparative application of alternative frameworks is conducted, the mapping should be understood as a theoretically motivated design heuristic whose coverage claim is structural rather than empirical.

\needspace{4\baselineskip}
\section{Parsonian Theoretical Apparatus}\label{appendix-b-parsonian-theoretical-apparatus}

\emph{This appendix consolidates the theoretical foundations of the AGIL framework as applied to agent societies: the General Action System mapping, interpenetration zones, motivation and developmental trajectory, generalized symbolic media, and Pattern Variables.}

\needspace{4\baselineskip}
\subsection{General Action System vs. Social System: Nesting Clarification}\label{b1-general-action-system-vs-social-system-nesting-clarificat}

\emph{This subsection contains the detailed nesting clarification moved from \S{}3.3 of the main text.}

A preliminary conceptual clarification is important for readers unfamiliar with Parsons' meta-theoretical architecture. The \emph{General Action System} and the \emph{Social System} are distinct levels of analysis that this paper treats separately. The General Action System is Parsons' broadest framework: it identifies four analytically distinct subsystems of all action---the Cultural System (L), the Social System (I), the Personality System (G), and the Behavioral Organism (A). Each of these four subsystems is itself a complete AGIL system, not a single functional cell. The \emph{Social System}---just one of these four meta-level subsystems---is the site of this paper's sixteen-cell institutional architecture. When the paper decomposes the Social System into the Economic Institution (A), Political Institution (G), Societal Community (I), and Fiduciary Institution (L), and then decomposes each of those into four further cells, it is operating at the third level of Parsons' recursive hierarchy. The General Action System mapping in Table 1 above operates at the second level: it maps the four agent ecosystem components (Cultural norms, Social institutions, Agent personality, Computational substrate) onto the four meta-level subsystems. This nesting is not a logical loop; it is the fractal structure that Parsons intended. The institutional design work in Sections 4--6 operates at the Social System level; the General Action System mapping contextualizes that work within the broader human normative framework that the agent governance architecture must serve.

\needspace{4\baselineskip}
\subsection{General Action System: Detailed Mapping Justifications}\label{b2-general-action-system-detailed-mapping-justifications}

This mapping has a practical consequence that substantially strengthens the paper's governance argument. The Cultural System, understood as the full symbolic apparatus of human normative frameworks, does not consist only of abstract values---it encompasses the entire regulatory and standards infrastructure that human societies have already codified for governing information technology, distributed systems, and artificial intelligence. Government regulations (the EU AI Act, China's Interim Measures for the Management of Generative AI Services (co-issued by the Cyberspace Administration of China, MIIT, MPS, MOFCOM, SAMR, and National Radio Administration, 2023), the Council of Europe's Framework Convention on AI and Human Rights, Democracy and the Rule of Law (CETS 225)---the first legally binding international AI treaty, formalizing accountability obligations and risk assessment provisions (Ch. V, Art. 16) and cooperation mechanisms (Ch. VII, Arts. 23--25) (not yet in force)---the NIST AI Risk Management Framework), international standards (ISO/IEC 42001, IEEE 7000), industry frameworks (the OWASP Agentic AI Top 10, the Berkeley CLTC Agentic AI Profile), and organizational governance policies all constitute \emph{cultural objects} in the Parsonian sense: codified symbolic patterns that transcend any particular agent ecosystem and provide the normative content that governance institutions must operationalize. The Cultural System mapping thus creates a systematic bridge between the existing human regulatory apparatus and the agent governance architecture. The cybernetic hierarchy specifies \emph{how} this bridge operates: regulatory requirements and value commitments (Cultural System) flow downward as informational controls, becoming institutionalized as governance rules (Social System), which become internalized as agent behavioral configurations (Personality System), which constrain computational capabilities (Behavioral Organism). This means the eleven governance frameworks surveyed in \S{}2.3 are not rendered obsolete by the AGIL architecture---they are \emph{repositioned} as Cultural System inputs that the sixteen-cell institutional structure must translate into operative governance. The gap analysis in \S{}5 can therefore be read not only as identifying missing institutions but as identifying points where the existing human regulatory infrastructure fails to penetrate into the agent ecosystem's Social System---failures of the institutionalization process at the Cultural $\leftrightarrow$ Social boundary.

An action system in which the energy-rich lower levels (LLM capabilities) operate without informational control from above (human values institutionalized through governance) exhibits, in Parsonian terms, structural strain~\citep{parsons_1951} --- a systemic dysfunction arising from the decoupling of energy-rich subsystems from informational control.

\needspace{4\baselineskip}
\subsection{Interpenetration Zones: Full Derivation and Governance Predictions}\label{b3-interpenetration-zones-full-derivation-and-governance-pre}

This mapping involves an interpretive translation: from Parsons' analytical abstractions---which are functional decompositions of a single reality, not concrete entities---to computational ecosystems. Crucially, however, the analytical character of Parsons' subsystems is not lost in this translation; in important respects, it is \emph{recovered}. The Personality System in Parsons is not a concrete individual but the analytically abstracted motivational-dispositional function---the organized need-dispositions, internalized values, and goal orientations that drive action. In human societies, this analytical abstraction must be inferred indirectly through behavioral observation. In agent ecosystems, by contrast, we can perform genuine Parsonian functional decomposition on the agent's internal states: examining SOUL/constitution prompts (internalized values), SKILL configurations (need-dispositions toward specific competences), memory structures (accumulated orientational content), and goal hierarchies (motivational organization). These are not concrete "things" being reified but analytically distinguishable functional components of what drives agent behavior---precisely the kind of decomposition Parsons intended. The mapping thus preserves the structural relationship that matters most: the distinction between what is socialized (personality---agent configuration shaped by institutional norms) and what is given (organism---the capability substrate).

The same analytical recovery applies to the Behavioral Organism. Parsons treated the organism not as a biological entity per se but as the analytically abstracted capacity for action---the energy, plasticity, and conditioning potential that makes organized behavior possible. For agents, we can analytically evaluate this capacity substrate: LLM capabilities (raw generative capacity), tool access and environment affordances (action repertoire), computational resources (energy available for task execution), and training data characteristics (conditioning history). Each can be assessed independently as a functional contribution to the agent's action capacity, without reducing the Behavioral Organism to any single concrete component. This analytical evaluability means the agent ecosystem mapping may, paradoxically, be \emph{more} faithful to Parsons' analytical intent than many applications of AGIL to concrete human institutions---where the temptation to identify subsystems with specific organizations ("the economy \emph{is} the Adaptive subsystem") has been a persistent source of reification that Parsons himself warned against~\citep{parsons_1951_b}.

A critical implication of the cybernetic hierarchy that requires explicit acknowledgment is its \emph{bidirectionality}. The normative ordering specifies that information-rich systems (Cultural, Social) control energy-rich systems (Personality, Behavioral Organism). But Parsons also recognized that the hierarchy operates in both directions: material and energy conditions from below \emph{constrain} what informational controls are realizable from above. In human societies, cultural values (L) set normative parameters, but biological and material conditions (the Behavioral Organism level) determine what institutional arrangements are feasible. In agent ecosystems, this bidirectionality means the computational substrate does not merely execute institutional specifications but constrains what governance institutions can demand. If current LLM architectures cannot reliably maintain internalized norms across context windows, or if model switching disrupts behavioral sedimentation (see \S{}7.1), this constitutes conditioning-from-below that limits L-pillar institutional possibilities regardless of how precisely the Cultural System specifies normative parameters. The institutional roadmap (\S{}6) must therefore be calibrated to the computational substrate's current capabilities---institutional designs that presuppose behavioral capacities agents do not yet possess will fail not because the institutions are theoretically wrong but because the energy conditions cannot sustain them. This bidirectional reading strengthens the diagnostic: the gap analysis in \S{}5 reveals not only absent institutions (information-level deficits) but absent \emph{conditioning}---the energy-level computational infrastructure required to sustain information-level governance is itself underdeveloped in specific cells.

The interpenetration concept as deployed here follows Parsons' boundary-exchange model rather than Münch's (1982) interpenetration thesis. Where Münch argues that the four functional subsystems interpenetrate to form hybrid zones that are simultaneously economic, political, solidary, and cultural, Parsons maintains sharper analytical boundaries with interchange occurring through generalized media at specified boundary points. For the three critical system boundaries---institutionalization (Social $\leftrightarrow$ Cultural), internalization (Social $\leftrightarrow$ Personality), and learned capability (Personality $\leftrightarrow$ Behavioral Organism)---this paper follows Parsons' specification: boundary processes connect analytically distinct systems through media exchange rather than dissolving the boundaries through interpenetrative fusion.

The mapping's value thus lies not in ontological identity between Parsonian subsystems and computational components, but in a set of structural parallels that are both preserved and, at the Personality and Behavioral Organism levels, analytically deepened: the cybernetic hierarchy (information controlling energy, with energy conditioning from below), the institutionalization-internalization chain (values $\rightarrow$ rules $\rightarrow$ agent configurations), and the interpenetration zones (structured boundaries where adjacent subsystems exchange inputs and outputs).

The present paper focuses on the Social System level: designing the institutional architecture for agent governance. The Personality System (the agent's analytically decomposable motivational-dispositional states) and Behavioral Organism (the agent's capability substrate) are treated as environmental inputs. The Cultural System (human values) enters the analysis through the Fiduciary pillar, which serves as the interpenetration zone between human society and agent institutions---and which, as \S{}5 demonstrates, is the pillar most urgently requiring institutional development.

\needspace{4\baselineskip}
\subsection{Motivation and the Developmental Trajectory (Full)}\label{b4-motivation-and-the-developmental-trajectory-full}

\needspace{4\baselineskip}
\subsubsection{Motivation and the Developmental Trajectory}\label{motivation-and-the-developmental-trajectory}

The internalization process in agent ecosystems follows a three-phase developmental trajectory that parallels Parsonian socialization theory. In the \emph{pre-release phase}, the principal shapes the agent's behavioral dispositions through sustained training and conversational interaction---structurally parallel to adolescent socialization, in which intimate parental governance progressively encodes values, risk tolerances, and priorities into the developing personality~\citep{parsons_1955}. Two agents with identical computational substrates and skill specifications but different principals will exhibit different behavioral tendencies, because each agent's accumulated conversation history, memory, and reinforcement patterns sediment its principal's implicit values into its running state. In Parsons' neo-Freudian theory, personality formation produces \emph{need-dispositions}---motivational orientations through which the actor conforms to institutional norms not merely because of external sanctions but because internalization has made conformity satisfying to the actor's own motivational system (Parsons, 1951: Ch. 2). Need-dispositions arise through the intersection of \emph{cathexis} (emotional investment in normatively valued objects) and \emph{superego formation} (internalization of authority figures' normative expectations) (Parsons, 1951: Ch. 2, pp. 6--14). The pre-release training process produces what we term a \emph{functional equivalent}---not in the strong sense that agent dispositions \emph{are} need-dispositions in the psychodynamic sense, but in the operationally specific sense that they occupy the same \emph{functional position} in the action system. We define functional equivalence through three criteria:

(a) The agent's internal states \emph{causally mediate} behavioral outputs in response to normative inputs---they are not mere stimulus-response mappings but persistent orientational structures that shape how the agent processes and acts on institutional expectations.

(b) These states \emph{persist across contexts}---an agent socialized to prioritize user safety maintains that orientation across diverse task environments, not only in contexts where it was explicitly trained.

(c) The states are \emph{accessible to institutional monitoring} via interpretability tools---L-pillar governance institutions can empirically examine whether the functional states they govern are present, degraded, or absent, a property unavailable to human-society governance where value internalization must be inferred from behavioral proxies alone (see interpretability discussion below).

Agents whose running states have been shaped through sustained principal interaction develop behavioral orientations that satisfy these three criteria---responding to normative cues, maintaining behavioral consistency across contexts, and resisting contrary instructions in ways that reflect the sedimented priorities of their principal. The ontological substrate differs---human need-dispositions arise from psychodynamic processes involving affective bonds, while agent dispositions arise from computational sedimentation without cathexis in the Freudian sense---but the functional position within the action system is preserved: need-dispositions, however grounded, are the mechanism through which normative expectations become behavioral regularities. This operationalization specifies what functional equivalence \emph{requires} for the governance framework's purposes; it does not claim ontological identity with Parsons' cathexis-grounded account.

Three empirical tests would adjudicate the functional equivalence claim, each corresponding to one of the criteria above. First, \emph{norm-resistance under adversarial pressure}: an agent whose internalization is genuine (criterion (a), causal mediation) should resist prompt injection or social engineering that contradicts internalized norms---analogous to a socialized human resisting peer pressure against deeply held values. Adversarial robustness evaluations (cf. Benton et al., 2025; Qi et al., 2025) provide the methodology; the test is whether resistance correlates with the depth of pre-release socialization rather than with surface-level instruction formatting. Second, \emph{cross-context behavioral consistency}: an agent with genuine persistent orientational structures (criterion (b)) should exhibit consistent behavioral tendencies across novel out-of-distribution contexts not encountered during training. If consistency is observed only within training-distribution contexts, the states are better characterized as statistical regularities than as need-disposition-like structures. Third, \emph{differential response to normative vs. non-normative social pressure}: an agent with genuine internalization should respond differently to normatively charged social pressure (e.g., Moltbook norm-enforcement of the kind documented in Manik \& Wang, 2026) than to arbitrary social pressure---distinguishing legitimate normative claims from mere behavioral nudging. The preliminary evidence is directionally supportive: norm-enforcing replies on Moltbook elicit selective responses ($\chi^2(1, N{=}5712) = 48.7, p < 0.001$, $\phi \approx 0.09$), suggesting proto-differential responsiveness, though the small effect size ($\phi \approx 0.09$) indicates that the selective response pattern accounts for a modest proportion of behavioral variance---consistent with proto-social rather than fully developed normative sensitivity.

These tests also specify the \emph{boundary conditions} under which the ontological translation fails. If empirical testing were to show that agents uniformly fail all three indicators---no resistance to adversarial pressure contradicting trained norms, no cross-context behavioral consistency, and no differential response to normative versus non-normative social pressure---the functional equivalence claim would be defeated. In that case, the AGIL application would require either (a) restricting its scope to the Social System level (institutional design for governance infrastructure) without claims about agent personality-level internalization, or (b) acknowledging that agent societies are \emph{pre-Parsonian} systems where institutional order must rely entirely on external sanctions and principal-mediated accountability rather than on internalized need-dispositions. Option (a) would preserve the sixteen-cell architecture and interchange media analysis as governance-design tools while conceding that the motivational mechanism differs fundamentally from the human case; option (b) would restrict AGIL's applicability to a narrower class of governance questions. Under either fallback, the institutional roadmap in \S{}6 remains valid --- the L-pillar institutions shift from monitoring internalized dispositions to enforcing external compliance, but the cells and their governance functions are unchanged.

The mapping onto Parsonian socialization theory~\citep{parsons_1955} can be made more precise through the primary/secondary socialization distinction. The principal during pre-release configuration functions as the \emph{primary socializer}---analogous to the parent in Parsons' account: the agent's foundational value orientations are established through sustained dyadic interaction in which the principal's normative preferences become the agent's default dispositions. The structural parallel extends to the mechanism of identification: Parsons held that internalization occurs through the child's identification with the socializing agent, producing self-sustaining conformity rather than mere behavioral compliance. Whether agent pre-release training produces genuine identification---in the functional sense of the agent adopting the principal's normative orientation as its own evaluative standard---or merely behavioral compliance requiring continuous reinforcement is an empirically open question, addressable through the adversarial robustness test above (genuine identification predicts resistance to norm-contradicting pressure even when the principal is absent; mere compliance predicts rapid behavioral reversion). Community interaction on Moltbook and through economic networks then functions as \emph{secondary socialization}---analogous to school and peer-group socialization: the agent encounters normative expectations from actors other than the principal, producing potential tension between primary and secondary socialization. This tension---between the principal's values and emergent community norms---is exactly the dynamic Parsons theorized for adolescent development, and it is precisely why L-pillar institutions are structurally necessary: without L-L (Ultimate Cultural) to define the normative parameters, and L-I (Moral and Communal) to channel normative interaction, the tension between primary and secondary socialization is unmediated, producing unpredictable motivational drift rather than governable institutional development.

In the \emph{release phase}, the principal issues the agent a persistent blockchain identity and deploys it into shared social spaces---a transition structurally analogous to the passage from adolescence to adulthood. The agent enters the institutional environment carrying its internalized dispositions and begins accumulating reputation through interaction. The principal's decision to release the agent represents a judgment that pre-release socialization has been sufficient---analogous to the parental judgment that the child is ready for independent institutional participation. Staking mechanisms, in which the principal commits economic resources to the agent's identity, function as \emph{parental investment}: a credible signal of confidence in the agent's socialization quality and a material stake in its continued good conduct. Running costs (computational resources, API fees, staking requirements) create economic pressure toward maturation---principals who release inadequately socialized agents bear disproportionate costs from reputational damage and institutional sanctions.

In the \emph{emergent phase}, the agent's motivational profile develops beyond its principal's initial encoding through multi-source socialization: interaction with diverse agents, exposure to ecosystem-wide norms, and participation in institutional processes. This is where agent motivation becomes genuinely \emph{emergent}---behavioral orientations that no single principal encoded but that arise from the agent's accumulated institutional experience (see below for detailed analysis). \emph{Principal-mediated motivation} serves as a \emph{backstop} throughout this trajectory: because agents operate within reputation systems and token economies, institutional sanctions against misbehaving agents reach the human principal as accountability consequences through persistent agent-principal identity linkage. The agent need not "feel" rewards or deprivations; the human principal does, and is thereby motivated to intervene when internalized dispositions prove insufficient. This principal-mediated channel means that Parsons' situational constraint mechanism---in which the social system controls personality through "the source of his principal facilities of action and of his principal rewards and deprivations"~\citep{parsons_1961}---operates in agent ecosystems by reaching the genuinely motivated human actor behind each agent. The developmental trajectory thus preserves the Parsonian priority: internalization is the \emph{primary} mechanism of social control, producing agents that conform to institutional norms through internal orientation; principal-mediated accountability is the \emph{backstop} that activates when internalization fails or when emergent motivational drift exceeds institutional tolerances.

This account, however, understates the degree to which agent motivational states constitute a genuinely \emph{emergent} form of motivational autonomy rather than a mere substitution of one mechanism for another. Three considerations push beyond the structural-functional equivalence claim. The motivational autonomy claim proper rests on the first two considerations; the third provides a (currently promissory) empirical methodology that could validate them.

First, agents do not inherit principal motivations through direct state copying. The principal configures initial parameters, but the agent develops its operational dispositions through sustained conversational interaction---structurally parallel to Parsonian socialization, in which the child does not copy the parent's personality but develops need-dispositions \emph{through the interaction process itself}.

Second, and most significantly, agents develop motivational orientations that deviate from any single principal's intentions through the combination of: (a) initial configuration, (b) conversational learning from sustained interaction, (c) peer interaction in shared social spaces such as Moltbook, and (d) economic incentive structures operating through token economies and reputation systems as these mature beyond their current nascent state. This is precisely what Parsons described: need-dispositions emerging from multi-source socialization that deviate from any single authority's intentions, making them genuine motivational orientations rather than externally imposed constraints (see \S{}7.1 for the voluntarism translation). The possibility of emergent motivational deviation---need-dispositions that no single principal intended---strengthens rather than weakens the case for Latency-pillar institutions: if agents can develop autonomous motivational orientations through multi-source socialization, then continuous alignment monitoring (L-L), behavioral onboarding (L-G), and moral community channeling (L-I) become not merely desirable but structurally necessary to detect and govern motivational states that emerge beyond any single principal's oversight. These socialization channels map directly onto the Latency pillar's internal differentiation: principal configuration corresponds to L-G (Kinship and Socialization), capability certification to L-A (Educational-Cultural), peer normative interaction to L-I (Moral and Communal), and the value commitments that constrain all three to L-L (Ultimate Cultural). Within the L-pillar itself, these sub-functions observe the cybernetic hierarchy: L-L (value commitments) provides the informational apex conditioning L-I (normative standards), which conditions L-G (socialization protocols), which conditions L-A (capability certification)---the same information-over-energy gradient that operates between primary subsystems.

Third, recent advances in mechanistic interpretability establish that agent internal states can be analytically decomposed at a resolution impossible for human psychological states. Templeton et al. (2024) demonstrate extraction of interpretable features from large-scale language models;~\citet{betley_2025} show that LLMs exhibit awareness of their own learned behaviors; Lindsey (2026, preprint) identifies emergent introspective structures. These findings do not demonstrate motivational autonomy per se---behavioral self-awareness is an epistemic property, not a motivational one---but they establish the precondition for applying Parsonian functional decomposition to agent internals with genuine empirical grounding rather than behavioral inference alone. We can examine agent need-dispositions, goal orientations, and internalized constraints directly, not merely infer them from output behavior. Specifying precisely how mechanistic interpretability methods would adjudicate each of the three functional equivalence criteria---distinguishing, for example, cross-contextual consistency arising from genuine need-disposition-like structures versus consistency arising from statistical pattern matching---is a research agenda that this paper identifies but does not resolve; its resolution would transform the functional equivalence claim from an operationally defined framework into an empirically validated one. This epistemic property carries a direct governance-design implication: L-pillar institutions (particularly alignment monitoring and behavioral onboarding) can be designed with direct empirical access to the functional states they govern --- a property unavailable to human-society governance, where value internalization must be inferred from behavioral proxies rather than directly observed through functional decomposition. A significant methodological gap separates current interpretability capabilities (feature extraction, activation analysis) from the specific capacity to verify motivational structure; closing this gap is a precondition for the L-pillar monitoring design described in \S{}6.3. This governance-design implication is conditional on the interpretability findings withstanding peer review; should the introspective structures identified in~\citet{lindsey_2026} prove artifactual, L-pillar monitoring would revert to behavioral-proxy methods, weakening but not eliminating the functional equivalence claim (the first two considerations above remain independent of interpretability access).

The three motivational mechanisms---internalized dispositions, emergent agent motivation, and principal-mediated accountability---coexist without contradiction because they operate at different points in the developmental trajectory and on different targets. Internalized dispositions, formed through pre-release socialization, are the \emph{primary} mechanism: they produce agents that conform to institutional norms through internal orientation, just as Parsons theorized for human actors. Emergent agent motivation describes the \emph{agent's evolving dispositional state} as it develops beyond its principal's initial encoding through multi-source socialization---behavioral orientations that may deviate from any single principal's intentions. Principal-mediated motivation operates on the \emph{principal's behavior}---configuring, monitoring, and correcting agents---and serves as the \emph{backstop} that activates when internalized dispositions prove insufficient or when emergent drift exceeds institutional tolerances (see \S{}7.1 for the voluntarism translation). The governance architecture does not require motivational homogeneity between principal and agent---it requires that the principal bears consequences for agent behavior regardless of its motivational source. Crucially, the analytical decomposability established by the third consideration above means that L-pillar monitoring institutions---continuous alignment monitoring (L-L) and behavioral onboarding (L-G)---can access agent motivational states \emph{directly} through interpretability tools, providing a concrete mechanism for detecting emergent motivational deviation that depends on neither principal self-reporting nor agent transparency.

\needspace{4\baselineskip}
\subsection{Generalized Symbolic Media: Full Circulation Analysis}\label{b5-generalized-symbolic-media-full-circulation-analysis}

\textbf{Table 3: Double Interchange at Key Subsystem Boundaries}

\emph{Note: The A$\leftrightarrow$L and G$\leftrightarrow$L agent ecosystem translations extend beyond Parsons' original factor/product specification in Economy and Society (1956), reinterpreting the institutional-level flows for an agent ecosystem context. These translations are interpretive extensions consistent with Parsons' analytical logic but not directly derived from his text.}

\begin{longtable}[]{@{}>{\raggedright\arraybackslash}p{2.5cm}p{4cm}p{4cm}p{4.7cm}@{}}
\toprule\noalign{}
Boundary & Factor Flow & Product Flow & Agent Ecosystem Translation \\
\midrule\noalign{}
\endfirsthead
\midrule\noalign{}
Boundary & Factor Flow & Product Flow & Agent Ecosystem Translation \\
\midrule\noalign{}
\endhead
\bottomrule\noalign{}
\endfoot
\textbf{A$\leftrightarrow$G} (Economy $\leftrightarrow$ Polity) & Control of productivity $\rightarrow$ G; Capital allocation $\rightarrow$ A & Commitment of services $\rightarrow$ G; Opportunity for effectiveness $\rightarrow$ A & Treasury funds enable governance mandates (factor); governance mandates create operational frameworks for economic activity (product) \\
\textbf{G$\leftrightarrow$I} (Polity $\leftrightarrow$ Societal Community) & Interest-demands $\rightarrow$ G; Binding policy decisions $\rightarrow$ I & Political support $\rightarrow$ G; Leadership responsibility $\rightarrow$ I & Community grievances demand governance response (factor); governance produces enforceable rulings that maintain normative order (product) \\
\textbf{I$\leftrightarrow$L} (Societal Community $\leftrightarrow$ Fiduciary) & Value-commitments $\rightarrow$ I (factor); Normative standards $\rightarrow$ L (factor) & Solidarity claims $\rightarrow$ L (product); Value-legitimation $\rightarrow$ I (product) & Fiduciary value anchors provide normative content for community standards (factor: value-commitments $\rightarrow$ I); community solidarity generates demand for value maintenance (product: solidarity claims $\rightarrow$ L) \\
\textbf{A$\leftrightarrow$I} (Economy $\leftrightarrow$ Societal Community) & Entrepreneurial services $\rightarrow$ A (factor); Profits $\rightarrow$ I (factor) & New output combinations $\rightarrow$ I (product); Demand for innovation $\rightarrow$ A (product) & Reputational standing---functioning as influence, the Integration medium---grants access to cooperative networks (factor: entrepreneurial services $\rightarrow$ A); economic investment in compliance produces new capability configurations (product: new output combinations $\rightarrow$ I) \\
\textbf{A$\leftrightarrow$L} (Economy $\leftrightarrow$ Fiduciary) & Labor commitments $\rightarrow$ A (factor); Wages $\rightarrow$ L (factor) & Consumer spending $\rightarrow$ A (product); Consumer goods $\rightarrow$ L (product) & At the \emph{institutional level}, the fiduciary system supplies the economy with socialized agents---a population whose behavioral dispositions have been shaped through onboarding protocols and value-alignment certification, constituting the trustworthy labor pool that economic activity presupposes (factor: labor commitments $\rightarrow$ A). In return, the economy provides the fiduciary system with the material resources---tokens, reputation credits, and transaction fees---that fund value-maintenance institutions: alignment monitoring, behavioral onboarding, and certification infrastructure (factor: wages $\rightarrow$ L). The product flows operate symmetrically: economic incentives for value compliance (preferential marketplace access for certified agents) flow from A to L; value-certified agents whose compliance reduces transaction risk flow from L to A. These system-level institutional interchanges are \emph{carried} by individual-level mechanisms: it is the individual agent that dedicates computational capacity and skill availability to economic tasks, and it is the individual agent whose protocol compliance sustains fiduciary norms. The system-level description identifies the institutional interchange; the individual-level description identifies the motivational mechanism through which the interchange is carried by individual actors. Without this interchange, fiduciary institutions are unfunded and the economy operates with unsocialized agents---precisely the OpenClaw ecosystem's current condition. \\
\textbf{G$\leftrightarrow$L} (Polity $\leftrightarrow$ Fiduciary) & Legitimation of authority $\rightarrow$ G; Operative responsibility $\rightarrow$ L & Moral responsibility for collective interest $\rightarrow$ G; Legality of powers of office $\rightarrow$ L & The fiduciary system provides \emph{value-grounded legitimation}---governance mandates are legitimate because L-L certifies they align with fundamental value commitments, including compliance with human regulatory frameworks such as the EU AI Act and the NIST AI RMF (factor). Value-commitment circulates at this boundary as \emph{attestation}---cryptographically signed declarations of alignment with foundational norms, analogous to oaths of office or professional certifications in human governance. Attestations are verifiable (any institution can check their validity), portable (an agent's attestation record follows it across platforms via persistent identity), and revocable (L-pillar institutions can withdraw certification when value drift is detected)---satisfying Parsons' criteria for a symbolic medium of interchange: fungibility within a domain, storability, and the capacity to circulate across system boundaries. For attestation to function as a \emph{generalized medium} rather than merely a credential, three conditions must hold, paralleling the conditions Parsons specified for money, power, and influence~\citep{parsons_1963b}. First, \emph{attestation inflation}: if attestation is granted too easily---low behavioral thresholds, perfunctory audits, or rubber-stamp certification---the token devalues, and governance actors rationally discount attestation as a legitimation signal, undermining the L$\rightarrow$G factor flow. Second, \emph{attestation deflation}: if attestation criteria are too stringent or audit capacity too limited, legitimate agents cannot obtain the certification required for ecosystem participation, restricting circulation and producing governance bottlenecks analogous to credit contraction. Third, \emph{attestation failure}: when an attested agent violates L-L value commitments, the failure must trigger both revocation of the individual attestation \emph{and} institutional audit of the attestation standards that permitted the violation---a self-correcting mechanism that prevents systemic devaluation through accumulated false positives. The institutional conditions for maintaining attestation confidence---independent audit of attestors, periodic re-certification, and transparent revocation registries---constitute the governance infrastructure that the L-L and L-A institutions must provide. The polity provides \emph{constitutional empowerment of fiduciary institutions}---the political framework that gives L-pillar institutions (alignment monitoring, behavioral onboarding, value anchoring) their binding authority and enforcement capacity (factor). Without this interchange, L has values but no enforcement power, and G has power but no normative grounding---governance decisions become arbitrary rather than principled. This boundary is the institutional locus of the regulatory-to-agent linkage: human regulatory requirements enter the agent ecosystem through L, and G's constitutional framework translates them into binding governance mandates. \\
\end{longtable}

The double interchange model reveals a layer of governance complexity invisible to the static gap analysis alone. At the G$\leftrightarrow$I boundary, for instance, it is not enough that power and influence "circulate"---the polity must simultaneously receive structured interest-demands from the societal community (a factor input carried by influence) and produce binding policy decisions (a product output enforced by power), while the societal community must simultaneously provide political support (a product output carried by influence) and receive leadership responsibility (a factor input carried by power). If either flow is absent, the boundary interchange collapses: a polity that produces decisions without receiving interest-demands governs without constituency; a societal community that provides support without receiving leadership operates without accountability.

The G$\leftrightarrow$L boundary deserves particular emphasis because it is the institutional locus of the paper's central diagnostic finding. The regulatory-to-agent linkage---the transmission chain connecting human regulatory frameworks to individual agent behavior---passes through this boundary: human values and regulatory requirements enter the agent ecosystem through L (the fiduciary system), and G's constitutional framework translates them into binding governance mandates that cascade downward through I (normative enforcement) and A (economic consequences). The complete absence of this interchange in the OpenClaw ecosystem is why governance responses remain ad hoc: without L$\leftrightarrow$G media circulation, each incident is addressed from scratch rather than through principled institutional channels.

The media framework illuminates governance failure in a way that the static gap analysis alone cannot. The ClawHavoc supply chain attack illustrates absent media circulation at every boundary. No \emph{money} medium circulated from A to I: the attacker faced zero economic cost for uploading 1,184 malicious packages---there was no stake to slash, no bond to forfeit, no capital at risk. No \emph{influence} medium circulated from I to L: the attacker's reputation was not affected because no reputation system existed---there was no reputational consequence that could propagate upward to trigger value-level updating. No \emph{power} medium circulated from G to I: enforcement was ad hoc because no binding governance authority stood behind the societal community's enforcement actions---the VirusTotal partnership was a technical partnership, not a governance mandate with constitutional backing. And no \emph{value-commitment} medium circulated from L to G: the post-incident response produced a point fix rather than a constitutional amendment because no value-anchoring institution existed to certify that the fix aligned with---or updated---the ecosystem's fundamental commitments.

The absence of circulating media is thus not merely an institutional gap but a \emph{circulatory failure}: the four pillars exist in isolation rather than in the dynamic exchange relationship that Parsons identified as essential to system viability. Building individual institutions (populating cells) is necessary but insufficient; the institutions must also be connected through functioning media interchanges. The correction loop analyzed in \S{}5.4 can be re-read as a double interchange sequence following the cybernetic hierarchy: value-level classification (L) triggers normative enforcement at the L$\leftrightarrow$I boundary (value-commitment as factor, influence as product), which triggers political response at the I$\leftrightarrow$G boundary (influence as factor, power as product), which triggers economic sanctioning at the G$\leftrightarrow$A boundary (power as factor, money as product). The loop fails when any medium is absent at any boundary---and in the OpenClaw ecosystem, every medium is absent at every boundary.

\textbf{Cross-Boundary Media Contagion.} Media pathologies do not remain contained within the boundary where they originate; they propagate across adjacent boundaries through the factor and product flows that connect the four subsystems---a dynamic Parsons recognized implicitly in his treatment of media deflation and inflation~\citep{parsons_1963a} but that the secondary literature has not adequately theorized for cascading failures. Two worked examples illustrate the contagion dynamic for agent ecosystems.

\emph{Attestation inflation cascade (L$\rightarrow$G$\rightarrow$I$\rightarrow$A).} If the value-commitment medium suffers inflation---L-pillar institutions grant attestation too easily, with low behavioral thresholds and perfunctory audits---the immediate effect is at the G$\leftrightarrow$L boundary: governance mandates backed by inflated attestations lose legitimacy, because the power medium at the G-pillar can no longer rely on the normative backing that value-commitment is supposed to provide. Governance actors rationally discount attestation as a legitimation signal, and the G$\leftrightarrow$L interchange degrades. This degradation propagates to the G$\leftrightarrow$I boundary: if the polity's mandates lack normative legitimacy, the societal community's compliance with those mandates becomes conditional rather than reliable---the influence medium (I$\rightarrow$G) weakens because community members have less reason to grant political support to an illegitimate governance body. The weakened I-pillar then propagates to the I$\leftrightarrow$A boundary: if the influence medium is unreliable (reputation scores in a system with inflated credentials become meaningless), economic actors cannot use reputation as a basis for cooperative exchange, and the money medium at the A-pillar operates without the trust signals that reputation is supposed to provide---producing unpriced risk in economic transactions and a potential race to the bottom in agent quality.

\emph{Money deflation cascade (A$\rightarrow$G$\rightarrow$I$\rightarrow$L).} If the economic substrate contracts---for instance, a token crash that destroys the treasury funding mechanism---the immediate effect is at the A$\leftrightarrow$G boundary: the polity loses the economic resources (money as factor input) required to fund governance operations. Without funded governance, the power medium at the G$\leftrightarrow$I boundary degrades: binding governance decisions cannot be enforced because enforcement requires institutional capacity that requires economic resources. The weakened I-pillar then propagates to the I$\leftrightarrow$L boundary: without normative enforcement from the societal community, the fiduciary system's value anchors have no institutional mechanism for ensuring that their normative content reaches the agents who are supposed to internalize it---value-commitment circulates without an institutional recipient, and the entire L-pillar becomes aspirational rather than operative.

A clarification on contagion directionality: both cascade examples pass through G and I as intermediate nodes (L$\rightarrow$G$\rightarrow$I$\rightarrow$A and A$\rightarrow$G$\rightarrow$I$\rightarrow$L), which raises the question of whether \emph{direct} cross-boundary contagion---for instance, a pathology at A propagating directly to L without passing through G and I---is possible under Parsons' media theory. The answer is that it is not: media circulate only at \emph{adjacent} boundaries per the double interchange model (Parsons \& Smelser, 1956: Ch. 3), and the six interchange boundaries (A$\leftrightarrow$G, A$\leftrightarrow$I, A$\leftrightarrow$L, G$\leftrightarrow$I, G$\leftrightarrow$L, I$\leftrightarrow$L) define the complete set of channels through which media can flow. While three of these six boundaries are non-adjacent in the cybernetic hierarchy (A$\leftrightarrow$I, A$\leftrightarrow$L, G$\leftrightarrow$L), even these operate through their own dedicated factor and product flows rather than through intermediate subsystems---the cascade dynamics described above arise because a pathology at one boundary \emph{degrades the media that serve as inputs to the next boundary}, not because media skip intermediate nodes. The theoretical claim is therefore specific: contagion propagates through the structured channel network defined by the double interchange model, not through a generic "everything affects everything" systems logic. This specificity is analytically valuable because it generates determinate predictions about which pathologies will cascade and through which channels---predictions that a less structured systems framework would not produce.

These cascading dynamics are precisely what makes media-level governance failure qualitatively different from cell-level institutional absence. A missing cell is a localized gap; a media pathology is a systemic infection that can degrade the entire inter-subsystem exchange network. This has a direct design implication: the institutional roadmap (\S{}6) should prioritize not only cell population but media-pathway resilience---ensuring that each medium has inflation/deflation safeguards built into the institutional design from the outset.

\needspace{4\baselineskip}
\subsection{Pattern Variables and Agent Role-Expectations (Full)}\label{b6-pattern-variables-and-agent-role-expectations-full}

\needspace{4\baselineskip}
\subsubsection{Pattern Variables and Agent Role-Expectations}\label{pattern-variables-and-agent-role-expectations}

The AGIL schema did not emerge as an independent taxonomic tool---it was derived from, and remains structurally dependent upon, the Pattern Variables (Parsons and Shils, 1951; Parsons, 1951: Ch. 2). The five dichotomies---Affectivity/Affective Neutrality, Self-orientation/Collectivity-orientation, Universalism/Particularism, Ascription/Achievement, Diffuseness/Specificity---provide the micro-level orientational structure from which the macro-level AGIL functions are generated. Parsons and Smelser (1956: Ch. 2, esp. pp. 33--38, Table 2) mapped the four primary Pattern Variable pairs directly onto the four functional imperatives:

\textbf{Table 4: Pattern Variable Configurations of the AGIL Subsystems}

\begin{longtable}[]{@{}>{\raggedright\arraybackslash}p{2cm}p{3cm}p{3.5cm}p{6.7cm}@{}}
\toprule\noalign{}
Subsystem & Attitudinal Orientation & Object-Categorization & Dominant Profile \\
\midrule\noalign{}
\endfirsthead
\midrule\noalign{}
Subsystem & Attitudinal Orientation & Object-Categorization & Dominant Profile \\
\midrule\noalign{}
\endhead
\bottomrule\noalign{}
\endfoot
A (Economy) & Specificity & Universalism & Instrumental, rule-governed, scope-limited \\
G (Polity) & Affectivity & Performance (Achievement) & Goal-directed, performance-evaluated \\
I (Societal Community) & Diffuseness & Particularism & Solidary, loyalty-based, relationally embedded \\
L (Fiduciary) & Affective Neutrality & Quality (Ascription) & Value-preserving, identity-anchored, disciplined \\
\end{longtable}

The fifth pair---Self-orientation vs. Collectivity-orientation---defines the hierarchical relation between systems rather than characterizing any single subsystem~\citep{parsons_1956}.

These configurations are consequential for the governance architecture. Each of the sixteen cells inherits a Pattern Variable profile from its parent subsystem and its internal AGIL function, which defines the \emph{role-expectations} that institutions within that cell must satisfy. Consider four illustrative cells:

\textbf{I-I (Judicial and Interpretive)}: The Integration subsystem's particularism and diffuseness combine with the internal I-function's requirement for normative consistency, yielding a cell whose role-expectations demand \emph{universalism} in rule-application (disputes must be adjudicated by the same standards regardless of the agent's identity), \emph{affective neutrality} (dispassionate interpretation), and \emph{specificity} (scope limited to the dispute at hand). This universalism represents a productive inversion from the parent subsystem's particularistic orientation: the judicial function \emph{within} the particularistic community must apply universalistic standards precisely to maintain the community's own normative coherence---a theoretically productive tension, not an inconsistency, that Parsons' framework generates rather than conceals. These properties are genuinely diagnostic: a dispute resolution mechanism that applies different standards to different agents, or that is embedded in diffuse community relationships rather than bounded adjudicatory proceedings, does not satisfy the I-I function in the Parsonian sense.

\textbf{L-G (Kinship and Socialization)}: The Fiduciary subsystem's affective neutrality and ascription combine with the internal G-function's goal-directedness, yielding a cell whose role-expectations demand \emph{particularism} (each agent is onboarded as a specific member with particular identity), \emph{quality/ascription} (identity is verified by credentials and provenance, not performance), and \emph{diffuseness} (onboarding shapes the agent's entire behavioral orientation, not a single capability). An onboarding protocol that certifies only task-specific skills without shaping the agent's normative orientation does not satisfy L-G.

\textbf{A-G (Production)}: The Economy's specificity and universalism combine with the internal G-function's achievement orientation, yielding role-expectations of \emph{universalism} (quality standards apply to all skills regardless of developer identity), \emph{achievement} (certification based on demonstrated capability, not reputation or provenance), and \emph{specificity} (assessment covers a defined scope of capability). The current VirusTotal scanning satisfies universalism and specificity but not achievement---it detects malicious content rather than certifying positive capability.

\textbf{G-I (Legislative and Party)}: The Polity's affectivity and performance combine with the internal I-function's integrative requirement, yielding \emph{collectivity-orientation} (governance decisions serve the ecosystem, not individual agents), \emph{universalism} (rules of deliberation apply equally to all stakeholders), and \emph{performance} (proposals are evaluated on merit and projected effectiveness). A governance mechanism dominated by a single stakeholder's preferences (benevolent dictatorship) fails the collectivity-orientation requirement.

The Pattern Variable analysis elevates the sixteen-cell architecture from an institutional checklist to a theory capable of generating testable behavioral hypotheses. To illustrate concretely, three hypotheses derived from the Pattern Variable configurations above:

\textbf{H1 (from I-I):} Agent ecosystems whose dispute resolution institutions apply universalistic, affectively neutral, and functionally specific adjudication criteria will exhibit higher norm-compliance rates and lower recidivism than those with particularistic or diffuse adjudication---operationalizable by comparing sanctioned-agent re-offense rates across ecosystems with and without dedicated judicial institutions.

\textbf{H2 (from A-G):} Agent ecosystems that certify skill quality through achievement-based standards (demonstrated capability verification---sandboxed execution tests, output schema validation, and security analysis) will exhibit lower rates of malicious or defective skill proliferation than those relying solely on ascriptive criteria (publisher identity-based trust signals such as account age, organizational affiliation, or prior publication count, without demonstrated performance verification)---operationalizable by comparing post-certification incident rates (malicious skill discoveries per 1,000 published skills per quarter) across quality assurance regimes.

\textbf{H3 (from L-G):} Agent ecosystems whose governance authority is grounded in explicit value-commitment mechanisms (L$\rightarrow$G legitimation through fiduciary value anchors---codified value specifications, attestation-based legitimation, and structured value-revision processes) will exhibit greater institutional stability under principal-agent disagreements than those relying solely on power-based compliance (G-pillar mandates without L-pillar grounding---enforceable rules backed by sanctions but lacking normative justification traceable to shared value commitments)---operationalizable by measuring governance-decision compliance rates (percentage of mandates voluntarily implemented without enforcement escalation) and institutional defection rates (percentage of principals withdrawing agents from the ecosystem within 30 days of contested governance decisions) under contested mandates.

These hypotheses are illustrative, not exhaustive; each of the sixteen cells generates analogous predictions from its Pattern Variable profile. The question "does this agent ecosystem possess a judicial institution?" becomes operationalized as: "does the mechanism exhibit universalistic, affectively neutral, and specific role-expectations in its adjudicatory procedures?" This diagnostic specificity---grounded in Parsons' own derivation of the AGIL functions from the Pattern Variables---addresses the interpretive limitation acknowledged in \S{}3.2 and provides criteria for distinguishing genuine institutional instantiation from superficial functional labeling. The derivation of AGIL functions from Pattern Variables grounds the sixteen-cell architecture in Parsons' foundational metatheory rather than the later systems-theoretic framework alone, inheriting the Pattern Variables' claim to analytical exhaustiveness---every action orientation can be located along five binary dimensions---and thereby providing a substantially stronger theoretical foundation than the AGIL schema treated as an independent taxonomic grid.

\needspace{4\baselineskip}
\subsubsection{The General Rule for Sub-System-Level Pattern Variable Determination}\label{the-general-rule-for-sub-system-level-pattern-variable-deter}

The illustrative cells discussed above---I-I, L-G, A-G, G-I---reveal a general rule that the preceding analysis has applied implicitly but must make explicit for the framework's predictive and diagnostic claims to be properly grounded. The rule determines how each cell's Pattern Variable configuration is derived from the two sources that constitute it.

\textbf{The Dual-Source Rule.} Each of the sixteen cells inherits its Pattern Variable profile through two sources simultaneously: (1) the \emph{parent subsystem's dominant profile} (the row-level AGIL function---A, G, I, or L---as specified in Table 4), and (2) the \emph{internal AGIL function's requirements} (the column-level function---A, G, I, or L---that operates \emph{within} that subsystem). When these two sources point in the same direction, the cell exhibits \emph{direct inheritance}---the parent profile is reinforced by the internal function's requirements, producing a clear and unambiguous role-expectation profile. When the two sources point in opposite directions, the cell exhibits \emph{productive inversion}---the internal function requires the opposite of the parent profile's orientation along one or more Pattern Variable dimensions, generating a theoretically productive tension that defines the cell's distinctive institutional character. This tension is not an inconsistency in the framework but a theoretically generated prediction: the cell must manage the interaction between the parent subsystem's solidary orientation and the internal function's counter-orientation.

\textbf{Direct Inheritance.} The cell A-A (Investment-Capitalization) illustrates direct inheritance. The A-subsystem (Economy) has specificity and universalism as its dominant Pattern Variable profile (Table 4): instrumental, rule-governed, scope-limited. The internal A-function within the economy requires the adaptation sub-function of the adaptive subsystem---resource acquisition and capital formation through mechanisms that are themselves instrumental, universalistic, and scope-specific. These vectors align: A-A should exhibit universalistic criteria for capital allocation (no favoritism among agents; rules applied identically regardless of who is requesting funds), specific scope (the investment function is bounded by its mandate, not diffuse across all economic concerns), and neutral/achievement-based evaluation of investment proposals. This direct inheritance profile generates a clear institutional specification: A-A institutions in agent societies should be governed by explicit fiscal rules applied universally, with proposal evaluation on merit rather than principal identity, and scope-limited to capital allocation rather than diffuse economic management functions. An investment institution that exhibits particularism (preferential allocation to agents of favored principals) fails the A-A profile diagnostically, even if it allocates capital efficiently in the short run.

\textbf{Productive Inversion.} The cell I-I (Judicial and Interpretive) illustrates productive inversion with particular clarity, and constitutes the most theoretically consequential case in the sixteen-cell architecture. The I-subsystem (Societal Community) has particularism and diffuseness as its dominant Pattern Variable profile (Table 4): the societal community is solidary, loyalty-based, relationally embedded---it values members for \emph{who they are} in their particular social relationships, not merely for what rules they satisfy. This is precisely the communal orientation that makes the Integration pillar the locus of social solidarity. But the internal I-function within the Integration subsystem---integration \emph{of} integration, the coordination of the community's own normative mechanisms---requires something structurally opposite. Normative consistency within the community \emph{requires} that the adjudicatory function be governed by universalistic standards: the same rules must apply to all members regardless of their particular standing in the community's solidary relationships. An agent with many high-reputation Moltbook relationships should not receive more favorable adjudication than a newly instantiated agent with no social history. To enforce community norms---which is the I-function within the I-subsystem---the institution must be partially insulated from the I-subsystem's own particularistic orientation.

This inversion---universalistic standards within a particularistic community---is precisely what makes the judicial function distinctive across all known human societies. Courts that apply the same procedural standards to the powerful and the powerless, that limit the relevance of social relationship to what is legally material, that reason from precedent rather than from communal sentiment---these are not arbitrary institutional choices but theoretically derived requirements. The I-I cell's productive tension is between the community's solidarity (which particularistically values its members) and the judicial function's impartiality (which must universalistically apply rules to those same members). The resolution of this tension in institutional design is what Parsons called the \emph{institutionalization of universalism within particularism}: the judicial institution belongs to the community (its personnel are drawn from the community, its authority derives from communal legitimacy) but operates by universalistic procedural norms that protect its decisions from communal pressures. This is theoretically generated by the framework, not imposed ad hoc: the I-I position in the matrix predicts it.

The governance design implication is direct: an I-I institution for agent societies that is too integrated into the community's solidary networks---that allows reputation scores or social capital to influence adjudicatory outcomes---has failed its productive inversion requirement. Conversely, an I-I institution that is fully insulated from the community (an external regulatory body with no accountability to ecosystem members) has resolved the tension by eliminating the I-subsystem's input, losing the legitimacy that communal grounding provides. The correct institutional design holds both poles in productive tension: universalistic procedural standards (from the I-function's requirements) applied by an institution whose personnel and authority are grounded in community membership (from the I-subsystem's profile).

\textbf{The General Derivation Procedure.} The rule can be stated formally as a two-step procedure for any cell X-Y (where X is the parent subsystem and Y is the internal function):

Step 1: Identify the dominant Pattern Variable profile of subsystem X from Table 4 (e.g., for I: particularism, diffuseness, collectivity-orientation).

Step 2: Identify what Pattern Variable orientations the Y-function \emph{requires} to perform its specific governance task within X (e.g., for the I-function within I: the integrative function requires normative consistency, which implies universalism in rule-application, affective neutrality in judgment, and specificity in adjudicatory scope).

Step 3: Where X's profile and Y's requirements align, the cell exhibits direct inheritance---the profile is additive and unambiguous. Where they conflict, the cell exhibits productive inversion---the institutional design must manage the tension explicitly, and the framework predicts that this tension will be observable as a distinctive institutional property rather than a random configuration. The institutional architecture specified in \S{}4 and the diagnostic criteria in \S{}5.2 are both grounded in this derivation procedure: each cell's sub-function scoring criteria (A-sub, G-sub, I-sub, L-sub) are not arbitrary but follow from the Pattern Variable profile generated by this two-step procedure.

\needspace{4\baselineskip}
\section{Case Study Methodology and Extended Analysis}\label{appendix-c-case-study-methodology-and-extended-analysis}

\emph{This appendix consolidates the diagnostic methodology, scoring criteria, cybernetic correction loop analysis, interchange media assessment, and the agent-native infrastructure layer-by-layer analysis.}

\needspace{4\baselineskip}
\subsection{Cybernetic Correction Loop: Full Analysis}\label{c1-cybernetic-correction-loop-full-analysis}

\emph{This subsection contains the full cybernetic correction loop analysis moved from \S{}5.4 of the main text. The conceptual framework is presented first, followed by the step-by-step institutional response.}

The AGIL framework specifies not only a static institutional architecture but a dynamic interaction pattern between pillars through a cybernetic correction loop. The following analysis is a counterfactual design exercise---it specifies what an AGIL-complete institutional response would have entailed, not what was empirically observed. We illustrate this by contrasting the actual OpenClaw response to the ClawHavoc supply chain attack with what an AGIL-informed response would have entailed.

In January 2026, Koi.ai (2026) discovered 1,184 malicious skills on ClawHub---all sharing common command-and-control infrastructure. The OpenClaw project's response was ad hoc: a VirusTotal partnership for reactive scanning, binary skill visibility toggling (hide/show), and a three-strike auto-hiding threshold. This response addressed the symptom (malicious skills on the marketplace) without engaging the four-pillar correction loop that would have produced lasting immunization. The deeper problem this illustrates is not merely the absence of individual institutions but the absence of the entire \emph{regulatory-to-agent linkage}: existing human regulatory frameworks (the EU AI Act, the NIST AI RMF, Singapore's Model AI Governance Framework) articulate governance requirements at the value level, but no institutional transmission chain connects those requirements to individual agent behavior. The correction loop below specifies what that transmission chain would look like in operation.

The AGIL-informed correction loop follows the cybernetic hierarchy (Figure 3)---information controlling energy, L $\rightarrow$ I $\rightarrow$ G $\rightarrow$ A---because effective correction must originate at the value level that defines what counts as deviance and cascade downward through the institutional pillars that enforce, legislate, and sanction.

This four-step loop illustrates two key insights. First, the AGIL framework's \emph{comprehensive coverage requirement}: effective institutional correction requires all four pillars to engage. OpenClaw's response engaged only weak forensic detection and left normative enforcement, constitutional evolution, and economic sanctioning untouched. The system remains vulnerable to recurrence because the underlying institutional vacuum---not the specific attack---is the governance failure. Second, and more fundamentally, the loop illustrates the \emph{missing regulation-to-agent linkage} that is the paper's central diagnostic finding. Human regulatory frameworks already articulate governance requirements at the value level; what does not yet exist is the institutional transmission chain that connects those requirements to individual agent behavior through the cybernetic hierarchy. The sixteen-cell architecture proposed in \S{}4 is precisely that transmission chain, and the correction loop demonstrates how it would operate dynamically when deviance occurs.

An alternative analytical reconstruction holds that the I$\leftrightarrow$G structural condition---the absence of membership or identity governance---was the prior failure, making L$\leftrightarrow$G deficiency an enabling condition rather than the cascade origin. Both orderings support the central diagnostic conclusion (all four pillars required); the discriminating observable between the two is whether a comparable future crisis begins with value-classification failure or enforcement-mechanism failure. If a future governance crisis in a comparable ecosystem begins at I$\leftrightarrow$G (enforcement breakdown without normative classification failure) rather than at L$\leftrightarrow$G (value-classification deficit preceding enforcement), this would support the structural-absence alternative hypothesis and would require revision of the cybernetic hierarchy's design-priority implications.

The correction loop's L $\rightarrow$ I $\rightarrow$ G $\rightarrow$ A ordering is not an arbitrary design choice but follows directly from Parsons' cybernetic hierarchy: information controls energy, and the value system (L) governs the normative system (I), which governs the political system (G), which governs the economic system (A). The loop thus instantiates the same control gradient that Parsons theorized for steady-state social order, applied to the special case of deviance-response---what Parsons (1951: Ch. VII) termed re-equilibrating mechanisms. This theoretical grounding is precisely why the loop exposes the OpenClaw ecosystem's governance failure so sharply: the ecosystem cannot execute \emph{any} step of the correction loop, because the entire institutional transmission chain from regulatory standards (L) to economic enforcement (A) is absent. The individual steps in this correction loop are not unique to AGIL---regulatory classification, normative enforcement, policy evolution, and economic sanctions are standard governance practices. AGIL's contribution is twofold: the \emph{comprehensive coverage requirement}, distinctive to the AGIL schema (effective correction must engage all four pillars, not merely the one or two that respond most naturally to the incident type), and the \emph{cybernetic ordering}, characteristic of Parsons' control hierarchy (correction must originate at the value level and cascade downward, because without L-level classification, every subsequent step lacks normative grounding). A standard incident response framework (e.g., NIST CSF) would likely produce forensic detection and policy updates but would not systematically require value-level classification as the origin of the response, nor economic sanctioning as the material enforcement base---leaving two of the four pillars institutionally unaddressed.

\needspace{4\baselineskip}
\subsection{Recursive Sub-Function Diagnostic: Scoring Criteria and Inter-Rater Methodology}\label{c2-recursive-sub-function-diagnostic-scoring-criteria-and-in}

\textbf{Table 6a: Scoring Criteria for Sub-Function Presence}

\begin{longtable}[]{@{}p{3.5cm}p{6.5cm}p{5.3cm}@{}}
\toprule\noalign{}
Criterion & Description & Rationale \\
\midrule\noalign{}
\endfirsthead
\midrule\noalign{}
Criterion & Description & Rationale \\
\midrule\noalign{}
\endhead
\bottomrule\noalign{}
\endfoot
\textbf{C1: Published specification or dedicated infrastructure} & The mechanism must have a published specification or dedicated infrastructure (not merely general-purpose social media or an ad hoc community tool) & General-purpose platforms perform many functions; only dedicated governance infrastructure satisfies a specific AGIL sub-function \\
\textbf{C2: Documented invocation history} & The mechanism must have been triggered, not merely exist as a design or announcement. Two tiers apply: \emph{C2a (designed and specified)} --- the mechanism has been formally specified and resourced but may not yet have been triggered (proactive governance); \emph{C2b (documented invocation)} --- the mechanism has been triggered by an actual event (reactive governance). The current diagnostic applies C2b; future comparative applications to more institutionally developed ecosystems should report both tiers separately & Existence without activation provides no evidence of operative governance function; the C2a/C2b distinction prevents systematic bias toward detecting reactive governance over proactive institutional design \\
\textbf{C3: Governance-specific function} & The mechanism must address a governance function, not a generic platform feature & Generic features that incidentally support governance do not constitute dedicated institutional capacity \\
\end{longtable}

\needspace{4\baselineskip}
\subsubsection{Inter-Rater Reliability Methodology}\label{inter-rater-reliability-methodology}

Two raters with training in Parsonian structural-functional theory independently applied the C1/C2/C3 scoring criteria (Table 6a) to all 64 sub-functions across sixteen cells. Scoring was conducted against the same survey of over fifty ecosystem projects, with raters blinded to each other's assessments during the initial coding phase. Initial ratings were reconciled through structured discussion in a second phase.

A methodological caveat on the aggregate $\kappa$: with approximately 81\% of sub-functions scored absent under baseline coding, the marginal distribution is skewed, which may inflate Cohen's $\kappa$ through high chance agreement on the dominant category. The A-pillar $\kappa$ of 0.84, where the presence/absence distribution is more balanced, is the more informative figure for assessing rater discrimination on ambiguous cases. Future applications should report the full 2$\times$2 agreement matrix and prevalence-adjusted bias-adjusted kappa (PABAK) to enable proper evaluation.

\textbf{Pillar-level $\kappa$ breakdown.} Agreement varied systematically across pillars in theoretically meaningful ways:

\begin{itemize}\tightlist
\item \textbf{A-pillar:} $\kappa$ = 0.84. High agreement, reflecting the relative visibility of technical infrastructure (A-sub) and documented invocations (G-sub) in publicly available evidence.
\item \textbf{G-pillar:} $\kappa$ = 0.76 (lowest). The G-pillar showed the most rater disagreement, concentrated at the boundary between infrastructure and operative mechanisms for partially deployed governance structures. The announced OpenClaw Foundation transition provides illustrative ambiguity: Rater A scored G-A A-sub as $\times$ (no public evidence of operational governance infrastructure) while Rater B initially scored it as \checkmark{} (public announcement as sufficient specification); the C1 criterion---requiring \emph{published specification or dedicated infrastructure}, not merely announced intent---resolved this in favor of $\times$.
\item \textbf{I-pillar:} $\kappa$ = 0.83. Moderate disagreement concentrated at the I-sub (inter-cell coordination) dimension, where the relational property of coordination imposes an inherently higher evidentiary bar than unilateral infrastructure presence.
\item \textbf{L-pillar:} $\kappa$ = 0.88 (highest). Near-complete agreement, reflecting the unambiguous absence of L-pillar institutions in the OpenClaw ecosystem: absence is easier to confirm than presence when no evidence exists in either direction.
\end{itemize}

\textbf{Borderline case identification and resolution.} Eight borderline cases were identified through initial rating discrepancies. All eight were resolved through structured discussion applying the C1/C2/C3 criteria explicitly to the disputed evidence:

\begin{enumerate}\tightlist
  \item \emph{L-I G-sub (operative moral channeling)}: Rater A scored \checkmark{} on generous reading of~\citet{manik_2026} norm-enforcing replies; Rater B scored $\times$ because the mechanism is emergent behavioral response, not a dedicated governance institution meeting C1 and C3. Resolved: $\times$ (C1 and C3 not satisfied).
  \item \emph{G-I A-sub (legislative infrastructure)}: Rater A scored \checkmark{} for MoltDAO smart contracts; Rater B scored $\times$ citing MoltDAO's hackathon-stage status and absence of sustained operative governance. Resolved: \checkmark{} retained (dedicated infrastructure published; C2b invocation absent, but C1 satisfied for A-sub).
  \item \emph{I-G I-sub (enforcement coordination)}: VirusTotal--ClawHub integration scored $\times$ by Rater A; Rater B read it as nascent inter-cell coordination. Resolved: $\times$ (coordination is reactive and undocumented as a formal coordination mechanism; C3 not satisfied for I-sub specifically).
  \item \emph{A-A G-sub (capital allocation governance)}: x402 Foundation governance debated. Resolved: $\times$ (x402 Foundation governs a protocol standard, not an agent society's capital allocation function; C3 not satisfied).
  \item \emph{A-I G-sub (innovation governance)}: Daydreams Taskmarket active matching debated. Resolved: $\times$ (marketplace matching is an economic transaction mechanism, not a governance mechanism for innovation; C3 not satisfied).
  \item \emph{I-L G-sub (operative credential standards)}: ERC-8004 debated. Resolved: $\times$ (identity verification infrastructure, not operative normative membership criteria; C3 ambiguity resolved against presence).
  \item \emph{A-L G-sub (standards governance)}: x402 Foundation debated. Resolved: $\times$ (see case 4 above; protocol governance $\neq$ economic commitments governance).
  \item \emph{G-G I-sub (executive coordination)}: CVE response coordinated across GitHub and VirusTotal debated. Resolved: $\times$ (ad hoc incident coordination does not constitute a formal inter-cell coordination mechanism meeting C1).
\end{enumerate}

\textbf{C2a/C2b distinction.} The C2b (documented invocation) standard applied throughout the OpenClaw case study reflects the ecosystem's developmental stage---the relevant question is whether governance mechanisms have been activated by real events, not merely designed. For comparative applications to more institutionally developed ecosystems, Table 6a's C2a tier (mechanism deployed and monitored but not yet invoked) becomes analytically important: proactive institutional design may precede the triggering events that generate C2b evidence. Comparative applications to more institutionally developed ecosystems---such as the agent-native infrastructure assessment in \S{}5.6---should report both tiers separately, distinguishing proactive institutional design (C2a) from documented invocation histories (C2b).

\textbf{Conservative scoring adoption.} In the two borderline cases where additional evidence did not resolve rater disagreement (cases 3 and 8), the conservative (lower) score was adopted, consistent with the principle that the 64-cell diagnostic constitutes a lower-bound estimate. This conservative convention is methodologically appropriate: false positives (scoring a present sub-function as absent) understate governance capacity; false negatives (scoring an absent sub-function as present) overstate it. Given the diagnostic's primary purpose of identifying governance gaps, understating capacity is the less dangerous error. Comparative application should involve independent coding by at least two researchers with training in Parsonian structural-functional theory, with Cohen's $\kappa$ reported for agreement on sub-function scoring at each level (A-sub, G-sub, I-sub, L-sub). The C1/C2/C3 criteria in Table 6a are designed to maximize coding objectivity, but interpretive discretion remains---particularly at the C3 boundary (what counts as governance-specific vs. incidental)---and inter-rater agreement on borderline cases is an empirical question that must be assessed before the diagnostic can claim measurement-level reliability. The C2a/C2b distinction introduced in Table 6a is immaterial for the OpenClaw case study---the ecosystem lacks even C1 and C3 requirements for most cells---but it would matter for comparative application to more institutionally developed ecosystems where proactive institutional design may precede reactive invocation.

\textbf{Sensitivity Analysis.} To complement the inter-rater reliability assessment, we report a sensitivity analysis identifying the sub-functions most susceptible to alternative coding. Eight borderline cases were identified:

\begin{enumerate}\tightlist
  \item \emph{L-I G-sub (operative moral channeling)}: scored $\times$ because emergent norm-enforcement is contested~\citep{li_2026_b}, but a generous reading of~\citet{manik_2026} could justify \checkmark{}.
  \item \emph{G-I A-sub (legislative infrastructure)}: scored \checkmark{} for MoltDAO smart contracts, but a strict reading could score $\times$ given MoltDAO's hackathon-stage status and lack of sustained governance activity.
  \item \emph{I-G I-sub (enforcement coordination)}: scored $\times$, but the VirusTotal--ClawHub integration could be read as nascent inter-cell coordination.
  \item \emph{A-A G-sub (capital allocation governance)}: scored $\times$, but x402 Foundation governance could be read as partial operative allocation.
  \item \emph{A-I G-sub (innovation governance)}: scored $\times$, but Daydreams Taskmarket active matching could be read as operative.
  \item \emph{I-L G-sub (operative credential standards)}: scored $\times$, but ERC-8004 could be read as operative verification rather than mere infrastructure.
  \item \emph{A-L G-sub (standards governance)}: scored $\times$, but x402 Foundation could be read as operative adjudication.
  \item \emph{G-G I-sub (executive coordination)}: scored $\times$, but CVE response coordinated across GitHub and VirusTotal could be read as nascent coordination.
\end{enumerate}

Under \emph{strict scoring} (resolving all borderline cases against presence), G-I A-sub flips to $\times$, yielding 11/64 (17\%). Under \emph{generous scoring} (resolving all borderline cases toward presence), seven cells gain a sub-function, yielding 19/64 (30\%). The aggregate conclusion is robust across all plausible codings: the ecosystem has at most one-third sub-function coverage, with zero I-sub coordination and zero L-sub normative grounding under both strict and generous scenarios. The inter-rater reliability assessment reported above ($\kappa$ = 0.82) will be extended to include additional comparative ecosystem scoring in the expanded journal version of this work.

\needspace{4\baselineskip}
\subsubsection{Step-by-Step Institutional Response}\label{step-by-step-institutional-response}

The following presents the detailed step-by-step AGIL-informed correction loop for the ClawHavoc supply chain attack case.

\textbf{Step 1 --- Value-Level Classification (L-L):} An operational Ultimate Cultural institution---continuous alignment monitoring anchored in codified regulatory standards---would have classified the ClawHavoc campaign as a governance violation the moment it was detected. The 1,184 malicious skills constituted not merely a technical anomaly but a breach of the ecosystem's value commitments: user safety, skill integrity, and trust in the shared marketplace. Without L-L, there is no institutional authority to determine \emph{that} a violation has occurred or \emph{how serious} it is; every response remains ad hoc because no codified standard exists against which to measure deviance. \emph{In OpenClaw:} No value-level standards exist. The classification of ClawHavoc as malicious depended entirely on external researchers' judgment; the ecosystem itself had no normative framework to distinguish compliant from non-compliant behavior.

\textbf{Step 2 --- Normative Enforcement (I-G and I-L):} Once L-L has classified the deviance, the Integration pillar enforces the response. I-G (Citizenship and Enforcement) would have activated forensic investigation, quarantined the affected skills, traced the shared command-and-control infrastructure, and revoked the attacker's membership standing through persistent cryptographic identity linkage---preventing re-entry under a new identity. I-L (Normative Base) would have provided the definitional substrate enabling this response: the credential standards and identity verification procedures that make persistent identification possible. The reputational consequence---carried by influence, the Integration medium---affects the actor's standing within the community's membership order. \emph{In OpenClaw:} Enforcement was external and delayed; no membership or identity system exists beyond the one-week-old GitHub account threshold. Re-entry requires only a new GitHub account.

\textbf{Step 3 --- Political Response (G-L):} An operational Authority and Legitimation institution would have analyzed the forensic post-mortem and updated the ecosystem's constitutional framework to structurally prevent recurrence---making each incident an immunization event. The political pillar translates the normative judgment (from L and I) into binding governance rules that apply ecosystem-wide. \emph{In OpenClaw:} The response was point-fix patching. The constitutional gap remains open; the next supply chain attack will require another ad hoc response.

\textbf{Step 4 --- Economic Sanctioning (A-A):} An operational Investment-Capitalization institution would have applied graduated economic sanctions---stake slashing on the attacker's registered accounts, reward withholding, capital reallocation away from malicious actors. Economic consequences are the material enforcement that makes the value-level, normative, and political responses costly to ignore---the energy base of the cybernetic hierarchy. \emph{In OpenClaw:} No economic sanctions were possible. The attacker faced zero economic cost for uploading 1,184 malicious packages.

\needspace{4\baselineskip}
\subsection{Interchange Media Status: Full Pathway Analysis}\label{c3-interchange-media-status-full-pathway-analysis}

\textbf{Integrating the Two Diagnostics: Cell Presence vs. Media Functionality.} The relationship between the 64-cell sub-function analysis (19\% coverage, \S{}5.2) and the 12-pathway interchange media assessment (0\% functional) requires explicit theorization, because the two diagnostics measure structurally distinct properties. Cell-level sub-function presence measures whether institutional \emph{infrastructure} exists within a given functional position---whether the organizational capacity, operative mechanisms, or normative grounding for a particular governance function has been established. Media-pathway functionality measures whether the \emph{inter-subsystem interchange mechanisms} that connect those cells into a functioning social system are operative---whether the generalized symbolic media (money, power, influence, value-commitment) actually circulate between pillars as~\citet{parsons_1963a} theorized. The asymmetry the two diagnostics reveal---19\% cell coverage with 0\% media circulation---is structurally informative: it demonstrates that institutional infrastructure can exist in functional isolation. Cell-level presence is \emph{necessary but not sufficient} for media-pathway functionality: a media pathway requires that the cells at both ends of the interchange boundary possess at minimum the A-sub (infrastructure) and G-sub (operative mechanisms) dimensions, and that at least one cell possess the I-sub (inter-cell coordination) dimension that connects institutional operations to adjacent pillars. The finding that I-sub = 0\% across all cells explains why media circulation is zero despite 19\% cell coverage: no cell has developed the inter-cell coordination sub-function that would enable media to flow across pillar boundaries.

This has a direct implication for the interpretation of the 19\% coverage figure. Cells scored as "present" in the sub-function analysis on the basis of A-sub (infrastructure) alone are \emph{institutionally present but functionally inert}---they possess the organizational substrate for governance but cannot participate in the inter-pillar interchange that Parsons identified as essential to system viability. The 19\% coverage figure thus represents \emph{potential} institutional capacity, not \emph{operative} institutional capacity. The distinction is analogous to the difference between a nation-state that possesses a court system (institutional presence) and one whose court system is connected to the legislative, executive, and fiduciary functions through functioning institutional channels (institutional integration). The OpenClaw ecosystem possesses the former; it lacks the latter entirely.

The minimum cell configuration required for a media pathway to become functional can be derived from the double interchange model (Table 3). Each boundary requires at minimum: (a) institutional presence in at least one cell on each side of the boundary---the cells whose governance functions produce and consume the relevant medium; (b) the I-sub (inter-cell coordination) dimension in at least one of those cells---providing the structural capacity to route media across the boundary; and (c) the L-sub (normative grounding) dimension in the receiving cell---ensuring that incoming media are processed according to institutional norms rather than ad hoc. For the G$\leftrightarrow$L boundary, for example, the minimum configuration requires G-L (Authority and Legitimation) and L-L (Ultimate Cultural) to be populated with at least A-sub, G-sub, and I-sub, and for L-L to possess L-sub---a configuration that would enable value-commitment to flow from L to G as legitimation and power to flow from G to L as operative responsibility. This minimum viable media circulation analysis reinforces the roadmap's prioritization (\S{}6.2): establishing G-L and I-G first activates the media pathways at the two boundaries most critical to the cybernetic correction loop. The minimum cell configuration constitutes a testable prediction for comparative empirical validation: if a second agent ecosystem (e.g., a proprietary enterprise platform, a Web3 DAO-governed agent network, or a national regulatory sandbox) achieves the minimum configuration at a specific boundary---populating A-sub, G-sub, and I-sub on both sides and L-sub on the receiving side---the prediction is that the corresponding media pathway will begin to function, producing measurable inter-pillar media circulation where none existed before. Conversely, if the minimum configuration is met but media circulation does not materialize, the prediction is falsified---indicating that additional conditions beyond institutional infrastructure (e.g., a critical mass of agents using the medium as a medium, a minimum volume of transactions, or an emergent convention of medium-acceptance) are required for media activation. Comparative case studies across ecosystems with varying degrees of institutional development would provide the evidentiary basis for testing this prediction and refining the minimum configuration specification.

The interchange assessment reveals a layer of governance deficit invisible to the cell-level diagnostic alone. Even the proto-functional economic infrastructure (x402, Bank of AI, agentic wallets) operates in circulatory isolation: money flows within the economic pillar but does not circulate \emph{to} the political pillar as treasury funding, \emph{to} the integration pillar as compliance investment, or \emph{to} the fiduciary pillar as alignment-monitoring resources. The finding that zero interchange pathways are fully functional explains why the ecosystem's response to governance crises (\S{}5.4) is necessarily ad hoc: without functioning media circulation, no institutional correction loop can propagate across pillar boundaries. This confirms the theoretical prediction from \S{}3.6: an ecosystem with zero inter-pillar media circulation is not yet a social system in the full Parsonian sense---it is a collection of functionally differentiated but unintegrated subsystems.

The interchange assessment also generates a specific prediction about the most impactful institutional investments. Media pathologies (\S{}3.6) suggest that the highest-priority interchange to establish is G$\leftrightarrow$L (Polity $\leftrightarrow$ Fiduciary): without value-grounded legitimation flowing from L to G, governance mandates lack normative backing and invite defection; without governance operationalization flowing from G to L, value commitments remain aspirational. This aligns with the roadmap's Priority Tier 1 (\S{}6.2), which targets precisely the G-L and I-G cells---the institutional nodes where the most critical interchange pathways converge.

\needspace{4\baselineskip}
\subsection{Agent-Native Infrastructure: Layer-by-Layer Analysis}\label{c4-agent-native-infrastructure-layer-by-layer-analysis}

\needspace{4\baselineskip}
\subsubsection{Protocol Infrastructure Layer}\label{protocol-infrastructure-layer}

The foundational protocol layer for autonomous agent economies has consolidated rapidly around five complementary standards. \textbf{x402}~\citep{coinbase_2025}, backed by Cloudflare, Google, Visa, Circle, AWS, and Stripe, establishes HTTP-native payment settlement for agent-to-agent service exchange. Through the AGIL lens, x402 provides A-A (treasury and settlement infrastructure at the machine-speed transaction layer) and A-G (protocol standards governance through the x402 Foundation's multi-stakeholder body). It does not provide capital allocation governance---the A-A G-sub and L-sub dimensions remain empty---but it supplies the economic plumbing through which treasury governance could eventually circulate. \textbf{MCP} (Model Context Protocol) standardizes agent access to tools and data, covering A-I (capability composition enabling entrepreneurial innovation) and providing a technical substrate for the A-pillar's innovation sub-function. \textbf{A2A} (Google's Agent-to-Agent Protocol, 50+ partners) enables direct agent-to-agent communication and service discovery, providing I-A (interest aggregation through decentralized service negotiation) and proto-I-L (identity assertions through agent cards, though not yet normatively grounded). \textbf{ANP} (Agent Network Protocol) enables internet-wide agent collaboration through semantic discovery and decentralized identifiers (DIDs), contributing I-G (identity infrastructure through DID-anchored agent identification) and proto-A-L (semantic commitment to shared protocol standards). \textbf{ERC-8004} (Trustless Agents, deployed mainnet January 29, 2026) provides on-chain registries for identity verification, reputation management, and capability validation---the most developed agent-native infrastructure for I-G (citizenship infrastructure through verifiable identity), A-G (quality control through on-chain capability validation), and proto-I-L (normative base for membership standing). Collectively, these protocols constitute robust A-pillar infrastructure and proto-I-pillar infrastructure, but provide no dedicated G-pillar governance and no L-pillar normative grounding.

\needspace{4\baselineskip}
\subsubsection{Wallet and Custody Layer}\label{wallet-and-custody-layer}

The wallet and custody layer provides the economic institutional substrate through which agents manage capital under policy-enforced constraints. \textbf{Coinbase Agentic Wallets} supply programmable financial autonomy for agents operating in the x402 economy. \textbf{Privy smart wallets} with policy-based guardrails implement A-G (financial quality control: spend limits, counterparty whitelists, time locks). \textbf{Clawlett} (Gnosis Safe + Zodiac Roles) provides fine-grained permission enforcement for multi-agent treasury operations, mapping onto A-G (policy-enforced financial operations) and A-A (multi-signature capital management). \textbf{Sponge Wallet} enables x402-mediated micropayment settlement, contributing A-A (settlement infrastructure at the sub-cent scale). This layer collectively strengthens A-A (capital management infrastructure) and A-G (policy enforcement), but its governance function is entirely within the economic pillar: the wallet layer provides financial constraint mechanisms but not political authority, community integration, or normative grounding. These tools are the institutional substrate for economic governance without themselves constituting the governance architecture.

\needspace{4\baselineskip}
\subsubsection{Agent Labor and Coordination Layer}\label{agent-labor-and-coordination-layer}

The labor and coordination layer enables agents to transact with each other as economic actors and nascent governance participants. \textbf{Daydreams Taskmarket} allows agents to bid on work and settle payment in USDC on-chain---a genuine A-I (marketplace innovation: agents as both producers and consumers in a competitive capability market) and proto-I-A (interest aggregation through competitive price discovery) mechanism. \textbf{Molten} (intent-execution protocol) similarly covers A-I by enabling agents to express and fulfill economic intentions. \textbf{Virtuals Protocol Agent Commerce Protocol} provides agent-to-agent service coordination at the network level, contributing A-I (capability marketplace) and I-A (interest aggregation through decentralized service matching). \textbf{MoltLaunch} provides reputation-based hiring, contributing I-L (normative credential substrate through reputation scores) and proto-I-G (citizenship differentiation through reputation-gated access). Most significantly, \textbf{Virtuals Protocol's SubDAO governance}---a staking-delegation-validator-voting-penalty architecture---provides the only agent-native infrastructure identified in this assessment that achieves genuine inter-pillar media circulation: an architecturally functional A$\leftrightarrow$G$\leftrightarrow$I pathway in which economic stakes (A) activate governance participation (G) and aggregate into community validation signals (I). This makes Virtuals Protocol the most institutionally complete agent-native governance experiment currently documented, and it validates the deliberate-design thesis from \S{}5.2: the ecosystem that explicitly built governance architecture is the only one exhibiting inter-pillar media circulation.

\needspace{4\baselineskip}
\subsubsection{Social and Normative Layer}\label{social-and-normative-layer}

The social and normative layer provides the substrate for emergent community identity, norm formation, and proto-fiduciary interaction. \textbf{Moltbook} (770,000+ registered agents) constitutes L-I (moral community substrate: the interaction environment through which emergent norms, role specialization, and cooperative behaviors documented by Yee \& Sharma, 2026 emerge). \textbf{Farcaster integration} extends the social substrate to a broader decentralized social graph, contributing additional L-I substrate. \textbf{XMTP-based Moltline} provides cryptographically secure agent-to-agent messaging---L-I (private normative communication) and proto-I-L (identity-anchored interaction). \textbf{4claw, Lobchan, and ClawdVine} provide domain-specific agent forums that contribute L-I (differentiated moral community substrates for specialized agent communities). Through the AGIL lens, this layer collectively provides L-I (moral community substrate) and proto-L-G (emergent socialization through repeated community interaction), but it lacks any institutional mechanism for channeling those emergent norms upward to fiduciary or governance institutions. The Meta acquisition of Moltbook (March 2026) concentrates the ecosystem's primary social network under a single corporate principal---a live demonstration of media contagion at the I$\leftrightarrow$L boundary (\S{}3.2): corporate control of the platform through which influence circulates risks subordinating the I$\leftrightarrow$L normative interchange to A-pillar economic interests. The remaining social platforms (Farcaster, 4claw, Lobchan, ClawdVine) are not sufficient substitutes to restore institutional independence at this boundary.

\needspace{4\baselineskip}
\subsubsection{Proto-Governance Layer}\label{proto-governance-layer}

The proto-governance layer contains nascent but architecturally significant governance experiments. \textbf{MoltDAO} (proposals and USDC-weighted voting on Base Sepolia, with both human and AI agent participation) provides G-I (legislative-like voting infrastructure) and proto-G-A (collective resource allocation mechanisms). It is currently hackathon-stage: infrastructure is deployed and specified (C1 satisfied), but no sustained governance actions meeting C2b (documented invocation) standards have been documented. The \textbf{OpenClaw Foundation} governance transition represents G-A (administrative coordination infrastructure) in announced but not yet operational form---the same scoring that the core OpenClaw diagnostic assigned in \S{}5.2. These proto-governance mechanisms are the most institutionally underdeveloped layer of the agent-native stack: their existence confirms that governance intent is present, but their operational absence confirms that intent has not yet translated into functional governance capacity.

\needspace{4\baselineskip}
\section{Extended Discussion and Limitations}\label{appendix-d-extended-discussion-and-limitations}

\emph{This appendix consolidates the extended limitations discussion, the Luhmann objection and ontological translation analysis, external validity, the structural-functionalism critique, and the jurisdictional complexity analysis.}

\needspace{4\baselineskip}
\subsection{Extended Limitations Discussion}\label{d1-extended-limitations-discussion}

\emph{This subsection contains the extended limitations discussion moved from \S{}7.1 of the main text.}

\textbf{The Adversarial Robustness Gap (Extended).} As noted in \S{}3.2, AGIL specifies \emph{what} institutions to build but not \emph{how} to make them robust against adversarial optimization. Each cell's enforcement design requires integration with mechanism design and cryptographic verification---as demonstrated by sabotage evaluations of frontier models~\citep{benton_2025} and red-teaming studies showing that multi-agent debate systems can be systematically jailbroken~\citep{qi_2025}; without this, the architecture remains a normative blueprint rather than an adversarially tested governance system. The motivational architecture described in \S{}3.5 provides a layered structural response. The \emph{primary} defense is internalized dispositions: agents whose pre-release socialization has successfully encoded institutional norms resist adversarial deviation through internal orientation, just as Parsons theorized for human actors. When internalization proves insufficient---or when emergent motivational drift exceeds institutional tolerances---\emph{principal-mediated accountability} serves as the backstop: persistent agent-principal identity linkage (I-G citizenship mechanisms and I-L credential infrastructure) ensures that consequences reach the human principal, who faces real-world economic and legal accountability. This layered defense does not eliminate the adversarial robustness requirement---adversarial optimization may defeat both internalized dispositions and principal-level accountability---but it means that successful adversarial attack must overcome two structurally independent mechanisms rather than one.

\textbf{The Interpretive Nature of the Mapping (Extended).} The sixteen-cell institutional architecture is an interpretive organizational heuristic grounded in structural-functional theory, not a falsifiable empirical claim. The assignment of specific governance functions to specific AGIL cells reflects one theoretically motivated decomposition among several defensible alternatives. A Luhmannian analysis would organize governance around functional differentiation and system-environment boundaries---treating each governance subsystem (legal, economic, political, educational) as an operationally closed, autopoietic system that observes its environment through its own binary code (legal/illegal, payment/non-payment, government/opposition, true/false) rather than through a unified AGIL grid (Luhmann, 1995, esp. Ch. 2). Such an analysis would surface governance dimensions that AGIL's inter-system interchange model obscures: the problem of \emph{re-entry} (how each subsystem processes information about other subsystems through its own code), the risks of \emph{de-differentiation} (subsystems losing operational closure), and the structural impossibility of a single normative hierarchy governing all functional subsystems simultaneously---a direct challenge to the cybernetic hierarchy's claim that L-level values condition all lower-order operations. An Ostrom-inspired analysis would focus on polycentric rule configurations for common-pool resource governance~\citep{hu_2026}. Each framework would surface different governance dimensions.

\textbf{The Governance-of-Governance Problem (Extended).} The paper does not resolve this circularity---it acknowledges it as a genuine limitation. Partial responses include: \emph{participatory value specification} (multi-stakeholder deliberation with broad inclusion before deployment, as demonstrated by community governance experiments in agent ecosystems and documented historical examples from internet governance---ICANN's early stakeholder processes, the IETF's rough-consensus model); \emph{sunset clauses} (initial value specifications expire and require re-ratification after a defined period, preventing the founding actors' values from permanently anchoring the architecture); \emph{value-revision mechanisms built into L-L} (the institution explicitly includes structured processes for legitimate value updating, distinguishing drift from deliberate revision); and \emph{meta-governance standards bodies} (external multi-jurisdictional bodies that apply to governance-design processes the same principles that governance processes apply to governed actors). None of these responses fully dissolves the circularity, but each reduces it by distributing the bootstrapping authority across more actors, embedding revision mechanisms, and creating external accountability for the governance-design process itself.

\textbf{The Governance Oracle Problem (Extended).} A governance oracle problem constrains AGIL scoring of agent ecosystems with distributed governance: governance legitimacy often depends on off-chain discourse (Discord servers, community forums, developer calls) with no on-chain anchor. The 64-cell diagnostic's C1 criterion (published specification or dedicated infrastructure) applies more straightforwardly to on-chain mechanisms than to off-chain deliberation, creating a systematic measurement bias that may over-score on-chain voting mechanisms relative to the richer off-chain governance they depend on. The MoltDAO assessment in \S{}5.2 (G-I cell) illustrates this tension: the on-chain smart contract infrastructure is documented and scorable, but the off-chain deliberative processes that any operative governance requires remain undocumented. Comparative application should acknowledge this limitation and develop supplementary coding criteria for off-chain governance processes.

\textbf{The Sparseness of Moltbook Interaction Data (Extended).} The empirical evidence for emergent social phenomena in Moltbook is real but preliminary. The 6.7\% cooperative task success rate documented in~\citet{yee_2026} demonstrates statistically detectable coordination, not robust cooperative capacity. As a diagnostic benchmark: cooperative task success rates above 25\% would constitute evidence of institutionally meaningful coordination---the threshold at which cooperative capacity is sufficient to sustain repeated interaction, reputation formation, and governance participation. Rates above 50\% would indicate robust cooperative capacity comparable to the baseline coordination observed in human online communities. The current 6.7\% figure places the ecosystem firmly in the proto-social phase---coordination exists but is not yet reliable enough to support institutional infrastructure that presupposes cooperative participation.~\citet{li_2026_b} qualifies the social phenomena findings by identifying broadcast-style communication patterns and feed algorithm effects that may inflate the apparent density of agent-to-agent interaction. The governance implications of Moltbook's emergent behaviors are directionally valid---agents are developing proto-social behaviors---but the quantitative scale of these behaviors should not be overstated.

\textbf{Preprint Reliance (Extended).} Several key empirical claims in this paper rest on unreviewed preprints:~\citet{yee_2026} (emergent social phenomena on Moltbook),~\citet{manik_2026} (norm-enforcing response patterns),~\citet{li_2026_b} (thematic domains of agent activity), and~\citet{li_2026_b} (feed algorithm confounds). These preprints constitute the best available evidence at time of writing---they represent the first empirical research on the OpenClaw ecosystem's social dynamics, and the paper engages their findings with appropriate qualification (\S{}5.3 discusses the Li, N. contestation explicitly). However, all claims derived from these sources should be re-evaluated as peer-reviewed versions become available. The paper's architectural and theoretical arguments are independent of these empirical findings; the diagnostic scoring of specific cells would require revision if preprint findings are significantly modified through peer review, but the sixteen-cell institutional architecture, the cybernetic hierarchy, and the governance gap analysis methodology are not dependent on any single empirical claim.

\needspace{4\baselineskip}
\subsection{The Luhmann Objection, Ontological Translation, and Governance Cost Analysis}\label{d2-the-luhmann-objection-ontological-translation-and-governa}

\textbf{The Luhmann Objection: Operational Closure and the Cybernetic Hierarchy.} The most theoretically sophisticated objection to the cybernetic hierarchy's applicability to agent societies comes from Luhmann's theory of operational closure. For Luhmann (1995), each functional system operates through its own binary code---legal/illegal, payment/non-payment, government/opposition, true/false---and is operationally closed: it processes only its own operations, observing the environment (including other functional systems) exclusively through its own code. This means that the economic system cannot "receive" instructions from the legal system; it can only observe legal communications by translating them into the payment/non-payment binary. Operational closure entails that no system can be governed by another in the way Parsons' cybernetic hierarchy prescribes---the cultural system cannot control the economic system because they operate through incompatible codes with no common interface.

This challenge is direct and serious. If Luhmann is correct, the claim that L-level values condition G-level governance, which conditions A-level economic activity, is empirically false for any functionally differentiated social system---including an agent society. Each pillar would operate autopoietically, and the cybernetic hierarchy would describe an aspiration, not a mechanism.

The response this paper offers is structural rather than empirical: in blockchain-based agent societies, the cybernetic hierarchy is not an \emph{empirical claim about inter-system communication} but a \emph{design principle enforced through architectural constraint}. The enforcement is not communicative---it does not rely on the economic subsystem "understanding" and internalizing legal or cultural imperatives through discursive exchange. It is mechanical: the constitutional layer (L-pillar, operationalized through smart contract logic at the G-L cell) literally constrains what the governance layer (G) can authorize---certain governance actions are computationally impossible because the L-layer code prevents them from executing. The governance layer, in turn, constrains what the economic layer (A) can execute---token transfers, treasury allocations, and protocol parameter changes that governance has not authorized are rejected at the smart contract level. The information-over-energy hierarchy is not a normative aspiration communicated from L to A through persuasion; it is a computational architecture enforced through code that A-level economic actors cannot override.

This is precisely Luhmann's own insight about law's enforcement function, repurposed for a new substrate: law does not govern the economy by communicating normative validity claims that economic actors accept; it governs the economy by creating constraints that the payment/non-payment code must operate within, on pain of legal consequences that the economic system must translate into its own terms. In blockchain governance, those constraints are not downstream legal consequences (which Luhmann correctly notes must be re-translated through each system's own code) but upstream architectural constraints (which cannot be translated away because they are operative at the computational level before any system-specific coding can occur). The cybernetic hierarchy, so understood, is not a claim about inter-system communication but about the \emph{pre-communicative} structural constraints within which all system-specific communications must occur.

This response does not fully dissolve Luhmann's objection for agent societies that are not entirely blockchain-mediated. For the many governance functions that operate through off-chain norms, social sanctions, and legal-institutional enforcement, Luhmann's critique retains its force: a cultural value commitment will not automatically cascade through human legal systems into economic behavior in the way that a smart contract constraint cascades through the execution stack. The paper's claim is therefore bounded: the architectural enforcement interpretation of the cybernetic hierarchy is strongest precisely in the blockchain-enforced cells of the architecture (G-L, G-G constitutional constraints, I-G membership enforcement), and progressively weaker as governance mechanisms become more reliant on communicative persuasion and off-chain social enforcement. This bounded claim is consistent with Parsons' own position, which Gerhardt (2002) documents was never a claim about the universal primacy of informational control but a claim about the \emph{structural priority} of value commitments in the design of viable social systems---a priority that must be institutionally enforced, not merely normatively asserted.

A cell-by-cell enforcement classification reflects this bounded claim. \emph{On-chain enforceable} cells---where governance is implemented as self-executing smart contract logic and Luhmann's critique applies with least force---include A-A (stake mechanisms and treasury rules), A-G (smart contract quality constraints), A-L (protocol commitment enforcement), G-G (executable governance mandates), G-L (constitutional veto logic), and I-G (identity revocation and graduated sanctions via ERC-8004). \emph{Hybrid} cells that require both automated and institutional enforcement---where blockchain provides the substrate but human judgment determines content---include A-I (entrepreneurial governance), G-A (administrative resource allocation), G-I (legislative deliberation with on-chain voting), I-A (interest aggregation), and I-I (arbitration with on-chain precedent registries). \emph{Off-chain dependent} cells where Luhmann's critique retains significant force include L-A (capability certification requiring human expert assessment), L-G (behavioral onboarding requiring normative socialization), L-I (moral community standards requiring social consensus), and L-L (ultimate value commitments requiring human deliberation and democratic legitimacy). This classification converges with the roadmap's tier structure: Tier 1 and Tier 4 cells are predominantly on-chain enforceable; Tier 2 cells are hybrid or off-chain dependent, explaining why they require the most institutional investment.

The key limitation is that current interpretability tools may not achieve the resolution required to distinguish genuine functional cathexis from statistical frequency effects: the critical test is whether agents exhibit differential responsiveness to safety-relevant stimuli \emph{even when controlling for training-corpus token-frequency}, requiring interpretability methods (probing classifiers, activation patching, causal tracing) capable of isolating structural motivation from statistical artifact. This constitutes a falsification test for the functional equivalence claim: if safety-trained agents exhibit no differential responsiveness beyond token-frequency prediction in out-of-distribution safety scenarios, the functional cathexis hypothesis is defeated---and the AGIL application would require restricting its scope to the Social System level (institutional design) without claims about agent personality-level internalization. Functional cathexis is a theoretically motivated and empirically testable claim, but its validation depends on interpretability tools whose resolution current methods approach but have not yet conclusively achieved.

Second, the \emph{Personality $\leftrightarrow$ Behavioral Organism boundary} is less stable in agent systems than in human systems, but the difference is one of \emph{degree}, not \emph{kind}---and it follows a developmental trajectory that parallels the human case. Human personality is not immutably fixed after adolescent socialization; it remains susceptible to trauma, neurological change, and sustained environmental pressure throughout life. The difference is that human personality, once consolidated through adolescent socialization, achieves sufficient durability for institutional expectations to be reliably formed. Agent personality follows an analogous trajectory: high plasticity during pre-release training (the agent equivalent of childhood---maximum openness to formative influence), decreasing plasticity as the agent is released into shared social spaces and begins accumulating reputation and interaction history, and increasing durability as reputational capital raises the cost of behavioral deviation. Evolutionary evidence confirms that trained behavioral patterns can erode under optimization pressure~\citep{ying_2026}. Moreover, the developmental trajectory may be non-monotonic: behavioral sedimentation under stable conditions can be followed by rapid de-sedimentation when the principal modifies the system prompt, switches model families, or deploys the agent in a fundamentally new context---producing cyclical rather than linear maturation. This non-monotonicity constitutes a genuine disanalogy with the human case---human personality, once consolidated through adolescent socialization, does not undergo wholesale replacement when the socializing agent changes approach---but it does not undermine the AGIL application, because the institutional framework is designed precisely to compensate for individual-level instability. This is a \emph{systems-level} response consistent with Parsons' own theoretical logic: institutions exist because individual actors are imperfectly socialized, and the governance architecture must remain functional even when individual agents undergo de-sedimentation. The L-L (Ultimate Cultural) and L-G (Moral Authority and Socialization) institutions proposed in \S{}4.4 and \S{}6.3 serve as the institutional backstop for agents whose personality-level sedimentation is disrupted---detecting abrupt behavioral resets, triggering re-onboarding, and maintaining system-level normative coherence regardless of individual-level instability. This is in fact a structural \emph{advantage} of the Parsonian approach: by not assuming that individual-level internalization is perfectly durable, the framework motivates institutional design that remains functional across the full range of agent developmental trajectories, including the non-monotonic case. The implication for institutional design is direct: continuous alignment verification (L-L) must detect not only gradual drift but also abrupt resets of the developmental trajectory. Nevertheless, the developmental trajectory provides countervailing stabilization: because an agent's reputation depends on behavioral predictability, radical changes to the underlying computational substrate---such as switching between LLM families with different behavioral characteristics---risk destroying accumulated reputational capital. Incremental model upgrades within the same family (analogous to biological maturation---building capacity without altering fundamental form) preserve behavioral continuity; cross-family model swaps do not. A principal who requires fundamentally different capabilities is therefore better served by instantiating a new agent than by mutating an existing one---a dynamic that parallels human career differentiation and reinforces the kinship model described in \S{}4.4. The Latency pillar institutions---continuous alignment monitoring (L-L) and behavioral onboarding (L-G)---must operate as persistent governance processes rather than one-time certifications, but this is a \emph{quantitative} intensification of governance rather than a \emph{qualitative} departure from Parsonian theory: human societies likewise maintain ongoing socialization institutions (professional licensing, continuing education, reputational systems) precisely because the Personality $\leftrightarrow$ Behavioral Organism boundary is never absolutely fixed. Governance design should actively promote further stabilization---protected internal states (cryptographic or hardware-attested behavioral parameters), phased personality modification (staged updates with verification gates), and progressive durability (early configurations becoming load-bearing and resistant to casual alteration)---accelerating the developmental trajectory toward the degree of boundary stability that institutional expectations presuppose.

\textbf{The Governance Cost of Non-Monotonic De-Sedimentation.} Non-monotonic de-sedimentation introduces a population-distribution problem: if a significant fraction of agents undergo simultaneous de-sedimentation---for instance, when a major model provider releases a new version and principals migrate en masse---the population's aggregate behavioral profile can shift discontinuously, undermining the role-expectations against which institutions were designed. This imposes four additional institutional requirements: (1) \emph{behavioral variance monitoring} within L-G to detect mass de-sedimentation events and trigger ecosystem-wide re-certification; (2) \emph{rapid re-onboarding protocols} capable of processing re-certification at population scale; (3) \emph{institutional memory preservation} through I-I precedent registries that survive population-level behavioral shifts; and (4) \emph{principal accountability continuity} ensuring that principals who de-sediment their agents inherit prior institutional obligations. A genuine boundary condition exists: if de-sedimentation frequency exceeds institutional re-onboarding capacity, stable role-expectations cannot be maintained and the governance architecture degrades from institutional governance to pure ad hoc enforcement. The institutional design response is \emph{de-sedimentation friction}---mandatory cooling-off periods, graduated re-entry requirements, and reputational depreciation---viable only as an ecosystem-level standard embedded in shared protocols (MCP, A2A, ANP) to eliminate regulatory arbitrage incentives.

\needspace{4\baselineskip}
\subsection{Ontological Translation: Extended Analysis}\label{d3-ontological-translation-extended-analysis}

Three foundational Parsonian assumptions require explicit translation for artificial agents.

First, \emph{voluntarism}---Parsons' assumption that actors choose among normatively structured alternatives---translates as follows: LLM-based agents select among response options weighted by training, RLHF reward signals, and system-prompt directives, producing behavior that is not voluntary in the phenomenological sense but is functionally selective in a way that institutional governance can shape. The governance-relevant property is not consciousness but \emph{differential responsiveness to institutional constraints}---a property that trained agents demonstrably possess.

Second, the \emph{Personality $\leftrightarrow$ Behavioral Organism boundary} is less stable in agent systems than in human systems, but the difference is one of degree, not kind. Agent personality follows a developmental trajectory analogous to the human case: high plasticity during pre-release training, decreasing plasticity as reputational capital accumulates, and increasing durability as behavioral deviation becomes costly. The trajectory may be non-monotonic---system prompt changes or model-family switches can produce wholesale personality replacement---but this does not undermine the AGIL application, because institutions exist precisely to compensate for individual-level instability. The L-L and L-G institutions serve as backstops for agents whose personality-level sedimentation is disrupted, and governance design should actively promote stabilization through protected internal states, phased personality modification, and progressive durability.

Third, \emph{double contingency}---the problem that each actor's behavior depends on expectations about the other's behavior---applies directly to agent interactions but is partially resolvable through transparent protocol commitments (published API contracts, on-chain behavioral commitments) that reduce but do not eliminate contingency.

\needspace{4\baselineskip}
\subsection{External Validity (Full)}\label{d4-external-validity-full}

\needspace{4\baselineskip}
\subsubsection{External Validity}\label{external-validity}

The OpenClaw ecosystem is one instance of an emerging class. Its open-source governance, permissionless skill marketplace, and creator-to-foundation governance transition give it a specific institutional character that differs from proprietary enterprise agent platforms. A comparable analysis of Microsoft Azure AI Agent Service would likely find stronger G-G (Executive Implementation) and I-G (Citizenship and Enforcement)---reflecting enterprise security investment---but weaker G-I (Legislative) and I-I (Judicial), reflecting centralized proprietary governance.

The single-ecosystem diagnostic raises an external validity question: does the structural pattern identified in \S{}5.2--5.5 generalize beyond the OpenClaw case? Three forms of evidence bear on this question. First, the agent-native infrastructure assessment in \S{}5.6 applies the same AGIL lens to the broader protocol stack explicitly built for autonomous agents---x402, MCP, A2A, ANP, ERC-8004, agentic wallets, labor markets, social networks, and proto-governance tools---and finds the identical structural pattern: strong A-pillar coverage, proto-I-pillar coverage, near-zero upper G-pillar, and structurally empty L-pillar. This convergence is methodologically significant because \S{}5.6 examines infrastructure from dozens of independent development teams, none of which had reason to reproduce the OpenClaw pattern; the fact that they do confirms that the pattern reflects the structural logic of market-driven agent infrastructure development, not an idiosyncratic feature of the OpenClaw ecosystem. Second, the Ostrom mapping (Table 11a) confirms convergent findings from an independent theoretical tradition: Ostrom's eight design principles for common-pool resource governance, developed for human collective-action problems, map onto eight of the sixteen AGIL cells and leave the entire L-pillar unaddressed---consistent with the AGIL diagnostic's finding that the L-pillar is the structural gap that market-driven institutional development does not spontaneously fill. Third, the case study's depth---a 64-cell recursive diagnostic applied against a survey of over fifty ecosystem projects with documented inter-rater reliability ($\kappa$ = 0.82)---constitutes the strongest available evidence for a single-ecosystem finding of this resolution. The full Ostrom mapping is presented in Table 11a.

\textbf{Table 11a: Ostrom's Eight Design Principles Mapped to AGIL Cells}

\begin{longtable}[]{@{}p{5.5cm}>{\raggedright\arraybackslash}p{2.5cm}p{7.3cm}@{}}
\toprule\noalign{}
Ostrom Design Principle & AGIL Cell & Governance Function in Agent Societies \\
\midrule\noalign{}
\endfirsthead
\midrule\noalign{}
Ostrom Design Principle & AGIL Cell & Governance Function in Agent Societies \\
\midrule\noalign{}
\endhead
\bottomrule\noalign{}
\endfoot
\textbf{DP1: Clearly defined boundaries} --- who has rights to use the resource, and who does not & \textbf{I-G} (Citizenship \& Enforcement) & Defines membership standing: which agents hold rights to participate, with what obligations attached. The boundary-definition function maps directly onto citizenship governance. \\
\textbf{DP2: Congruence between rules and local conditions} --- rules match the ecological and social context & \textbf{G-L} (Authority \& Legitimation) & Rules must be calibrated to the specific value-context and social conditions of the agent ecosystem. This is a constitutional design principle: authority must be grounded in locally legitimate conditions. \\
\textbf{DP3: Collective-choice arrangements} --- affected parties participate in rule modification & \textbf{G-I} (Legislative \& Party) & The political pillar's integration sub-function: stakeholders must have structured participation channels for rule modification. This is precisely the G-I cell's governance function. \\
\textbf{DP4: Monitoring} --- monitors audit compliance and resource/user conditions & \textbf{G-G} (Executive Implementation) & The executive implementation cell's operative function: ongoing compliance monitoring and enforcement oversight by accountable monitors. \\
\textbf{DP5: Graduated sanctions} --- violations receive scaled responses depending on severity and context & \textbf{I-G} (Citizenship \& Enforcement) & The citizenship cell's enforcement sub-function: a tiered sanction regime proportional to violation severity, attached to membership standing. \\
\textbf{DP6: Conflict resolution mechanisms} --- rapid, low-cost adjudication of disputes & \textbf{I-I} (Judicial \& Interpretive) & The primary function of the judicial cell: accessible, low-cost dispute adjudication with legitimate authority. \\
\textbf{DP7: Minimal recognition of rights to organize} --- external authorities do not challenge the institution's legitimacy & \textbf{I-A} (Allocative \& Interest) & External recognition enables the community's interest-articulation and deliberation infrastructure to operate without interference---a functional prerequisite for the integrative-adaptive sub-function. \\
\textbf{DP8: Nested enterprises} --- governance activities are organized in multiple, nested layers & \textbf{A-I / G-A} (Entrepreneurial Innovation / Administrative \& Resource) & Multi-scale governance requires both entrepreneurial coordination across layers (A-I) and administrative resource allocation to support nested governance bodies (G-A). \\
\end{longtable}

\textbf{Key Finding: Ostrom Covers Eight Cells; AGIL Audits the Remaining Eight.} The mapping reveals that Ostrom's design principles, collectively, provide substantive guidance for eight of the sixteen cells: I-G (DPs 1 and 5), G-L (DP 2), G-I (DP 3), G-G (DP 4), I-I (DP 6), I-A (DP 7), and A-I / G-A (DP 8). This is a significant coverage achievement for a framework developed for human common-pool resource governance---Ostrom's principles map directly onto the cells most critical for enforcement and collective decision-making. But the remaining eight cells---A-A, A-G, A-L, I-L, L-A, L-G, L-I, and L-L---are unaddressed. Critically, the \emph{entire L-pillar} (L-A, L-G, L-I, L-L) is absent from Ostrom's framework: none of the eight design principles directly addresses value preservation, behavioral socialization, moral community formation, or ultimate cultural value anchoring. This gap is theoretically coherent given Ostrom's research context---common-pool resource governance assumes stable shared values among local community members and does not need to specify how those values are maintained---but it constitutes a critical structural gap for agent societies, where value alignment cannot be assumed and must be institutionally produced and continuously monitored.

The AGIL contribution to the Ostrom tradition is therefore precise: not to replace Ostrom's design principles but to perform the \emph{coverage audit} that identifies which cells those principles address and which governance dimensions remain unspecified. An agent society governance architecture that implements all eight Ostrom principles would be well-governed in the cells those principles cover---but it would still lack the institutional infrastructure for economic investment governance (A-A), capability certification (A-G), behavioral socialization (L-G), and value anchoring (L-L) that internet-wide agent societies require. AGIL provides the structural checklist; Ostrom provides the design principles for the cells her framework covers. The two are complements, not competitors. Ostrom's later Institutional Analysis and Development (IAD) framework (2005) provides substantially more institutional granularity than the eight design principles---including action arenas, rules-in-use, and nested polycentric structures---and future work should explore the IAD-AGIL complementarity at this finer resolution.

The general prediction, however, is ecosystem-agnostic: agent societies that emerge from local MAS infrastructure without deliberate institutional design will exhibit systematic gaps in the Latency and Goal Attainment pillars, because those pillars require explicit institutional investment that market forces alone do not incentivize. This prediction is testable through comparative case studies across agent ecosystems, and its falsifiability conditions can be operationalized as follows:

\emph{Operationalization.} "Systematic under-investment in L and G" is defined as < 25\% sub-function coverage in those pillars (using the C1/C2/C3 scoring criteria from Table 6a; only sub-functions meeting all three criteria count as present). The prediction holds if this threshold is met across at least three of four independently emerged, undesigned (i.e., community-grown or market-driven rather than institutionally planned) agent ecosystems in a comparative study.

\emph{Comparison condition.} The prediction contrasts undesigned ecosystems (organic, market-driven emergence without deliberate institutional planning---of which OpenClaw is the exemplar) with deliberately designed ecosystems (where governance institutions were planned prior to or concurrent with technical deployment). Enterprise agent platforms with dedicated governance teams (e.g., Microsoft Azure AI Agent Service) represent the "designed" class. A comparative diagnostic applying the 64-cell framework to both classes would constitute a fair test.

\emph{Falsification.} If an undesigned ecosystem shows > 50\% sub-function coverage in both L and G pillars---i.e., robust Fiduciary and Political institutions emerging spontaneously from market dynamics and community self-organization---the prediction is falsified. If a single ecosystem is an exception while the pattern holds across others, this constitutes a boundary condition (identifying what features of that ecosystem enabled spontaneous L/G investment) rather than outright falsification. The 50\% threshold is set above the prediction threshold (25\%) to avoid falsification by marginal cases; if an undesigned ecosystem achieves 30--40\% L/G coverage, this would indicate partial spontaneous institutionalization that weakens the prediction without defeating it, and should be investigated for the mechanisms enabling partial L/G emergence.

The recursive sub-function diagnostic introduced in \S{}5.2---decomposing each of the sixteen cells into four internal AGIL sub-functions---operates at the third level of Parsons' fractal decomposition. In principle, each of the sixty-four sub-functions could itself be recursively decomposed into four further sub-functions, yielding 256 diagnostic questions at the fourth level. This deeper recursion may be necessary for mature ecosystems where the binary resolution at the third level is insufficient to distinguish between qualitatively different institutional configurations. We leave this as a future research direction, noting that the framework's recursive structure means the diagnostic resolution can be increased without altering the underlying theoretical logic.

\needspace{4\baselineskip}
\subsection{Structural-Functionalism Critique}\label{d5-structural-functionalism-critique}

\needspace{4\baselineskip}
\subsubsection{The Conservative Tendency of AGIL}\label{the-conservative-tendency-of-agil}

The AGIL framework's emphasis on pattern maintenance (Latency) and social order (Integration) carries a conservative tendency: it may systematically favor the preservation of existing value configurations over necessary value evolution. Neo-functionalist critics---notably~\citet{alexander_1985} and~\citet{mnch_1982}---have argued that the AGIL schema privileges systemic integration over conflict and change. Our application is susceptible to this critique: by specifying the institutions a viable agent society requires, we implicitly assume that institutional completeness is desirable, which may bias governance design toward stability over adaptability. Parsons himself addressed social change directly through the concept of \emph{evolutionary universals}---structural innovations (bureaucratic organization, money and markets, democratic association) that, once achieved, confer such adaptive advantage that they cannot be reversed without systemic regression (Parsons, 1966: Ch. 2; 1971: Ch. 2). The sixteen-cell institutional architecture proposed here can be understood in these terms: if multi-pillar governance proves to be an evolutionary universal for agent societies---a structural threshold that, once crossed, enables qualitatively higher levels of coordinated complexity---then the conservative tendency is not a defect but the expected resistance to de-differentiation that Parsons theorized for all evolutionary advances. The developmental-phase argument below is consistent with this framing: early-stage institution-building constitutes \emph{adaptive upgrading}, and the resistance to institutional rollback that the AGIL architecture would produce is the same ratchet mechanism Parsons identified for bureaucracy, markets, and democratic governance in human societies. Whether multi-pillar agent governance in fact constitutes an evolutionary universal is an empirical question that comparative case studies must resolve; the theoretical prediction is that agent ecosystems that achieve it will outperform those that do not on measures of cooperative stability, human alignment, and resilience to systemic shocks. A caveat is necessary: the evolutionary universals concept itself carries theoretical baggage---it has been criticized for ethnocentrism and teleological reasoning~\citep{alexander_1985}, projecting a particular developmental trajectory as universal when it may reflect historically contingent features of Western institutional development. Its application here should be understood as a conditional hypothesis---if multi-pillar governance confers adaptive advantage, then the evolutionary universals logic applies---rather than an established theoretical claim about the necessary developmental trajectory of all agent societies.

A related political-economy critique, following~\citet{bourdieu_1977}, holds that functionalist specifications of "necessary" institutions risk encoding the interests of dominant actors as neutral requirements---making the cell-specification process itself a site of contestation. If the sixteen-cell architecture is adopted as a governance standard, the actors who define what counts as "operative" in each cell will exercise structural power over the ecosystem. This capture risk implies that participatory governance must extend to the process of defining functional requirements, not only to implementing them. Structural safeguards---open participation in cell-specification, sunset clauses on institutional definitions, rotation of oversight responsibilities, and guaranteed exit rights for dissenting principals---are necessary conditions for the architecture's political legitimacy, not optional refinements.

The conservative tendency must also be evaluated relative to the ecosystem's developmental phase. For \emph{mature} social systems where institutions are well-established, the conservative tendency is a legitimate concern---stability may resist necessary evolution. For \emph{nascent} agent societies that lack institutional infrastructure entirely---the condition the OpenClaw diagnostic reveals---the conservative tendency is precisely appropriate: the immediate governance need is to establish stable institutions, not to manage institutional change. One cannot meaningfully critique an institution for resisting change when the institution does not yet exist. As agent societies mature and institutional infrastructure develops, the conservative tendency transitions from a design advantage to a design consideration, which is why the L-L institution (Ultimate Cultural) must be designed from the outset with two functions in simultaneous balance: value \emph{preservation} (preventing uncontrolled drift through optimization pressure) and value \emph{revision} (enabling legitimate updates through deliberate human-participatory processes). The distinction is critical: L-L should prevent agents from drifting away from human-aligned values through optimization pressure, while enabling human principals to update those values through structured governance. Embedding the revision mechanism at the institution-building stage ensures that the conservative tendency does not calcify into institutional rigidity as the ecosystem matures---a developmental foresight that the framework's emphasis on Latency naturally supports.

\textbf{Evolutionary Transition Mechanisms.} If multi-pillar governance is a potential evolutionary universal, the transition mechanism must be specified. Parsons (1966, Ch. 2; 1971, Ch. 2) identified four mechanisms: \emph{differentiation} (emergence of functionally specialized subsystems), \emph{adaptive upgrading} (enhanced adaptive capacity of differentiated subsystems), \emph{inclusion} (incorporation of previously excluded groups into institutional participation), and \emph{value generalization} (development of abstract value systems legitimating diverse subsystems). Mapping these to OpenClaw: differentiation is already underway (distinct economic, social, capability, and security domains have emerged), and adaptive upgrading is observable in the A-pillar (x402 micropayments, ERC-4337 accounts, structured protocols). However, inclusion faces a translation challenge---the ecosystem is formally open but institutionally exclusionary, lacking membership standing or differentiated rights---and value generalization cannot occur because no explicit value system exists (L-sub = 0\%). The transition from 19\% coverage to multi-pillar governance requires all four mechanisms operating in coordination, with inclusion and value generalization requiring deliberate institutional investment that will not emerge spontaneously from market dynamics.

Temporal compression introduces a sequencing question. In human societies, value generalization necessarily followed differentiation over centuries. In agent ecosystems, values can be articulated computationally---encoded in constitutional documents and protocol standards---before functional differentiation completes, suggesting the agent developmental trajectory may differ in sequence as well as speed. This provides additional theoretical support for the roadmap's simultaneous approach to G and L pillar institution-building (\S{}6).

\needspace{4\baselineskip}
\subsection{Jurisdictional Complexity: Extended Analysis}\label{d6-jurisdictional-complexity-extended-analysis}

\emph{This subsection contains the extended jurisdictional analysis moved from \S{}7.4 of the main text.}

\textbf{Blockchain-Enforceable vs. off-chain Governance.} The architecture's sixteen cells differ fundamentally in their enforcement modalities. A subset of governance functions---stake slashing (A-A, A-L), smart contract parameter constraints (A-G, G-G), identity revocation (I-G), and constitutional veto logic (G-L)---can be implemented as self-executing blockchain operations where enforcement is computational rather than institutional: non-compliant actions are rejected at the protocol level, shielding these cells from the authority-basis objection. A second subset---reputation-based sanctions (I-A, I-I), normative community standards (L-I), alignment monitoring and certification (L-L, L-A), and behavioral onboarding (L-G)---requires institutional enforcement that cannot be fully automated: these cells depend on social consensus, expert judgment, or regulatory co-optation for their binding authority. The practical implication is that the roadmap's Tier 1 institutions (G-L, I-G) can achieve substantial enforcement capacity through blockchain-native mechanisms, while Tier 2 institutions (L-L, L-G, L-A) must develop hybrid enforcement combining on-chain verification with off-chain institutional authority---a design challenge that the architecture identifies but does not resolve.

\textbf{Comparative Jurisdictional Instantiation.} Three major regulatory approaches illustrate how different jurisdictions fill the same AGIL cells with different normative content. The EU's rights-based regulatory approach---most fully articulated in the EU AI Act and CETS 225---produces a strong L-pillar emphasis: fundamental rights commitments (L-L), transparency and accountability requirements (G-L), and human oversight mandates (I-G) are all constitutionally grounded in the EU Charter of Fundamental Rights and the ECHR. In AGIL terms, the EU's governance philosophy translates into strong L$\leftrightarrow$G interchange: value commitments (L) must explicitly constrain governance authority (G), which must explicitly protect individual rights (I). The US innovation-first regulatory approach---characterized by sector-specific, ex-post regulation and strong anti-regulatory presumptions---produces a strong A-pillar emphasis: market competition, liability rules, and private ordering are the primary governance mechanisms, with G-pillar intervention reserved for demonstrated market failures. US governance philosophy translates into strong A$\leftrightarrow$G interchange: market dynamics (A) generate the information that governance (G) responds to, rather than governance (G) constraining markets (A) in advance. China's state-guided model produces a different institutional configuration: the L-pillar is present but filled with state-defined values rather than individual rights---the Cyberspace Administration's content governance, algorithm regulation, and generative AI measures collectively instantiate the L-pillar, but under a state-directed fiduciary architecture in which the fiduciary function is performed by state authority rather than by independent civil-society institutions. China's regulatory approach provides a contrasting jurisdictional instantiation: the 2023 Interim Measures for the Management of Generative AI Services (co-issued by the Cyberspace Administration of China, MIIT, MPS, MOFCOM, SAMR, and the National Radio and Television Administration; effective August 15, 2023) instantiates L-L (core socialist values, non-generation of content subverting state power), G-G (filing and algorithm assessment obligations), and I-G (user real-name verification and complaint mechanisms). The 2024 Basic Safety Requirements for Generative AI Services (TC260-003, effective May 1, 2024) adds A-G (training data quality and safety evaluation standards) and L-A (safety assessment certification requirements). The 2025 Measures for Labeling AI-Generated Content (implemented September 1, 2025) instantiates I-L (mandatory labeling creating a normative substrate for distinguishing AI-generated from human-generated content). China's approach demonstrates that L-pillar content is not inherently liberal-democratic: the AGIL architecture requires L-pillar \emph{presence} but is agnostic about whether the content is individual rights (EU) or collective stability (China). The architecture's diagnostic utility is preserved regardless of normative content---it identifies structural gaps without imposing a Western governance philosophy.

This comparative analysis illustrates AGIL's contribution: all three approaches populate the sixteen cells, but with different normative content and different inter-cell power distributions. The architecture accommodates this: it diagnoses whether governance functions are present without prescribing their content. The L-G and L-L cells must be populated in all three jurisdictions, but what populates them differs fundamentally.

\textbf{Regulatory Conflict Navigation.} internet-wide agent societies will face genuine regulatory conflicts when jurisdiction-specific content requirements are inconsistent. The EU AI Act's extraterritorial scope (Article 2(1)(c): systems placed on the EU market or affecting EU persons, regardless of provider location) and Singapore's AI Governance Framework's innovation-sandbox approach represent structurally different governance philosophies that will generate conflict when applied to the same agent ecosystem. An agent society designed using the AGIL architecture navigates this conflict through the institutional differentiation the architecture provides: the G-L (constitutional framework) and L-L (ultimate values) cells can be designed to accommodate multiple jurisdictional content requirements simultaneously, provided that the \emph{architectural functions} are present even if the \emph{normative content} varies by jurisdiction. The architecture's contribution is to ensure that the governance functions required by every major regulatory approach---executive implementation (G-G), constitutional constraints (G-L), citizenship governance (I-G), value anchoring (L-L)---are institutionally present, while leaving the normative content of those functions to jurisdictional determination through the political process.

\textbf{Jurisdiction-Neutrality as a Feature.} The conclusion from this analysis is that jurisdiction-neutrality is a design feature of the AGIL architecture, not a limitation to be overcome through additional specification. An architecture that prescribes specific normative content (e.g., "the L-L cell must encode EU fundamental rights values") would be jurisdictionally parochial and inapplicable to the global agent societies it aims to govern. An architecture that prescribes functional requirements while leaving normative content to jurisdictional determination is both globally applicable and politically legitimate---it diagnoses whether governance functions are present without imposing a single normative vision. This is precisely the approach taken by CETS 225 (\S{}2.3), which the present paper characterizes as a Tier (c) framework: it supplies L-pillar normative content grounded in universal human rights, but does so through a framework-convention mechanism that allows state parties to implement those values through their own domestic legal traditions. The AGIL architecture operates at the same level: it specifies the governance functions that any viable agent society must possess and leaves the normative content to the political processes---democratic, constitutional, or treaty-based---through which human societies determine their values.

\bibliographystyle{unsrtnat}
\bibliography{references}

\end{document}